\documentclass[reqno]{amsart}
\usepackage{style}
\usepackage{graphicx} % Required for inserting images
\usepackage[hidelinks]{hyperref} 
\usepackage[foot]{amsaddr}

\usepackage{todonotes} % remove later
\title[ML ENHANCED PARAMETRIC RANS AT FULL- AND REDUCED-ORDER LEVELS]{Machine Learning enhanced parametric Reynolds-averaged Navier-Stokes equations at the full- and reduced-order levels}
\author[D. Oberto, M. Strazzullo, S. Berrone]{Davide Oberto$^{1,*}$, Maria Strazzullo$^{2,*}$ and Stefano Berrone$^{2,*}$}
\date{\today}

\address{
$^1$ Mathlab, Mathematics area, SISSA, Via Bonomea 265, 34136, Trieste, Italy.}
\email{doberto@sissa.it, \{maria.strazzullo, stefano.berrone\}@polito.it}
\address{
$^2$ Politecnico di Torino, Department of Mathematical Sciences ``Giuseppe Luigi Lagrange'', Corso Duca degli Abruzzi 24, 10129, Torino, Italy.}
\address{
$^*$ INDAM, GNCS member.}

\begin{document}

\maketitle

\begin{abstract}
In this contribution, we focus on the Reynolds-\red{a}veraged Navier-Stokes (RANS) models and their exploitation to build reliable reduced\redred{-}order models to further accelerate predictions for real-time applications and many-query scenarios. Specifically, we investigate how machine learning can be employed to enhance the predictive capabilities of the model, both at the \red{full-order model} (FOM) and \red{reduced-order model} (ROM) levels.
    We explore a novel integration of these two areas. We generate the FOM snapshots, essential for ROM construction, using a data-driven RANS model: the $\nu_t$-Vector Basis Neural Network. This is the first time that \red{this} machine learning procedure \red{covers} a large parametric variation, and we propose tailored training strategies to increase the accuracy of the FOM model. At the ROM level, we compare the results obtained by standard \red{p}roper \red{o}rthogonal \red{d}ecomposition \red{(POD)} in an intrusive Galerkin setting (PODG) and POD \red{n}eural \red{n}etwork approach (PODNN). \red{The numerical validation is conducted on a classic turbulent square-duct flow benchmark, with the bulk Reynolds number as the sole varying parameter.}
    Our investigation reveals that the PODG method proves \red{to be} unstable and inaccurate for turbulent flow prediction, while PODNN demonstrates superior performance in terms of accuracy and computational efficiency.
\end{abstract}

\iffalse

\section*{Acknowledgements}
All authors are members of the Gruppo Nazionale Calcolo Scientifico-Istituto Nazionale di Alta Matematica (GNCS-INdAM). SB kindly acknowledges financial support by PNRR M4C2 project of CN00000013 National Centre for HPC, Big Data and Quantum Computing (HPC) CUP: E13C22000990001. DO acknowledges the support provided by the European Union-NextGenerationEU, in the framework of the iNEST-Interconnected Nord-Est Innovation Ecosystem (iNEST ECS00000043– CUP G93C22000610007) consortium.
Moreover, this study was carried out within the ``20227K44ME - Full and Reduced order modelling of coupled systems: focus on non-matching methods and automatic learning (FaReX)" project – funded by European Union – Next Generation EU  within the PRIN 2022 program (D.D. 104 - 02/02/2022 Ministero dell’Università e della Ricerca).% This manuscript reflects only the authors’ views and opinions and the Ministry cannot be considered responsible for them. 
DO and MS thank the INdAM-GNCS Project “Metodi numerici efficienti per problemi accoppiati in sistemi complessi” (CUP E53C24001950001). {MS thanks and the ECCOMAS EYIC Grant ``CRAFT: Control and Reg-reduction in Applications for Flow Turbulence".
\fi

\section{Introduction} \label{sec:Intro}

\red{The} Reynolds-averaged Navier-Stokes (RANS) equations are a common choice in industry to investigate turbulent flows \cite{Moin1997}, and they are deduced by formally applying a Reynolds operator to the Navier-Stokes (NS) equations \cite{Reynolds1895-pq}. RANS equations drastically decrease the massive computational cost required for the NS equations simulations based 
%by the discretization of the NS equations due to the 
on fine meshes induced by Kolmogorov dissipative scale \cite{Pope2000}. However, the application of the Reynolds operator yields a new unclosed tensor term in the RANS equations, the Reynolds \red{s}tress \red{t}ensor (RST), that needs to be modeled \cite{Wilcox} to close the system.

Despite having decades of history, classic RANS models are still inaccurate for some classes of flow due to the poor description of the effects of the RST on the mean ﬂow \cite{Craft1996}. Furthermore, they are facing a slow development period, if not even stagnation \cite{Oliver2011}. Additionally, in spite of the continuous growth of High Performance Computing, scale-resolving models, i.e., models capable of describing the main or all the flow features, such as \red{l}arge-\red{e}ddy \red{s}imulation (LES) and \red{d}irect \red{n}umerical \red{s}imulation (DNS), are computationally too expensive in industrial settings where tens or hundreds of simulations are required for the same flow typology to perform optimisation analysis.

Such motivations have pushed researchers to investigate the use \blue{of} \red{m}achine \red{l}earning (ML) techniques to define data-driven RANS models, aiming to be computationally cheap as RANS but having an accuracy that is comparable to LES or DNS. Within this framework, the standard setting for ML-based models is to train specific flow configurations using available LES or DNS data \cite{deZordoBanliat2024,McConkey_2021,Oulghelou2025,Pinelli2010,Xiao2020}. If on one side ML is a valuable asset for enhancing RANS simulations, on the other the generalizability of these techniques is still limited \cite{mcconkey2022generalizability}. This paper is a first step towards a broader parametric model structure where 
ML-based models can be effectively exploited %in a parametric setting 
where a flow configuration changes with respect to a physical parameter (for example the Reynolds number). In this framework, ML-models, once appropriately trained, are suited to generate hundreds of reliable simulations if compared to the high-fidelity data used during the training. 

{Another important technique when dealing with flows framed in a parametric setting is reduced order modelling (ROM), founded on the \emph{offline-online} splitting paradigm \cite{Quarteroni_ROM}. In the offline phase, many high-fidelity and computationally expensive full\red{-}order model (FOM) simulations are performed, spanning a chosen parametric space. Such simulations are generally obtained using classic discretization techniques such as \red{f}inite \red{v}olume (FV) or \red{f}inite \red{e}lement (FE) methods. From these simulations, usually referred to as \emph{snapshots}, it is possible to retrieve a parameter-independent basis that spans a low-dimensional space that approximates the full\red{-}order system. Many methods have been proposed to compute the reduced basis in the literature, and in this paper, we use the \red{p}roper \red{o}rthogonal \red{d}ecomposition (POD) \cite{Quarteroni_ROM}. Once the computational\red{ly} expensive offline phase is carried out, in the online phase, for any unseen parameter, an approximate solution belonging to the low-dimensional space is quickly computed in terms of coefficients of the reduced basis expansion.}

We investigate the possibility of integrating these two research areas. In particular, we generate the FOM snapshots required by ROM methods using the $\nut$-\red{v}ector \red{b}asis \red{n}eural \red{n}etwork ($\nut$-VBNN) data-driven RANS models  \cite{Oberto2025}, while we use ROM as a real-time tool capable to investigate novel parametric instance in a fast yet accurate way. To the best of the authors' knowledge, while the usage of ROM methods is well established in the computational fluid dynamics community (see, for instance, \cite{HIJAZI2020109513,IVAGNES2023127920,Ivagnes2025,Strazzullo2025,Zancanaro_2022_turbulent, Zancanaro_2022_laminar, Zoccolan2024, Zoccolan2025237} and the references therein), it is the first time that ML-based RANS and ROMs are leveraged together to generate solutions comparable to the DNS in an affordable computational time.

As a \red{proof of concept} numerical environment to test our procedure, we chose the flow in a square duct \red{with varying bulk Reynolds number}. Despite its geometric simplicity, standard eddy-viscosity RANS models fail in the description of the secondary motion that occurs in the cross-square section of the duct, while ML-based RANS models have been proven to achieve DNS-like accuracy for this flow phenomenology \cite{Fonseca_2022, Vinuesa2021, Rincon_2023, Wang2017}.
The common setting for training data-driven RANS models is to use data from simulations in the fully turbulent regime, only.
 In this work, however, we address the issue of dealing with a wider interval of Reynolds numbers 
coupling $\nut$-VBNN enhanced RANS simulations and the ROM methods.
In other words, the focus of the data-driven RANS model is \red{to provide accurate predictions not only in the turbulent regime, but also in the transitional regime occurring at lower Reynolds numbers.}

Once the snapshots for the ROM step are created by the $\nut$-VBNN enhanced RANS, we compare an intrusive, i.e., a projection-based model, and a non-intrusive, i.e., purely data-driven, ROM methodology. For the former, we use a POD-Galerkin (PODG) approach that mimics the SIMPLE splitting algorithm proposed at the FOM level, building on \cite{Stabile_2020, Zancanaro_2021}. For the latter, we employ the POD Neural Network (PODNN) approach, proposed in \cite{Hesthaven_2018}, to quickly infer the velocity field in the reduced space. This method directly predicts the parameter-dependent coefficients of the POD-based reduced basis expansion, without solving any reduced system.

The main novelties of this contribution are:
\begin{itemize}
    \item \red{the investigation of the effectiveness of ML-based RANS models in the transitional regime};
    \item the exploitation of the ML-based RANS snapshots to train ROM models, both intrusive and non-intrusive, to provide a general parametric representation;
    \item an experimental analysis on the effect of splitting the training data to focus separately on the two flow regimes, both at the full\redred{-} and reduced\redred{-}models.
\end{itemize}

Besides this introduction, the paper is structured as follows: in Section \ref{sec:RANS}, we briefly recall the closure problem for the RANS equations, present the $\nut$-VBNN method and discuss the numerical discretization of the FOM problem; in Section \ref{sec:ROM}, we introduce the ROM aspects that are exploited in this paper; Section \ref{sec:duct_flow} focuses on the main characteristics of flow in a square duct, with particular attention to the behaviour of the streamwise flow and the secondary motion for different parametric instances; this analysis is preparatory to Section \ref{sec:FOM_results}, where we describe how the $\nut$-VBNN model is used to generate the FOM simulations and the accuracy of the obtained velocity fields is carefully checked; in Section \ref{sec:ROM_results} we discuss the ROM results in terms of accuracy and computational time. Finally, in Section \ref{sec:Conclusions}, conclusions are drawn and possible perspectives discussed.

\section{RANS equations: closure and discretization} \label{sec:RANS}
This section describes the RANS model we employed to build the reduced surrogate, i.e.,
%Specifically, we introduce 
the steady RANS model, its discretization, and the ML technique used to enhance its accuracy. 
%Finally, we introduce the numerical discretization we rely on, together with the solver used.
\subsection{The Steady RANS equations}
The Steady RANS equations are a common choice when dealing with turbulent flows characterised by high Reynolds number $Re = U L / \nu$, being $U$ and $L$ characteristic velocity and length of the flow and $\nu$ the kinematic viscosity of the fluid. Considering the domain $\Omega$, Steady RANS equations read
\begin{equation} \label{eq:RANS}
\begin{dcases}
&\nabla \cdot \vel=0, \\
&\dive (\vel \otimes \vel) - \dive [ 2 \nu \bS] = - \nabla p -\dive \btau + \bm{b},
\end{dcases}
\end{equation}
being: $\vel$ and $p$ the Reynolds-averaged velocity and pressure fields, respectively, \cite{Pope2000,Wilcox}, from now on referred as velocity and pressure for the sake of brevity, $\bS=\frac{1}{2} [\nabla \vel + (\nabla \vel)^T]$ the symmetric part of the velocity gradient, $\btau = \overline{\vel' \otimes \vel'}$ the second order symmetric Reynolds stress tensor (RST), with $\vel'$ the fluctuating component of the velocity, and $\bm{b}$ any force term. We define the Reynolds \red{f}orce \red{v}ector (RFV) as $\bm{t} = \dive \btau$ \cite{CRUZ2019104258}. Depending on the flow, appropriate boundary conditions should be defined.

In a 3D setting, RANS equations contain ten unknowns and consist of four equations. Thus, some modelling to reduce the number of unknowns is required. To do so, the standard approach consists of writing the RST as a function of some averaged fields, i.e., $\btau = \btau(\vel,\nabla \vel,p,\nabla p, \dots)$ either explicitly, through an algebraic relation, or implicitly, as a solution of Partial Differential Equations (PDEs).

Among the algebraic models, we briefly discuss the \red{e}ddy-viscosity ones based on Boussinesq's hypothesis
\begin{equation} \label{eq:Boussinesq}
\btau = \frac{2}{3} k \ide - 2 \nut \bS,
\end{equation}
where $k = \frac{1}{2} \tr{(\red{\btau})}$ is the \emph{turbulent kinetic energy}, $\nut$ is the \emph{turbulent viscosity} scalar field, and $\ide$ is the identity \red{tensor}.
Once \eqref{eq:Boussinesq} is inserted into \eqref{eq:RANS}, one obtains
\begin{equation}\label{eq:RANS_nut}
\begin{dcases}
&\nabla \cdot \vel=0, \\
&\dive (\vel \otimes \vel) - \dive [ 2 (\nu + \nut) \bS] = - \nabla p + \bm{b},
\end{dcases}
\end{equation}
where, for the sake of brevity, we define the pressure as \red{$p \coloneqq p + \frac{2}{3} k$}.
Equations \eqref{eq:RANS_nut} are still not closed because the turbulent viscosity is unknown.

Several models have been proposed to overcome this issue, such as the Spalart-Allmaras \cite{SPALART_1992}, the $k-\omega$ \cite{Wilcox_2008} or the $k-\omega$ SST \cite{Menter1993} models. In this work, we employ the Launder and Sharma (LS) $k-\varepsilon$ model \cite{Launder1974}, based on the assumption
\begin{equation}
\nut = C_{\mu} \frac{k^2}{\varepsilon},
\end{equation}
being $C_\mu$ a known constant, classically $C_\mu=0.09$, and $\varepsilon$ the dissipation rate of the turbulent kinetic energy. Finally, the model is closed by writing two transport equations for $k$ and $\varepsilon$.
\subsection{The $\nu_t$-Vector Basis Neural Network}

In this work, we use the $\nu_t$-VBNN as ML model to close and enhance RANS equations. Initially proposed in \cite{Berrone_2022}, subsequently improved in \cite{Oberto2025}, and tested in \cite{OBERTO2025_PH}, it predicts through two feed-forward neural networks the turbulent viscosity $\nu_t$ and  the divergence of the orthogonal component of the RST with respect to $\bS$, %the component of the Reynolds force vector (RFV), defined as $\bm{t} = \dive \btau$ that is not taken into account in the turbulent viscosity term,
denoted by $\rfvNLdim = \dive (\btau + 2 \nut \bS)$. 

The idea of predicting an RFV-related field through the data-driven RANS model was proposed in \cite{CRUZ2019104258} and is justified by the better reliability of the RFV compared to the RST when using DNS results. As a matter of fact, in \cite{THOMPSON20161} it is shown that the RFV field retrieved from DNS simulations is less affected by statistical errors compared to the RST because the former can be obtained by first order statistics, namely through $\vel$, $p$ and their derivatives. Conversely, the RST tensor is less reliable because second order statistics must be computed.

Furthermore, we predict $\nut$ and $\rfvNLdim$ separately, inspired by \cite{Brener_Cruz_Thompson_Anjos_2021,Wu2019} to improve the conditioning of the obtained RANS system that, consequently, is less affected by errors induced by the model regression step. 

We infer $\tilde{\nut} = 1000\nut$ and $\rfvNL = k^{1/2} / \varepsilon \ \bm{t}^{\perp}$ to handle fields in the order of $O(1)$ or $O(10)$, helping, in this way, the learning procedure \cite{Ling2015,Wu2018}. Following \cite{Oberto2025}, $\tilde{\nut}$ is predicted by a neural network, while $\rfvNL$ is written as linear combinations of a specific vector basis $\{\bm{v}_k\}_{k=1}^{N_c}$, where the coefficients $\{c_k\}_{k=1}^{N_c}$ of such combination are the targets of the training process. Specifically, we have
\begin{subequations} \label{eq:nut_rfvNL}
\begin{align}
    &\tnut = \tnut(\lambda_1,\dots,\lambda_{N_i}), \label{eq:nut_lin_comb} \\ & 
    \rfvNL = \sum_{k=1}^{N_c} {c_k(\lambda_1,\dots,\lambda_{N_i}) \bm{v}_k},
    \label{eq:rfvNL_lin_comb}
\end{align}
\end{subequations}

where both $\tilde{\nut}$ and the coefficients $c_k$ depend on scalar fields $\lambda_i, \ i=1,\dots, N_i$, called \emph{invariants}, obtained by appropriate multiplications of the symmetric and antisymmetric parts of the velocity gradient, i.e., $\bS$ and $\bm{W} = \frac{1}{2} [\nabla \vel - (\nabla \vel)^T]$, respectively, and $\nabla k$. For a complete list of invariants and of the vector basis, we refer to Appendix \ref{sec:inv_vb}.\\

In the $\nut$-VBNN, the invariants are the inputs of the neural networks, while the outputs are either the turbulent viscosity or the coefficients. This specific setting ensures both Galilean invariance and independence of the chosen frame of coordinates \cite{Oberto2025, Berrone_2022}.
In practice, to use the $\nut$-VBNN model, first \red{a} RANS simulation \red{is performed using} a classic closure model, the LS $k-\varepsilon$ model in our case. Then, the invariants and the vector basis \red{are collected} to \red{be fed into} the $\nut$-VBNN model. Successively, once $\tilde{\nut}$ and $\rfvNL$ are predicted, the $\nut$ and $\rfvNLdim $ fields are retrieved and inserted into the RANS equations
\begin{equation}\label{eq:RANS_nut_rfvNL}
\begin{dcases}
&\nabla \cdot \vel=0, \\
&\dive (\vel \otimes \vel) - \dive [ 2 (\nu + \nut) \bS] = - \nabla p - \rfvNLdim + \bm{b},
\end{dcases}
\end{equation}
to obtain new velocity and pressure fields. \red{It is worth mentioning} that Equation \eqref{eq:RANS_nut_rfvNL} corresponds to \eqref{eq:RANS} with $\dive \btau = \rfvNLdim - \dive [2 \nut \bS]$, that is, a generalization of the Boussinesq's hypothesis in equation \eqref{eq:Boussinesq}. In this work, to speed up the online stage, we always set as initial condition of the $\nut$-VBNN based simulation the $\vel$ and $p$ fields obtained by the LS $k-\varepsilon$ model. Moreover, we train the $\nut$-VBNN in a parametric setting: the model is trained for different values of the \emph{bulk Reynolds number} ($Re_b$). A more detailed definition of $Re_b$ will follow in Section \ref{sec:duct_flow}. For now, we anticipate that,
    in the RANS model, the definition of the Reynolds number can be not trivial and different expressions are used in the literature for the same flow phenomenology. The Reynolds bulk number is the classical dimensionless quantity used to determine the regime of a flow in a square duct, i.e., the numerical test we focused on \cite{Pinelli2010,Pirozzoli2018}.
Once trained, the $\nut$-VBNN model is tested for $Re_b$ values not used for training.  Both the training stage and the inference stage are schematised in Algorithms \ref{alg:offline_nut_VBNN} and \ref{alg:online_nut_VBNN}, respectively.

\begin{algorithm}[h]
\caption{$\nut$-VBNN training}
\label{alg:offline_nut_VBNN}
\begin{algorithmic}[1]
\Require \textnormal{DNS data from $N$ simulations with $\{Re^p_b\}_{p=1}^N$ bulk Reynolds number values, RANS model to improve}
\Ensure \textnormal{trained $\nut$-VBNN model}
%\Statex
\For{$p = 1, \dots, N$}
\State{\textnormal{Perform RANS simulation with $Re_b^p$}}
\State{\textnormal{Compute $\{\lambda_i\}_{i=1}^{N_i}$ and $\{\bm{v}_j\}_{j=1}^{N_c}$}}
\State{\textnormal{Interpolate DNS target data on RANS meshes}}
\EndFor
\State{\textnormal{Split data into training and validation set}}
\State{\textnormal{Train for $\nut$ and $\rfvNLdim$ separately}}
\end{algorithmic}
\end{algorithm}

\begin{algorithm}[h]
\caption{$\nut$-VBNN inference}
\label{alg:online_nut_VBNN}
\begin{algorithmic}[1]
\Require \textnormal{$Re_b$ not seen in training stage}
\Ensure \textnormal{$\nut$-VBNN-enhanced simulation}
%\Statex
\State{\textnormal{Run LS $k-\varepsilon$ RANS simulation with $Re_b$}}
\Comment{\textnormal{Same RANS model of \emph{training} stage}}
\State{\textnormal{Compute $\{\lambda_i\}_{i=1}^{N_i}$ and $\{\bm{v}_j\}_{j=1}^{N_c}$}}
\State{\textnormal{Predict $\nut$ and $\rfvNLdim$ separately with the $\nut$-VBNN model}}
\State{\textnormal{Solve for \eqref{eq:RANS_nut_rfvNL} with LS $k-\varepsilon$ RANS solution as initial solution and $\nut$ and $\rfvNLdim$ fixed}}
\end{algorithmic}
\end{algorithm} 

\subsection{The Finite Volume discretization}
In this work, the RANS equations are discretised using the FV method \cite{FerzigerCFD,Moukalled_2016} using the OpenFoam library \cite{OpenFOAM}. 
The FV method deals with the integral form of the differential equations to obtain an algebraic expression in terms of cell-averages. Let us define the computational domain $\Omega$ and a \redred{tessellation} of it (i.e., a mesh), defined as $\mathcal{T} = \{\Omega_i\}_{i=1}^{N_h}$, where $\bigcup_{i=1}^{N_h} \overline{\Omega}_i = \overline{\Omega}$ and $\Omega_i \cap \Omega_j = \emptyset$ for $i \neq j$. Moreover, for the generic field $\varphi$, we define its cell-average over $\Omega_i$ as
\begin{equation*}
\varphi_i = \frac{1}{V_i} \int_{\Omega_i}{\varphi \ dV}.
\end{equation*}
The integral momentum equation of \eqref{eq:RANS_nut}, assuming $\bm{b} = \bm{0}$, reads:
\begin{equation} \label{eq:MomIntegral}
\sum_{i=1}^{N_h} \left [
    \int_{\Omega_i}{\dive (\vel \otimes \vel) \ dV} - 
    \int_{\Omega_i}{\dive [2 (\nu + \nut) \bS ] \ dV}  + 
    \int_{\Omega_i}{\nabla p \ dV}  \right ] = \bm{0},
\end{equation}
and, thanks to the Gauss theorem, the volumetric terms containing the divergence operator are recast as surface integrals over the border of the cells. For example, the convective term reads
\begin{equation}
    \int_{\Omega_i}{\dive (\vel \otimes \vel) \ dV} = \int_{\partial \Omega_i}{\vel \otimes \vel \cdot d\bm{S}} \approx \sum_{f}{\bm{S}_f \cdot \vel_f \ \vel_f}  = \sum_{f}{F_f \ \vel_f},
\end{equation}
where the summation is over the faces of $\Omega_i$, $\bm{S}_f = S_f \bm{n}_f$ is the vector with magnitude equal to the area of the $f$-th face and direction of its outward normal vector $\bm{n}_f$. Finally, $\vel_f$ is the value of the velocity at the centre of the face $f$. Consequently, $F_f$ is the velocity flux crossing the face $f$. 
Analogously, the viscous term reads%, defining as $\nu_{t}_{i}$ the cell-average over $\Omega_i$ of $\vu_{t}_i$,
\begin{equation*}
\int_{\Omega_i}{\dive [2 (\nu+\nut) \nabla \vel] \ dV} = \int_{\partial \Omega_i}{2 (\nu + \nut) \nabla \vel \ d\bm{S}} \approx (\nu+\nu_{{t}_i}) \sum_{f}{(\nabla \vel)_f \cdot \bm{S}_f}.
\end{equation*}
Finally, for the pressure-related term, the pressure gradient is computed using the Gauss theorem:
\begin{equation*}
\int_{\Omega_i}{\nabla p \ dV} = \int_{\partial \Omega_i}{p \ d\bm{S}} \approx \sum_{f}{p_f \bm{S}_f},
\end{equation*}
where $p_f$ \redred{is} the pressure at the centre of the face $f$.

We highlight that $\vel_f, (\nabla \vel)_f$ and $p_f$ are not available in a FV setting and, thus, must be approximated by interpolating the corresponding cell-averaged fields. In this work, all the chosen schemes are second-order accurate, and the used meshes are orthogonal and non-skewed, thus not affecting the order of the discretization of the method \cite{Bruno_2022,JasakPhD}. The number of degrees of freedom for the discretised RANS equations through the FV method is $N_h^{\vel} = 3 N_h$ for the velocity field and $N_h^p = N_h$ for the pressure field.

\subsection{The SIMPLE algorithm}
The discretised RANS equations can be written in matrix form as
\begin{equation} \label{eq:saddle_point}
    \begin{pmatrix}
        \bold{A}_h(\vel_h) & G_h \\ M_h & 0
    \end{pmatrix}
    \begin{pmatrix}
        \vel_h \\ p_h
    \end{pmatrix} = 
    \begin{pmatrix}
        \bm{b}_h \\ 0
    \end{pmatrix},
\end{equation} where the matrix $\bold{A}_h(\vel_h) \in \mathbb{R}^{N_h^{\vel} \times N_h^{\vel}}$ is associated to the discretization of the diffusive and convective terms of the RANS equations, $G_h \in \mathbb{R}^{N_h^{\vel} \times N_h^p}$ is the discretised pressure operator, $M_h \in \mathbb{R}^{N_h^p \times N_h^{\vel}}$ is the discretised divergence operator, and $\bm{b}_h \in \mathbb{R}^{N_h^{\vel}}$ represents any source term. %For the RANS counterpart, depending on the closure choice, it is sufficient to add the turbulent viscosity contribution in $\bold{A}_h(\vel_h)$ and/or consider additional source terms ($\dive \bm{\tau}$, $\rfvNLdim$, etc.).
Instead of dealing with the fully coupled system \eqref{eq:saddle_point}, we exploit the \red{s}emi-\red{i}mplicit \red{m}ethod for \red{p}ressure-\red{l}inked \red{e}quations (SIMPLE) algorithm \cite{Patankar_1972} to decouple the $\nut$-VBNN RANS system using an iterative strategy. \\
Let us assume that the target velocity field can be written as $\vel_h = \vel_h^* + \vel_h'$, where $\vel^*$ plays the role of a velocity predictor, that is not divergence-free, and $\vel_h'$ is interpreted as a correction to enforce the divergence-free condition. In addition, the pressure field is written as $p_h = p_h^* + p_h'$, where $p_h^*$ is the pressure value at the previous iteration. Finally, given the parameter instance $Re_b$, let us denote by $\nut$ and $\rfvNLdim$ the turbulent fields obtained by the $\nut$-VBNN model. Then, the $(k+1)$-th step of the SIMPLE algorithm for RANS equations can be schematised as follows:
\begin{enumerate} [label=\roman*)]
    \item the matrix $\bold{A}_h = \bold{A}_h(\vel_h^k)$ is split as sum of its diagonal part $\bold{D}_h$ and its extra-diagonal part $\bold{H}_h$, i.e., $\bold{A}_h = \bold{D}_h + \bold{H}_h$. We remark that $\bold{A}_h$ depends on $\nut$. The discrete velocity predictor $\vel^*_h$ is retrieved by solving
    \begin{equation} \label{eq:u_equation_simple}
        \left(\bold{H}_h + \frac{1}{\alpha_{\vel}} \bold{D}_h \right)\vel^*_h = - G_h p_h^k +\bm{b}_h + \left(\frac{1}{\alpha_{\vel}} - 1 \right) \bold{D}_h \vel_h^k,
    \end{equation}
    where $\alpha_{\vel} \in (0, 1]$, called under-relaxation factor, is usually kept fixed for all iterations and increases the diagonal dominance of the system. We remark that in \eqref{eq:u_equation_simple}, $\bm{b}_h$ depends on $\rfvNLdim$;
    \item all fluxes are computed at the cell boundaries using $\vel^*_h$ and the discrete operators are updated;
    \item assuming $\bold{H}_h \vel'_h = \bm{0}$, the velocity correction reads %\dob{Qua ho corretto un errore nella formula}
\begin{equation} \label{eq:u_correction}
    \vel_h' = -\bold{D}_h^{-1} G_h p_h'.
\end{equation}
To obtain the pressure correction $p_h'$ to enforce the diverge-free condition $M_h \vel_h = M_h (\vel_h^* + \vel_h') = 0$, the following Poisson equation is solved
\begin{equation} \label{eq:p_simple}
    - M_h (\bold{D}_h^{-1} G_h p_h') = - M_h \vel_h^*;
\end{equation}
    \item we define $p^{k+1} = p_h^k + \alpha_p p_h'$, being $\alpha_p \in (0 ,1]$ the under-relaxation factor for the pressure field defined to increase the stability of the method;
    \item exploiting \eqref{eq:u_correction}, the new velocity is
    \begin{equation} \label{eq:velocity_correction}
        \vel_h^{k+1} = \vel_h^* - \bold{D}_h^{-1} G_h p_h'.
    \end{equation}
\end{enumerate}

Algorithm \ref{alg:SIMPLE_FOM} synthesises the SIMPLE algorithm 
given some stopping criteria $sc$, usually based on the residual of  system and/or on the rate of change of the obtained velocity and pressure fields compared to the previous ones.
\begin{algorithm}[h]
\caption{The {SIMPLE} algorithm}
\label{alg:SIMPLE_FOM}
\begin{algorithmic}[1]
\Require \textnormal{$\vel_h^0$ and $p_h^0$, $Re_b$, $\alpha_{\vel}$, $\alpha_p$, number max of iterations $N_{max}$, $sc$}
\Ensure \textnormal{discrete solutions $\vel_h$, $p_h$}
\State{$k=0$}
\State{\textnormal{$\nut$, $\rfvNLdim \gets$ $\nut$-VBNN($Re_b$), see Algorithm \ref{alg:online_nut_VBNN}}}
\While{ \textnormal{not $sc$ and $k \le N_{max}$}}
\State{\textnormal{$u_h^* \gets$ Solve \eqref{eq:u_equation_simple} for the velocity predictor using $p_h^k$}}
\State{\textnormal{Obtain through interpolation the fluxes fields using $\vel_h^*$}}
\State{\textnormal{$p_h' \gets$ Solve \eqref{eq:p_simple} with $\vel_h^*$ and updated fluxes}}
\State{$p_h^{k+1} = p^k + \alpha_p p_h'$}
\State{\textnormal{$\vel_h^{k+1} \gets$ Correct the velocity using \eqref{eq:velocity_correction} }}
\State{$k \gets k + 1$}
\EndWhile
\State{\textnormal{Return $\vel_h = \vel_h^{k}$, $p_h = p_h^{k}$}}
\end{algorithmic}
\end{algorithm}

\section{Reduced\redred{-}order modelling for RANS Equations} \label{sec:ROM}
This section focuses on the ROM application to the proposed $\nu_t$-VBNN RANS model. First, the POD approach employed to build the low-dimensional framework \red{is described}. Next, the two algorithms \red{used} in the numerical results, i.e., the PODNN \cite{Hesthaven_2018} and the PODG SIMPLE \cite{Zancanaro_2021}\red{, are discussed}.

\subsection{The POD approach}
\label{sec:g-rom}
ROM techniques aim to create a surrogate model building on some high-fidelity information, i.e., FOM simulations. The core idea behind ROM is to initially invest computational effort in generating a low-dimensional model derived from FOM solutions. Despite its significantly reduced size, the system is still accurate and can be effectively and efficiently used to explore a wide range of parametric scenarios.
At the FOM level, we obtain the solutions by means of a trained $\nu_t$-VBNN RANS discretised through FV. Namely, we take $N_s$ snapshots of $\nu_t$-VBNN RANS simulations, for several $Re_b$ in the considered parametric range, denoted as $\mathcal P$, i.e., we consider $\{\bm{u}_h(Re_b^i)\}_{i=1}^{N_s}$ and $\{{p}_h(Re_b^i)\}_{i=1}^{N_s}$, for $Re_b^i \in \mathcal P$. For details on the POD algorithm, we refer to \cite{hesthaven2015certified, Quarteroni_ROM}. 
The POD yields a set of basis functions $\{\bm{\varphi}_j\}_{j=1}^{r_u}$ and $\{\psi_j\}_{j=1}^{r{_p}}$, for velocity and pressure, respectively. The reduced spaces spanned by these two sets of basis functions are $\mathbb U_r$ and $\mathbb Q_r$, and we write
$$
\mathbb U_{r} = \text{POD}(\{\bm{u}_h(Re_b^k)\}_{k=1}^{N_{s}}, r) \quad \text{and} \quad \mathbb Q_{r} = \text{POD}(\{p_h(Re_b^k)\}_{k=1}^{N_{s}}, r),
$$
assuming for the sake of presentation that {$r_u=r_p=r$}. \\
In the reduced setting, $\bm{u}_{r}(x,Re_b) = \bm{u}_{r} \in \mathbb U_{r}$ and ${p}_{r}(x,Re_b) = p_{r} \in \mathbb Q_{r}$ represent the reduced velocity and pressure fields, respectively, and they can be expanded as
$$ 
	\bm{u}_{r}
	= \sum_{j=1}^r a_j(Re_b) \bm{\varphi}_j(x) 
    \quad \text{and} \quad
  p_{r} 
	= \sum_{j=1}^r b_j(Re_b) \psi_j(x),
	\label{eqn:g-rom-1}
$$ 
for some reduced real coefficients $\{a_{j}(Re_b)\}_{j=1}^{r}$ and $\{b_{j}(Re_b)\}_{j=1}^{r}$, respectively. The vectors in $\mathbb R^r$, whose entries are defined by the reduced coefficients, are denoted by $a$ and $b$, for velocity and pressure. Finally, we define $\mathbf{U}_r \in \mathbb R^{N_h^{\boldsymbol{u}} \times r}$ and $Q_r \in \mathbb R^{N_h^p \times r}$ as the two matrices gathering as columns the reduced basis vector for the velocity and pressure field, respectively.

\subsection{PODNN method for RANS Equations}
Recently, thanks to the gain of popularity of ML and the increasing data-availability in computational sciences, pure data-driven approaches have been exploited. In this work, we use the PODNN \cite{Hesthaven_2018} as a non-intrusive approach. This method approximates the maps between the parameter space and the reduced coefficients through feed-forward neural networks. In our setting, the two approximated maps are
$$
    \pi_{\vel}: Re_b \mapsto a
\qquad \text{and} \qquad
    \pi_p: Re_b \mapsto b.
$$
Once the PODNN model is properly trained, the online phase consists of making inferences of the neural networks, completely disregarding the nature of the underlying physical model and avoiding all issues associated to nonlinear terms in the PDEs. \\
The reduced coefficients required for the training of the neural networks can be retrieved from the FOM snapshots by solving the following normal systems \cite{PICHI2023105813}
\begin{subequations} \label{eq:norm_system_podnn}
    \begin{equation} \label{eq:norml_system_u}
    \mathbf{U}_r^T \mathbf{X}^{\redred{\vel}}_h\mathbf{U}_r a(Re_b^k) = \mathbf{U}_r^T \mathbf{X}^{\redred{\vel}}_h \vel_h(Re_b^k),
    \end{equation}
    \begin{equation} \label{eq:norml_system_p}
        Q_r^T X^{\redred{p}}_h Q_r b(Re_b^k) = Q_r^T X^{\redred{p}}_h p_h(Re_b^k),
    \end{equation}
\end{subequations}
for $k = 1, \dots, N_s$, where $\mathbf{X}^{\redred{\vel}}_h \in \mathbb{R}^{N_h^{\vel} \times N_h^{\vel}}$ and $X^{\redred{p}}_h \in \mathbb{R}^{N_h^p \times N_h^p} $ are symmetric positive definite matrices \blue{\cite{Quarteroni2015-bh}}. In particular, in our work $X^{\redred{p}}_h$ is the $L^2$-norm inducing matrix, i.e., the diagonal matrix having as $i$-th diagonal component the cell volume $V_i$ \blue{\cite{Stabile2017CAF}}, while $\mathbf{X}^{\redred{\vel}}_h$ is the diagonal matrix having on the diagonal the same diagonal of $X^{\redred{p}}_h$ repeated three times. \blue{Equations \eqref{eq:norml_system_u} and \eqref{eq:norml_system_p} directly derive from solving for 
$$
a = \argmin_u\norm{\vel_h - \mathbf{U}_r u}_{\mathbf{X}^{\redred{\vel}}_h} \quad \text{and} \quad b = \argmin_q\norm{p_h - Q_r q}_{X^{\redred{p}}_h},
$$ respectively.
}\\
Further details on the loss definition for the PODNN are postponed to Section \ref{sec:ROM_results}.

\subsection{PODG SIMPLE method for RANS Equations}
A classic approach to obtain the reduced coefficients is to solve the reduced system deduced by Galerkin-projection of some full-order linear systems \cite{Quarteroni_ROM}. Several studies highlight the benefits of using consistent models at the FOM and ROM level \cite{INGIMARSON2022115620_consistency, Strazzullo_consistency}, and, thus, we adopt an iterative algorithm that mimics the full\redred{-}order SIMPLE strategy at the reduced level \cite{Zancanaro_2022_turbulent, Zancanaro_2022_laminar}, referred to as PODG SIMPLE. \\
Given an initial guess on the reduced coefficients vectors $a^0$ and $b^0$, the PODG SIMPLE algorithm follows the same steps of the SIMPLE algorithm by iteratively solving a linear system for the velocity predictor, followed by one for the pressure correction. %Such systems are obtained from the corresponding FOM ones assuming $\vel_h \approx \vel_r$ and $p_h \approx p_r$ and by projection using the $\mathbf U_{r}^T$ and $ Q_{r}^T$ matrices respectively. 
In particular, the reduced velocity predictor equation reads
\begin{equation} \label{eq:u_red_equation_simple}
    \mathbf{U}_r^T \left(\bold{H}_h + \frac{1}{\alpha_{\vel}} \bold{D}_h \right)\mathbf{U}_r a^* = \mathbf{U}_r^T \left[- G_h Q_r b^k +\bm{b}_h + \left(\frac{1}{\alpha_{\vel}} - 1 \right) \bold{D}_h \Big(\mathbf{U}_r a^k \Big) \right],
\end{equation}
where $a^*$ denotes the reduced coefficients of the velocity predictor, while the equation for the reduced coefficients of the pressure correction reads
\begin{equation} \label{eq:p_red_equation_simple}
    -Q_r^T M_h(\mathbf{D}_h^{-1}G_h Q_r b') = -Q_r^T M_h \mathbf{U}_r a^*.
\end{equation}
\begin{remark}
    One could use different under-relaxation factors in the reduced\redred{-order} model and the full\redred{-}order one. In this work, to mimic as much as possible the FOM framework, we use the same relaxation factors for the FOM and ROM.
\end{remark}
\begin{remark}
    To assemble the $\mathbf{H}_h$ and $\mathbf{D}_h$ matrices and the $\bm{b}_h$ vector, one should take into account the turbulent fields $\nut$ and $\rfvNLdim$. 
    In the numerical results, we opt for using the same turbulent fields of the FOM model. Namely, for each new parameter instance, the $\nut$-VBNN approach is used and, consequently, the FOM matrices are assembled. The rationale behind this choice relies on comparing the most consistent projection-based ROM model to the non-intrusive PODNN strategy. We anticipate that even this consistent choice will lead to poor velocity field approximations.    
\end{remark}
The whole PODG SIMPLE algorithm is presented in Algorithm \ref{alg:SIMPLE_ROM}.

\begin{algorithm}[h]
\caption{The {PODG SIMPLE} algorithm}
\label{alg:SIMPLE_ROM}
\begin{algorithmic}[1]
\Require \textnormal{$a^0$ and $b^0$, $\mathbf{U}_r$ and $Q_r$, $Re_b$, $\alpha_{\vel}$, $\alpha_p$, $N_{max}$, $sc$}
\Ensure \textnormal{reduced coefficients $a$, $b$}
\State{$k=0$}
\State{\textnormal{$\nut$, $\rfvNLdim \gets$ $\nut$-VBNN($Re_b$)}}
\While{\textnormal{not $sc$ and $k \le N_{max}$}}
%\State{Solve for turbulent fields ($\nut$, $\bm{\tau}$, \dots) using $u_h^k$ and $p_h^k$} 
\State{\textnormal{$a^* \gets$ Solve \eqref{eq:u_red_equation_simple} for the velocity predictor using $b^k$}}
\State{\textnormal{Obtain through interpolation the fluxes fields using $a^*$}}
\State{\textnormal{$b' \gets$ Solve \eqref{eq:p_red_equation_simple} with $a^*$ and updated fluxes}}
\State{$b^{k+1} = b^k + \alpha_p b'$}
\State{\textnormal{$a^{k+1} \gets$ Correct the velocity using $b'$}}
\State{$k \gets k + 1$}
\EndWhile
\State{\textnormal{Return $a = a^{k}$, $b = b^{k}$}}
\end{algorithmic}
\end{algorithm}

\section{The Duct Flow benchmark} \label{sec:duct_flow}
In this section, the essential insights on the phenomenology of the flow in a square duct \red{are provided} to better analyse the FOM and ROM results in the next sections.

\subsection{Flow phenomenology}
\begin{figure}[h]
\centering
\includegraphics[width=0.7\textwidth] {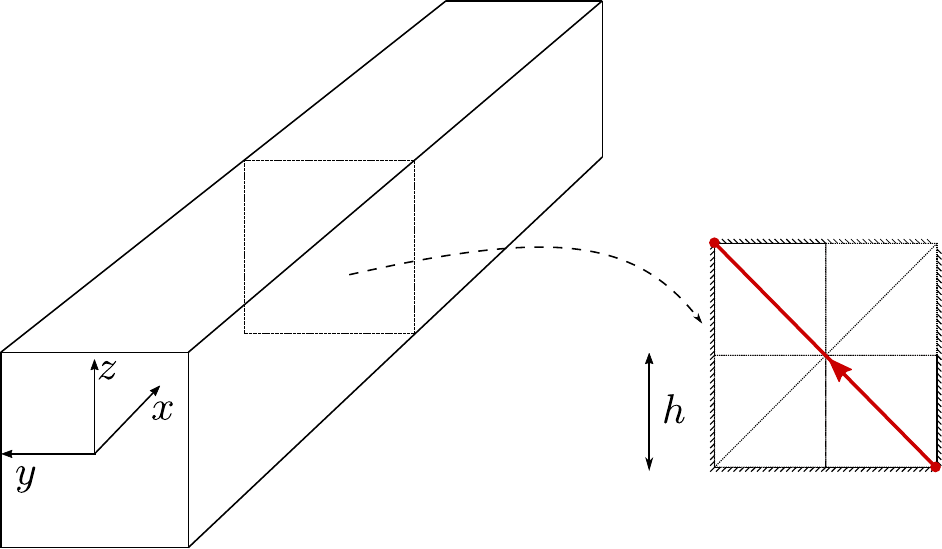}
\caption{Square duct geometry. Image adapted from \cite{Berrone_2022}.}
\label{fig:square_duct_geom}
\end{figure}

\begin{figure}[h!]
\centering
    \begin{subfigure}[t]{0.24\textwidth}
    \includegraphics[width=\textwidth] {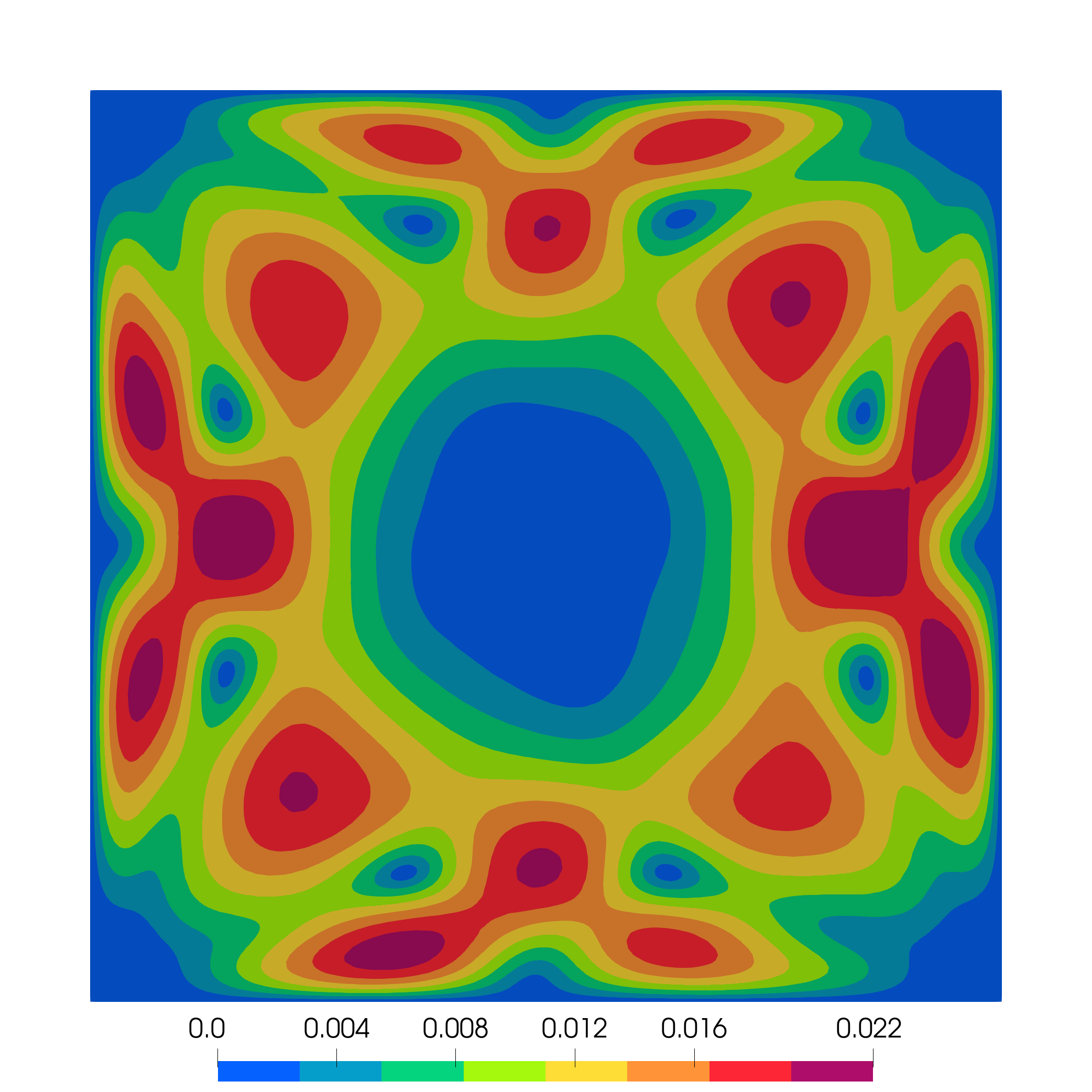}
    \caption{$Re_b = 1500$}
    \end{subfigure}
    %\hfill
    \begin{subfigure}[t]{0.24\textwidth}
    \includegraphics[width=\textwidth] {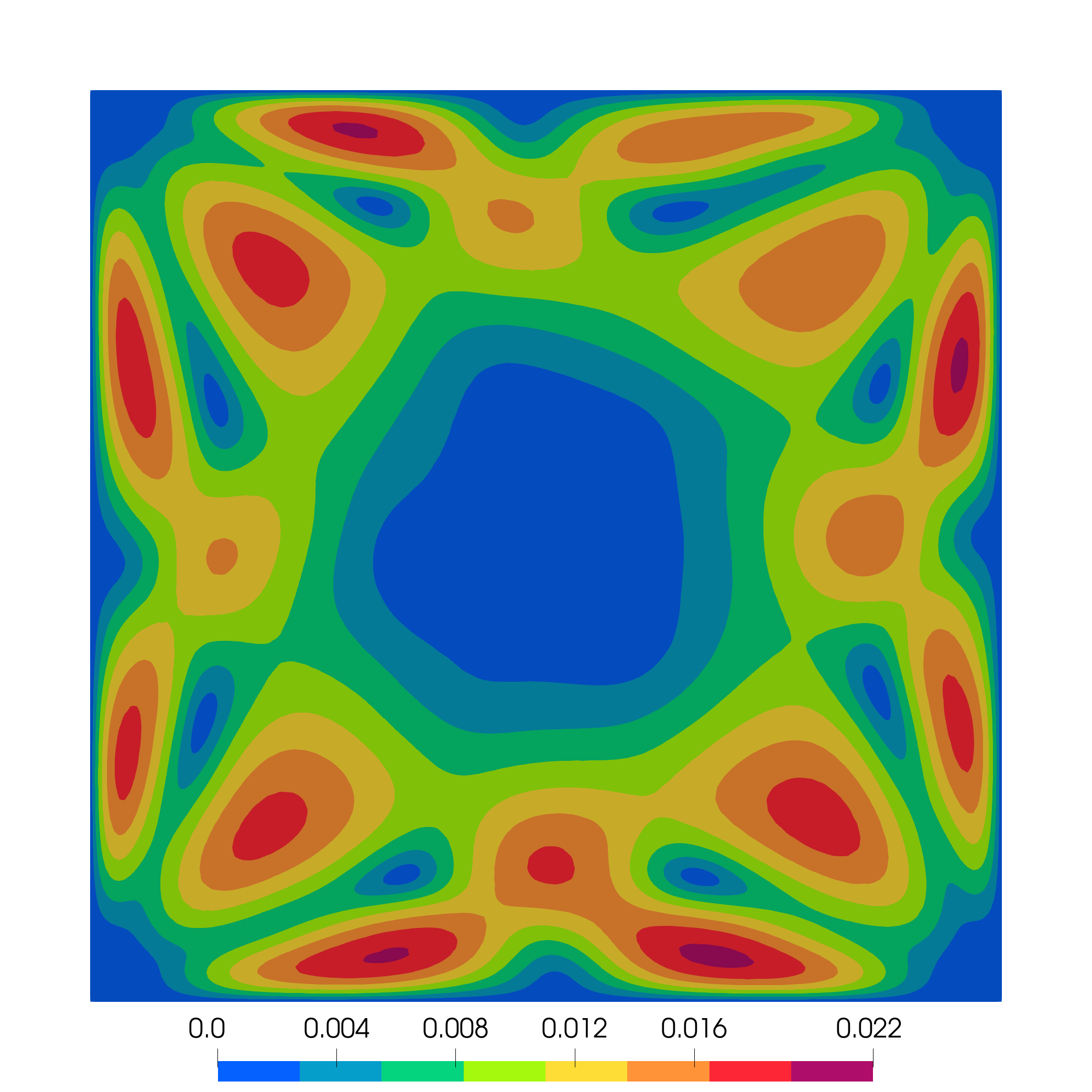}
    \caption{$Re_b = 1800$}
    \end{subfigure}
    \begin{subfigure}[t]{0.24\textwidth}
    \includegraphics[width=\textwidth] {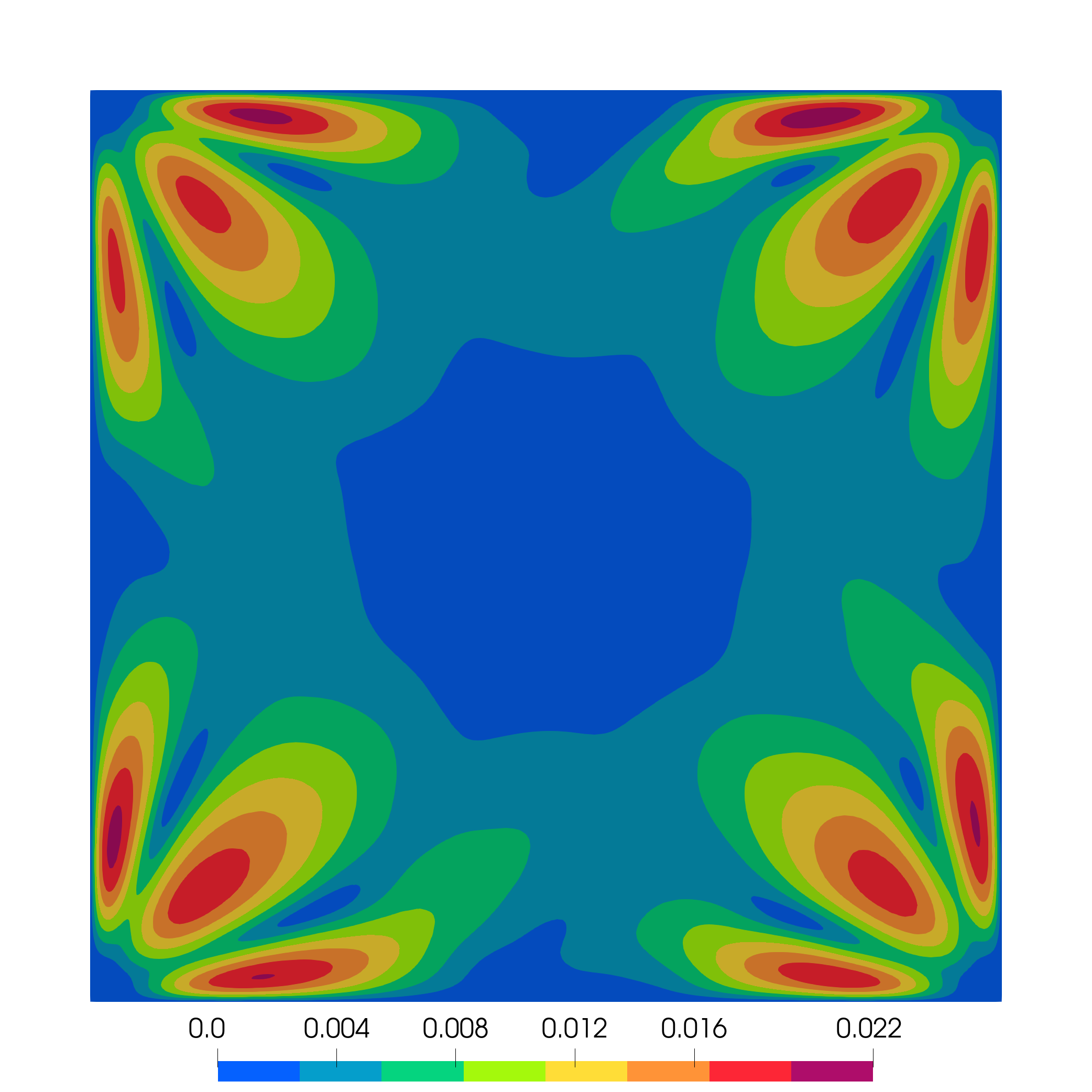}
    \caption{$Re_b = 2900$}
    \end{subfigure}
    %\hfill
    \begin{subfigure}[t]{0.24\textwidth}
    \includegraphics[width=\textwidth] {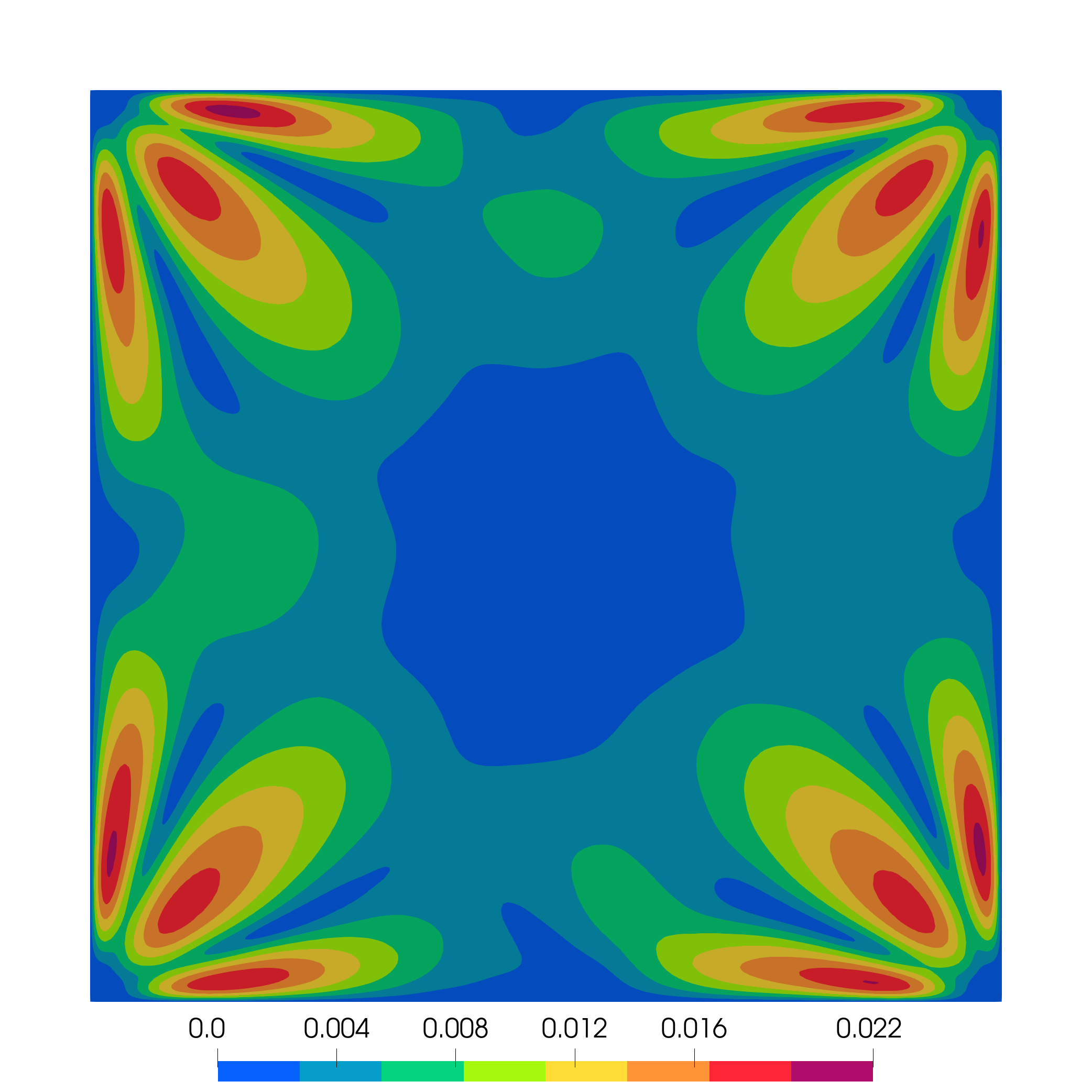}
    \caption{$Re_b = 3500$}
    \end{subfigure}  
\caption{$I_s$ for the flow in a square duct for different $Re_b$. The results come from the Pinelli et al. dataset \cite{Pinelli2010}.}
\label{fig:sec_mot_DNS}
\end{figure}

\begin{figure}[h!]
\centering
    \begin{subfigure}[t]{0.49\textwidth}
    \includegraphics[width=\textwidth] {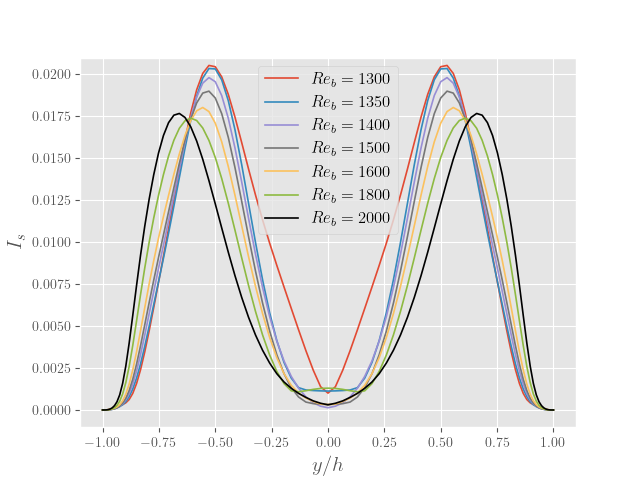}
    \caption{$Re_b \in [1300, 2000]$} \label{fig:sec_mot_DNS_diags_lowRe}
    \end{subfigure}
    \hfill
    \begin{subfigure}[t]{0.49\textwidth}
    \includegraphics[width=\textwidth] {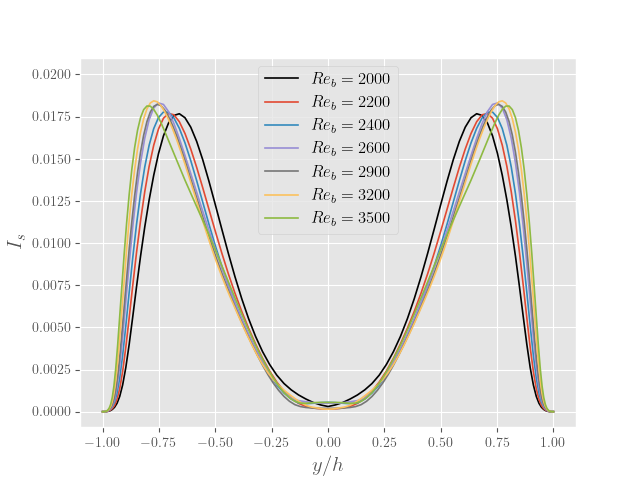}
    \caption{$Re_b \in [2000, 3500]$} \label{fig:sec_mot_DNS_diags_highRe}
    \end{subfigure}
\caption{$I_s$ over the red diagonal represented in Figure \ref{fig:square_duct_geom} for different $Re_b$ ranges: low and turbulent regimes in the left and right plot, respectively.}
\label{fig:sec_mot_DNS_diags}
\end{figure}

The flow in a square duct is a classic benchmark in the turbulence modelling community \cite{McConkey_2021}. DNS data from Pinelli et al. \cite{Pinelli2010} are available for several bulk Reynolds numbers $Re_b = U_b \ h / \nu$, being $U_b$ the mean cross-section velocity and $h$ half duct width, see Figure \ref{fig:square_duct_geom}. Periodic boundary conditions are imposed at the inlet-outlet faces while classic no-slip boundary conditions at wall are set. The $U_b$ value is imposed by applying the appropriate $\bm{b}$ value in \eqref{eq:RANS}. This flow is statistically uniform in the streamwise-direction and, thus, time-averaged velocity and Reynolds stresses on a cross-section only are provided in \cite{Pinelli2010}. Moreover, it exhibits a secondary motion, associated with non-negligible $u_y$ and $u_z$ components. We define its intensity as $$I_s = \Big(\sqrt{u_y^2 + u_z^2} \Big)/U_b.$$ 
This flow is of great interest for turbulence modelling because classic eddy-viscosity models, such as the LS $k-\varepsilon$ model used in this work, systematically fail in the secondary motion prediction \cite{Rincon_2023,Wang2017,CRUZ2019104258,Ling2015, Wu2018,Ling2016,Hoffmann_Muck_Bradshaw_1985}, see Appendix \ref{sec:role_nut_rfvNL} for a deeper analysis.\\
Pinelli et al.\red{'s}\ dataset consists of sixteen simulations with $Re_b$ ranging from 1100 to 3500. We decided to discard the $Re_b \in \{1100, 1150,1250\}$ results, since we observed that the averaged velocity fields were affected by strong non-physical asymmetries, possibly due to the choice of a not wide enough time-span averaging when computing statistics. A possible solution would have been to average the DNS data across the cross-section symmetry axes as proposed in \cite{Fonseca_2022}. However, we opted to deal with asymmetric fields to test the robustness of our model to unavoidable statistical and numerical errors coming from DNS data from generic simulations.\\
The flow behaviour changes significantly in the investigated $Re_b$ interval. For lower Reynolds numbers, the flow is in a transitional regime from laminar to turbulent, while for higher $Re_b$ values, the flow is fully turbulent. Pinelli et al. identified $Re_b = 2000$ as the threshold value of these two distinct regimes. The flow phenomenology depending on $Re_b$ can be observed by looking at $I_s$ in Figure \ref{fig:sec_mot_DNS}. It immediately appears that, at lower $Re_b$ regimes, secondary motion occurs not only near the corners but also near the walls' midpoints, contrary to the high Reynolds regime. In addition, the behaviour of $I_s$ along the diagonals changes with respect to the $Re_b$ value. In particular, Figure \ref{fig:sec_mot_DNS_diags} shows $I_s$ along the red cross-section diagonal displayed in Figure \ref{fig:square_duct_geom}. For the low $Re_b$ regime, depicted in Figure \ref{fig:sec_mot_DNS_diags_lowRe}, the $I_s$ maxima significantly increases with respect to $Re_b$, while it is not the case for $Re_b>2000$ where a marginal opposite trend can be observed. Finally, independently of the flow regime, the higher $Re_b$, the closer the maxima of $I_s$ are to the corners.


\section{FOM results}\label{sec:FOM_results}
In this Section, the $\nut$-VBNN model \red{is trained} to enhance the accuracy of RANS equations to generate FOM results for several values of $Re_b$. 

\subsection{$\nu_t$-VBNN training}
First, the neural networks of the $\nut$-VBNN model to predict $\tnut$ and $\rfvNL$ must be trained. We perform it on a wider $Re_b$ range compared to the majority of works dealing with the square duct case \cite{CRUZ2019104258,Berrone_2022,Wu2019,Wu2018}. \red{As a matter of fact}, we select $Re_b$ ranging from 1300 to 3500, for a total of thirteen simulations. The reason is twofold: on one hand, our model will be used to generate around one hundred of FOM simulations to build an effective ROM that should be able to span a significantly large parameter space; on the other hand, by covering two different and distinct flow regimes, we make the investigated problem more challenging. To access the effect of dealing with two flow regimes, we train, for both $\tnut$ and $\rfvNL$, three different neural networks with data coming from different $Re_b$ values: one with $Re_b \in [1300, 3500]$, one with $Re_b \in [1300, 2000]$ and one with $Re_b \in [2000, 3500]$, referred as $\all$, $\low$, and $\high$, respectively. For each case, data from at least one DNS simulation is kept out from training to test our network's ability on unseen data. Table \ref{tab:Reb_intervals} summarises all this information.\\
\begin{table}
\centering
\caption{Three different networks training settings. \texttt{$Re_b$ interval} indicates the $Re_b$ range used, \texttt{$\#$ simulations} specifies how many simulations have been used for training, \texttt{$Re_b$ kept out} reports the $Re_b$ value(s) kept out from training.}
\begin{tblr}{
  columns={halign=c},
  hline{1,1} = {-}{},
  hline{1,2} = {-}{},
  hline{1,5} = {-}{},
  vline{1,1} = {-}{},
  vline{1,2} = {-}{},
  vline{1,5} = {-}{},
}
  & \texttt{$Re_b$ interval} & \texttt{$\#$ simulations} & \texttt{$Re_b$ kept out} \\
 \all &  $[1300, 3500]$ & 11 & 1600, 2900\\
 \low &  $[1300, 2000]$ & 6 & 1600 \\
 \high &  $[2000, 3500]$ & 6 & 2900
\end{tblr}
\label{tab:Reb_intervals}
\end{table}
All models are implemented using the pyTorch python library \cite{pytorch}. 
Regarding the optimisation procedure, the $\tnut$ and $\rfvNL$ networks are trained using the Adam algorithm \cite{Adam} to minimise the following \emph{discretized loss functions}, respectively:
\begin{subequations}
\begin{align}
    &\mathcal{L}_{\tnut}(\cdot ; \theta_{\tnut}) = \frac{1}{n} \sum_{i=1}^n (\tilde{\nu}_{t,i}^{\DNS} - \tilde{\nu}_{t, i}^{\NN}(\cdot))^2 + \lambda_{\tnut} \sum_{i=1}^{n_{\tnut}} {(\theta_{\tnut}^i)^2}, \\ %\frac{1}{n} \| \tnut^{\DNS} - \tnut^{\NN}(\cdot) \|^2_2 + \lambda_{\tnut} \| \bm{\theta}_{\tnut} \|^2_2, \\ &
    & \mathcal{L}_{\rfvNL}(\cdot ; \theta_{\rfvNL}) = \frac{1}{3n} \sum_{i=1}^n \sum_{j=1}^3 (\tilde{t}^{\perp, \DNS}_{j, i} - \tilde{t}^{\perp, \NN}_{j, i}(\cdot))^2 + \lambda_{\rfvNL} \sum_{i=1}^{n_{\rfvNL}}(\theta_{\rfvNL}^i)^2, %\mathcal{L}_{\rfvNL}(\cdot ; \bm{\theta_{\rfvNL}}) = \frac{1}{n} \| \rfv^{\perp,\DNS} - \rfv^{\perp,\NN}(\cdot) \|^2_2 + \lambda_{\rfvNL} \| \bm{\theta_{\rfvNL}} \|^2_2,
    \label{eq:losses}
\end{align}
\end{subequations}
where we denote by $\tilde{\nu}_t^{\DNS}$ and $\tilde{\bm{t}}^{\perp, \DNS}$ the discrete DNS target fields interpolated on the RANS meshes, for a total of $n$ values, by $\tilde{\nu}_t^{\NN}$, and $\tilde{\bm{t}}^{\perp,\NN}$ the fields predicted \redred{through} neural networks using \eqref{eq:nut_rfvNL}, by ${\theta}_{\tilde{\nu}_t}$ and ${\theta}_{\rfvNL}$
 the vectors, of size $n_{\tnut}$ and $n_{\rfvNL}$ respectively, containing the weights of the neural networks to be tuned in the optimisation procedure, by $\lambda_{\tnut}$ and $\lambda_{\rfvNL}$ the so-called \emph{regularization factors}. By the \blue{subscript} $i$, we denote the $i$-th element of the discretised field or the $i$-th weight of the network, while $j$ spans the three components of $\rfvNL$.

For each network and each hyperparametric choice, fifteen training runs with different randomly sampled starting weights are performed. For each $Re_b$ value used to train, we extract the fields required for the $\nut$-VBNN training from the $N_h$ computational cells of the RANS simulation. The 80$\%$ of these data are used to train the networks, while the remaining $20\%$ are used for validation purposes during the training.\\  %\dob{ho spiegato qua cosa intendo con data}\\

During the hyperparameters tuning stage, we observed a strong dependence of the overfitting behaviour of the $\nut$-VBNN model from the regularization factors. Consequently, we rigorously carried out an analysis to find their optimal values in terms of testing errors. The results are detailed in Appendix \ref{sec:reg_factor_FOM}. 
The final choices for the regularization parameters are reported in Table \ref{tab:reg_factors}, while additional information about the networks' architecture are presented in Appendix \ref{sec:nut_VBNN_hyper}. 
\begin{table}[]
\centering
\caption{Selected $\lambda_{\nut}$ and $\lambda_{\rfvNL}$ values for the $\all$, $\low$ and $\high$ models.} % These choices are dictated by the average error behaviours in Figures \ref{fig:nut_err_vs_Reb_avg_std} and \ref{fig:rfvNL_err_vs_Reb_avg_std}.} 
\renewcommand{\arraystretch}{1.2}
\begin{tblr}{
  columns={halign=c},
  hline{1,1} = {-}{},
  hline{1,2} = {-}{},
  hline{1,4} = {-}{},
  vline{1,1} = {-}{},
  vline{1,2} = {-}{},
  vline{1,5} = {-}{},
}
& $\texttt{all}$ & $\texttt{low}$ & $\texttt{high}$  \\ 
$\lambda_{\tnut}$ & $10^{-3}$ & $10^{-3}$ & $10^{-4}$  \\ 
$\lambda_{\rfvNL}$ & $10^{-4}$ & $10^{-4}$ & $10^{-5}$   \\ 
\end{tblr}
\label{tab:reg_factors}
\end{table}
It appears that a smaller regularization parameter can be used when training on a unique flow regime without occurring in overfitting. This behaviour can be intuitively explained as follows: when inferring with a new $Re_b$ value belonging to the same flow regime as all the training $Re_b$ values, the expected $\tnut$ and $\rfvNL$ fields are close to all training fields, thus reducing the overfitting phenomenon. Finally, each training is stopped if the minimum of the validation loss does not decrease for fifty consecutive epochs for a maximum of $N_{e} = 500$ epochs. Generally speaking, the average training time per epoch for the $\all$ networks is the double of the $\low$ and $\high$ one, being the former around $60s$ and the latter around $30s$. This behaviour is due to the $\all$ networks being trained with the double of data compared to the $\low$ and $\high$ networks.
%\dob{Aggiunto commento sulle tempistiche di training}

\begin{figure}
\centering
    \begin{subfigure}[t]{0.42\textwidth}
    \includegraphics[width=\textwidth]{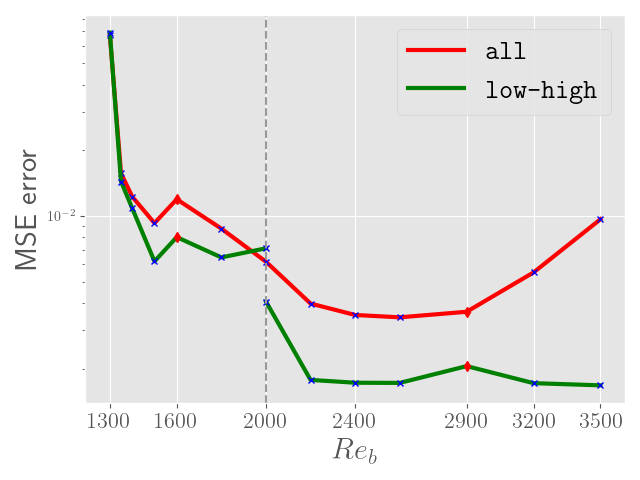}
    \caption{$\tnut$}\label{fig:nut_best_error_plot}
    \end{subfigure}
    \hfill
    \begin{subfigure}[t]{0.42\textwidth}
    \includegraphics[width=\textwidth]{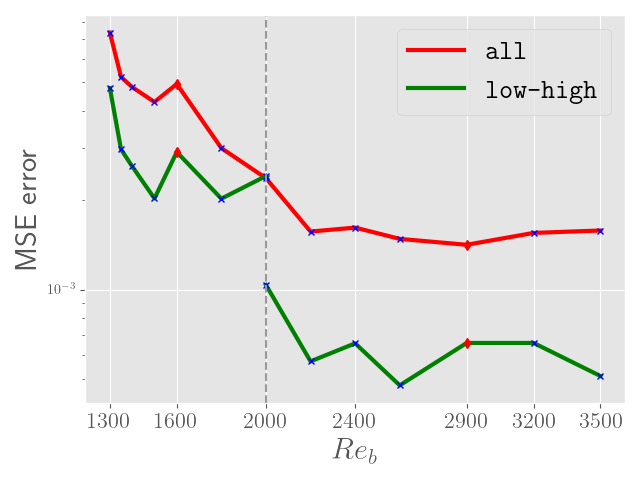}
    \caption{$\rfvNL$}\label{fig:rfvNL_best_error_plot}
    \end{subfigure}
    \caption{Mean Squared Error (MSE) with respect to $Re_b$ for the chosen NNs. The simulations used and discarded in the training are denoted by blue stars and red diamonds, respectively.}
    \label{fig:best_error_plot}
\end{figure}

To emphasise the benefits of splitting the $\nut$-VBNN models training into the $\low$ and $\high$ intervals, Figure \ref{fig:best_error_plot} shows the \red{m}ean \red{s}quared \red{e}rror (MSE) of the inferred fields with respect to $Re_b$ for the final selected networks. It immediately appears that the training interval splitting approach leads to lower errors because each neural network focuses on a specific flow regime.
In the following, we will refer to the setting with $\nut$-VBNN neural networks trained using data from $Re_b \in [1300, 3500]$ as $\all$, while to the setting with separate trainings depending on the $Re_b$ value as $\lowhigh$. In the latter, the $\low$ network is selected for inference if $Re_b < 2000$, otherwise the $\high$ one is used.

\subsection{FOM results analysis}
Once the models for both the $\all$ and $\lowhigh$ settings are trained, the $\nut$-VBNN enhanced simulations can be performed to generate high-fidelity FOM simulations avoiding the computational burden of DNS. 

\red{
\begin{remark}
 For the sake of brevity, the results obtained with the baseline LS $k-\varepsilon$ model will not be reported in this Section because consistent with the literature \cite{Wang2017,Berrone_2022}. Specifically, for all $Re_b$ values, no secondary motion is described, i.e. $I_s = 0$ across the square section, and the $u_x/U_b$ field has always circular isocontours typical of the laminar case. In the Appendix \ref{sec:role_nut_rfvNL}, the $Re_b=2900$ case is discussed. 
\end{remark}
}

\begin{figure}[]
\centering
    \begin{subfigure}[t]{0.25\textwidth}
    \includegraphics[width=\textwidth]{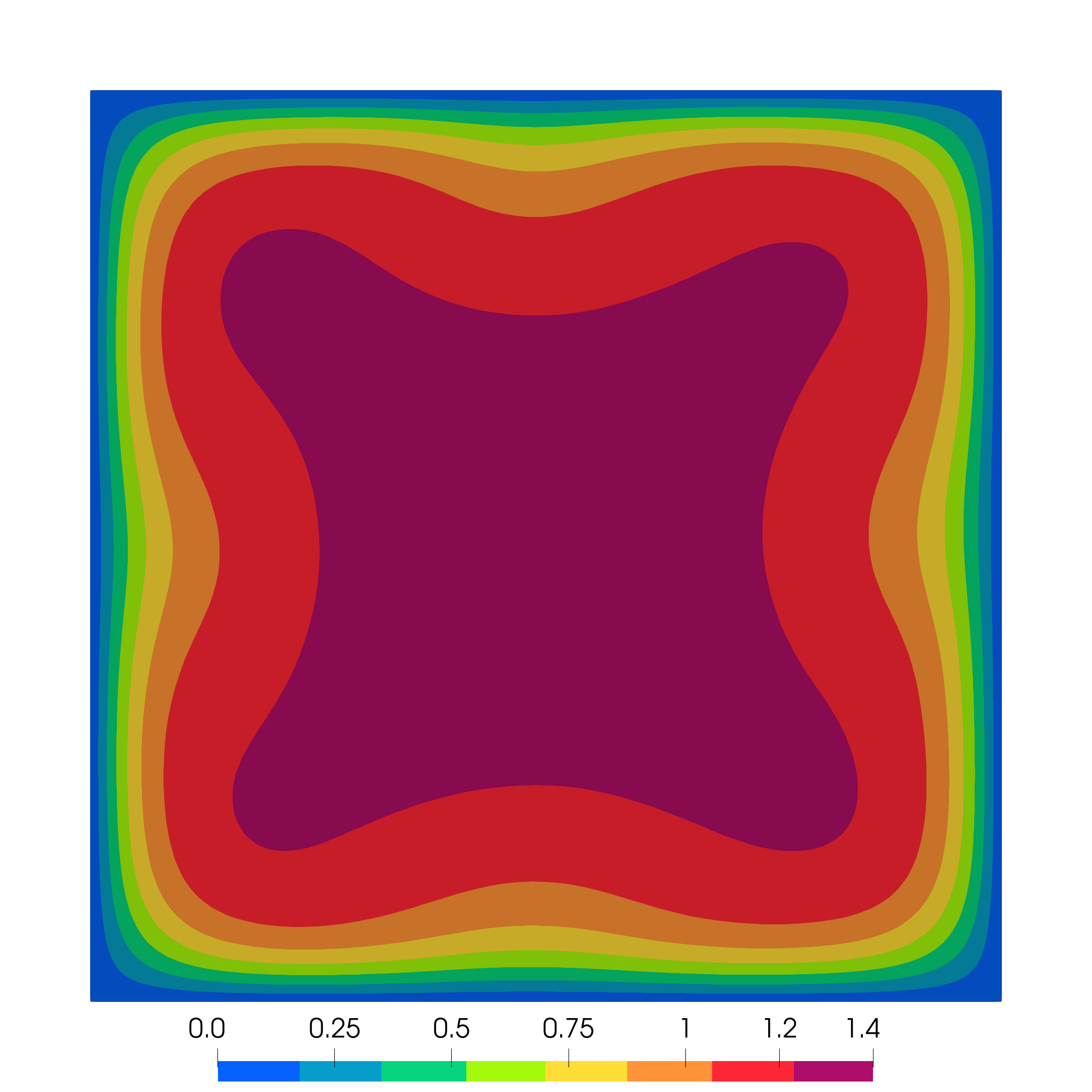}\label{fig:Ux_DNS_1600}
    \caption{$u_x$ DNS}
    \end{subfigure}
    \hfill
    \begin{subfigure}[t]{0.25\textwidth}
    \includegraphics[width=\textwidth]{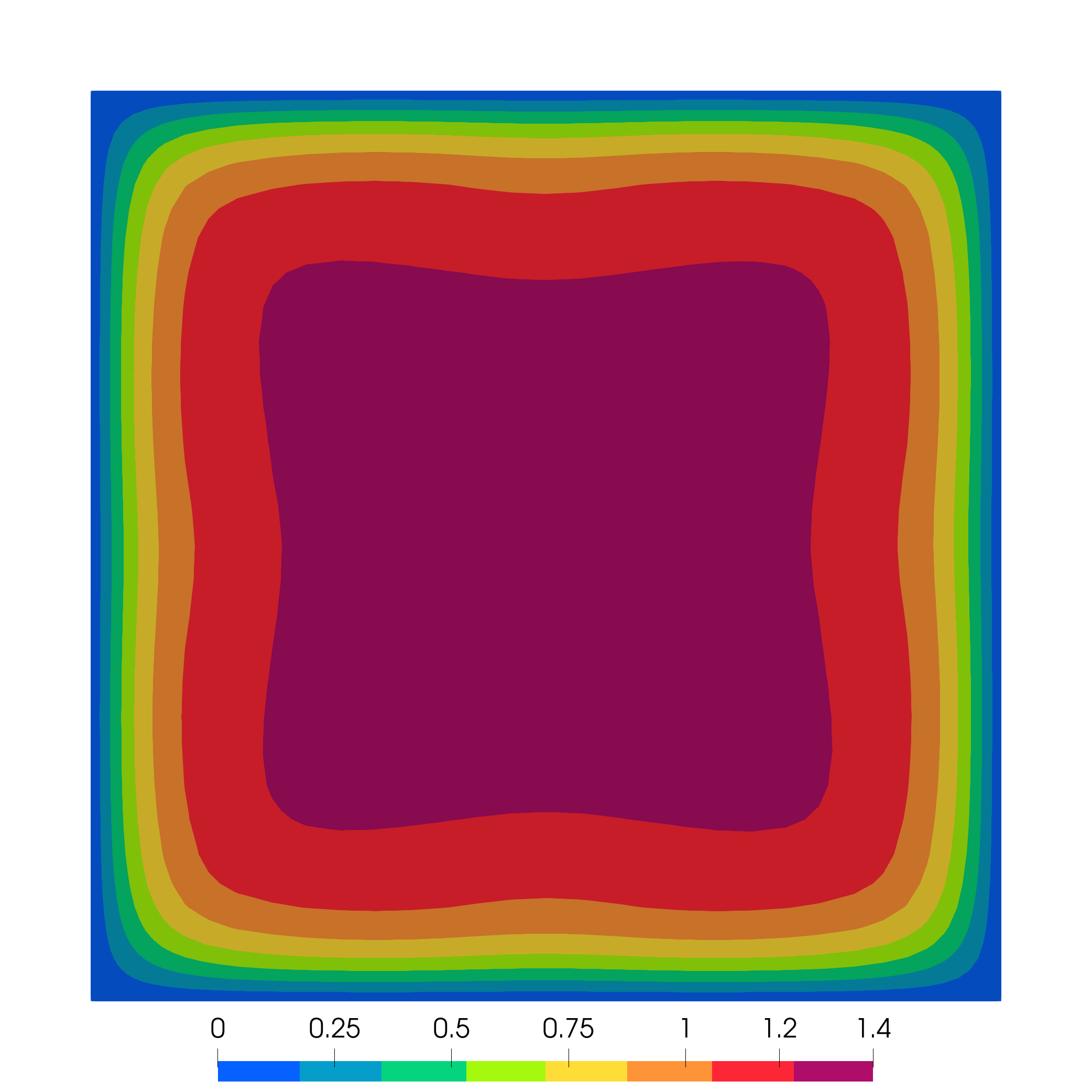}\label{fig:Ux_nutall_rfvNLall_Reb1600}
    \caption{$u_x \ \all$}
    \end{subfigure}
    \hfill
    \begin{subfigure}[t]{0.25\textwidth}
    \includegraphics[width=\textwidth]{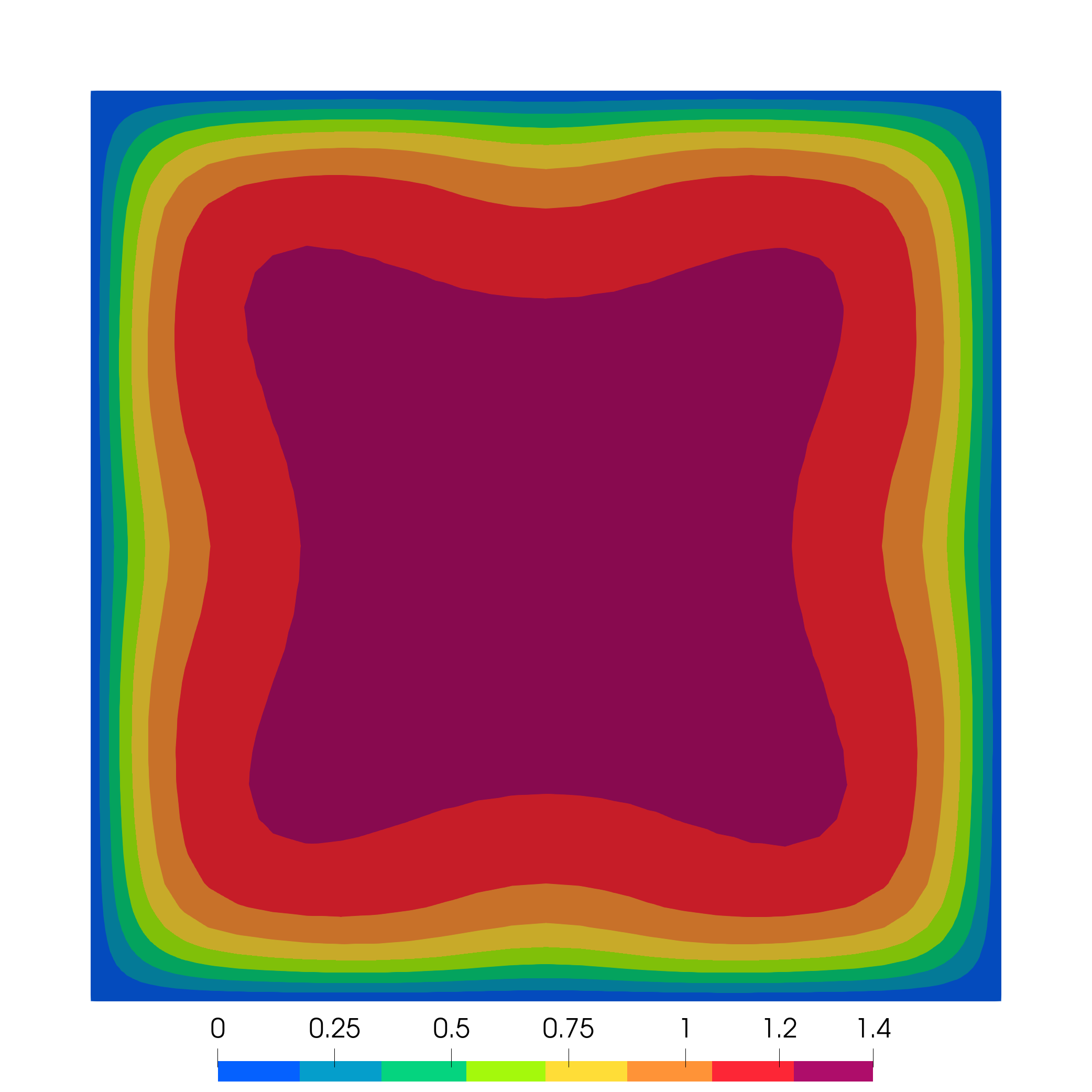}\label{fig:Ux_nutlowhigh_rfvNLlowhigh_Reb1600}
    \caption{$u_x \ \lowhigh$}
    \end{subfigure}

        \begin{subfigure}[t]{0.25\textwidth}
    \includegraphics[width=\textwidth]{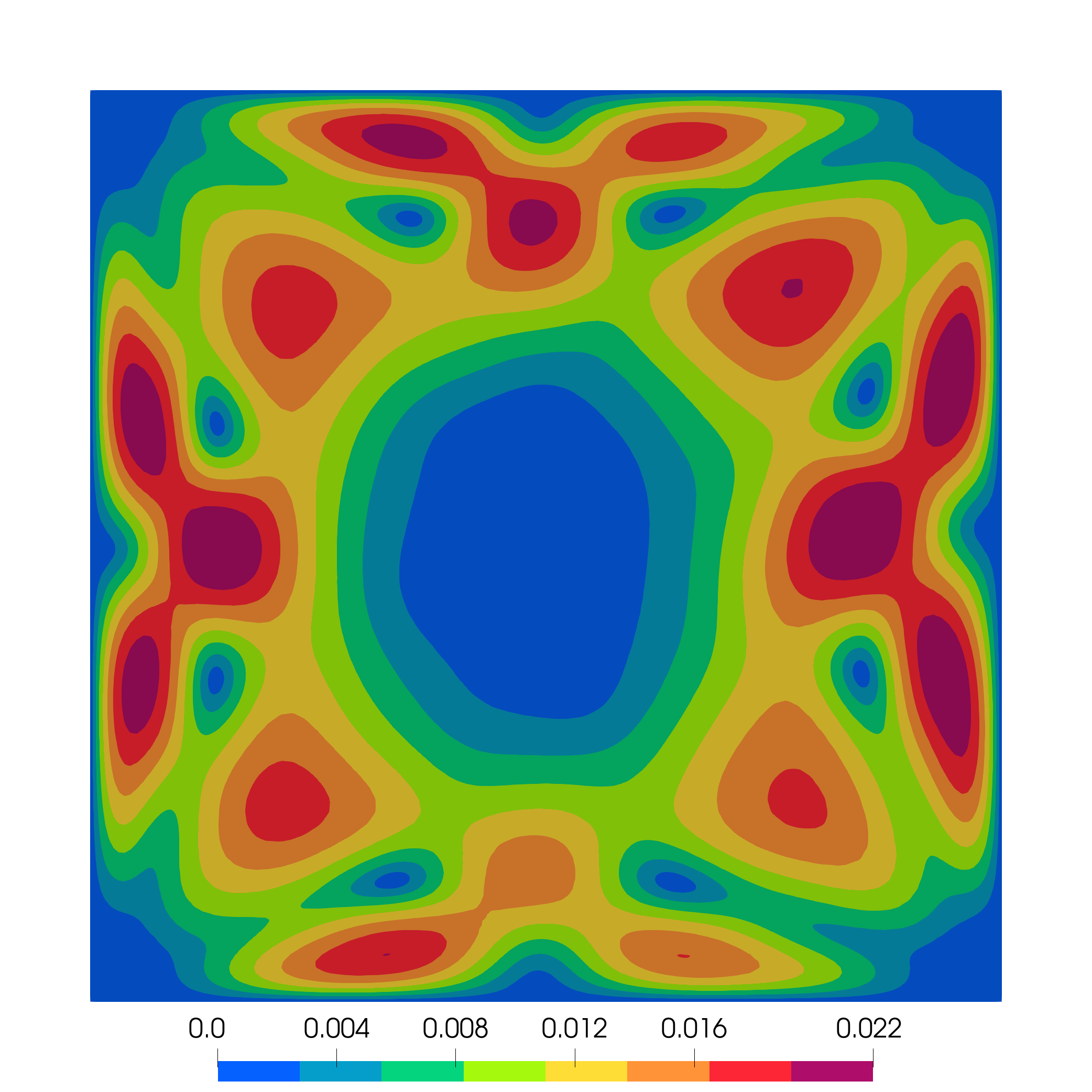}\label{fig:secmot_DNS_1600}
    \caption{$I_s$ DNS}
    \end{subfigure}
    \hfill
    \begin{subfigure}[t]{0.25\textwidth}
    \includegraphics[width=\textwidth]{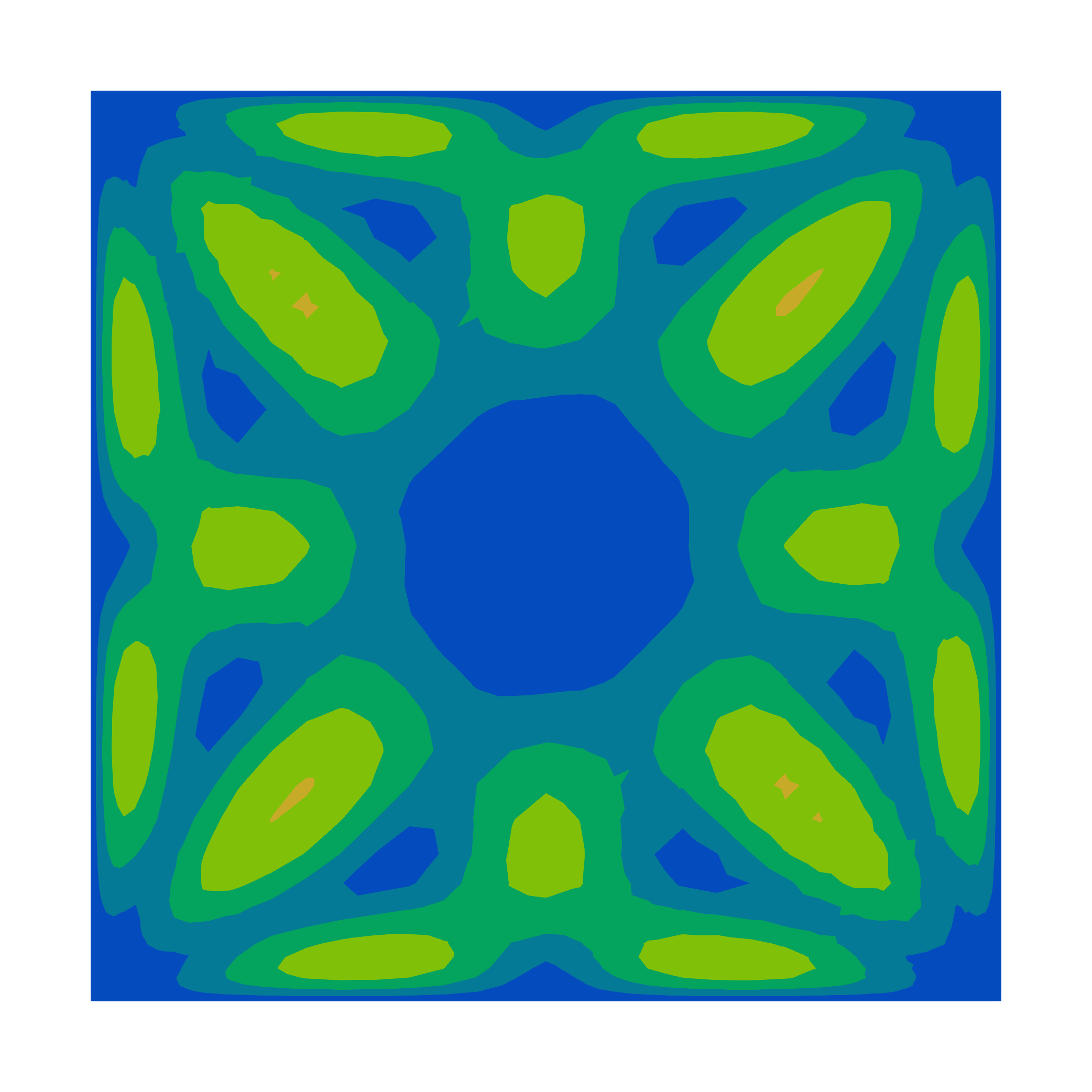}\label{fig:secmot_nutall_rfvNLall_Reb1600}
    \caption{$I_s \ \all$}
    \end{subfigure}
    \hfill
    \begin{subfigure}[t]{0.25\textwidth}
    \includegraphics[width=\textwidth]{images/secmot_OF/nutlowhigh_rfvNLlowhigh/Reb1600}\label{fig:secmot_nutlowhigh_rfvNLlowhigh_Reb1600}
    \caption{$I_s \ \lowhigh$}
    \end{subfigure}
    \caption{Comparison of $u_x/U_b$ and $I_s$ obtained with the $\all$ and $\lowhigh$ models at $Re_b = 1600$ and the DNS references.}\label{fig:Reb1600_DNS_allall_lowhigh}
\end{figure}
\begin{figure}
\centering
    \begin{subfigure}[t]{0.25\textwidth}
    \includegraphics[width=\textwidth]{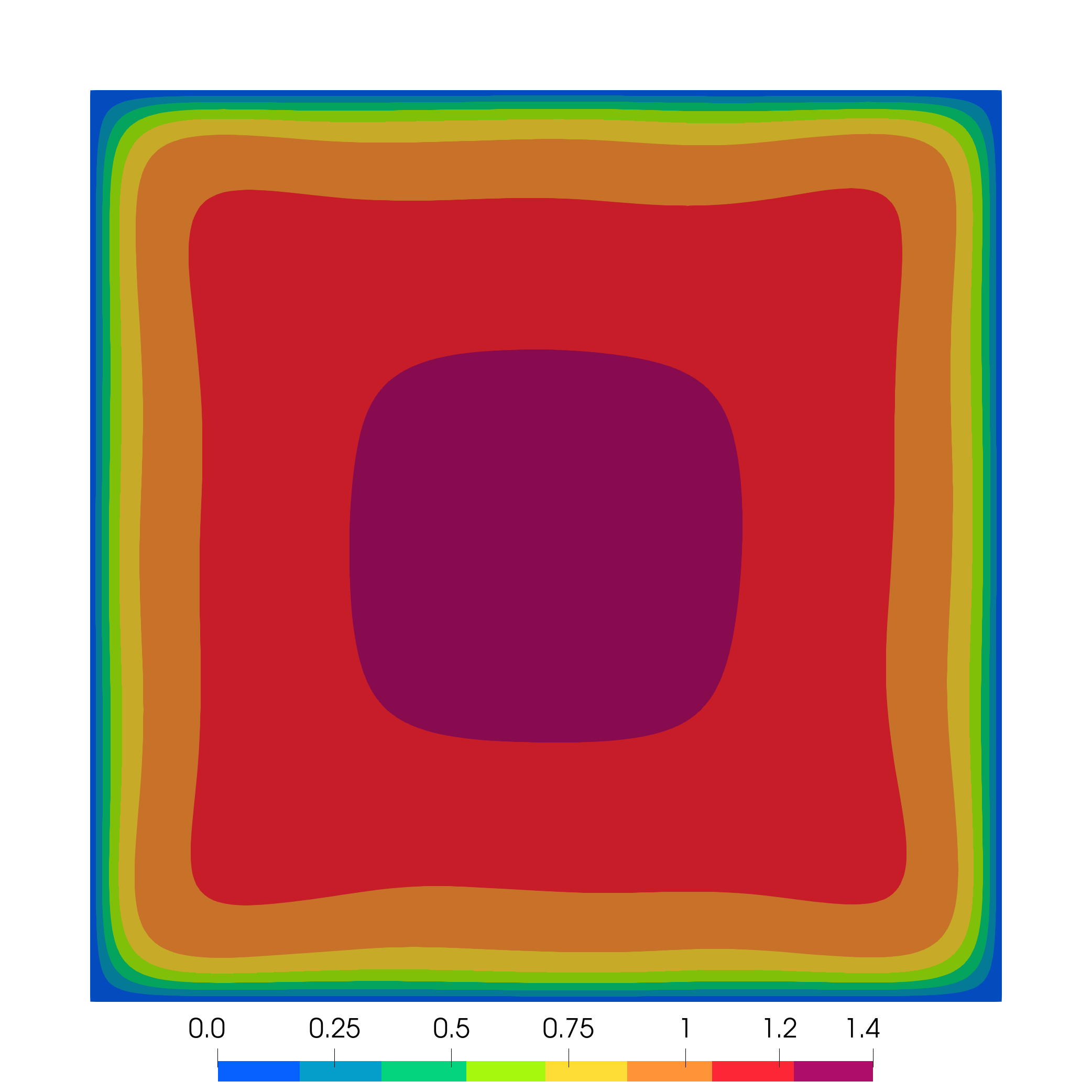}\label{fig:Ux_DNS_2900}
    \caption{$u_x$ DNS}
    \end{subfigure}
    \hfill
    \begin{subfigure}[t]{0.25\textwidth}
    \includegraphics[width=\textwidth]{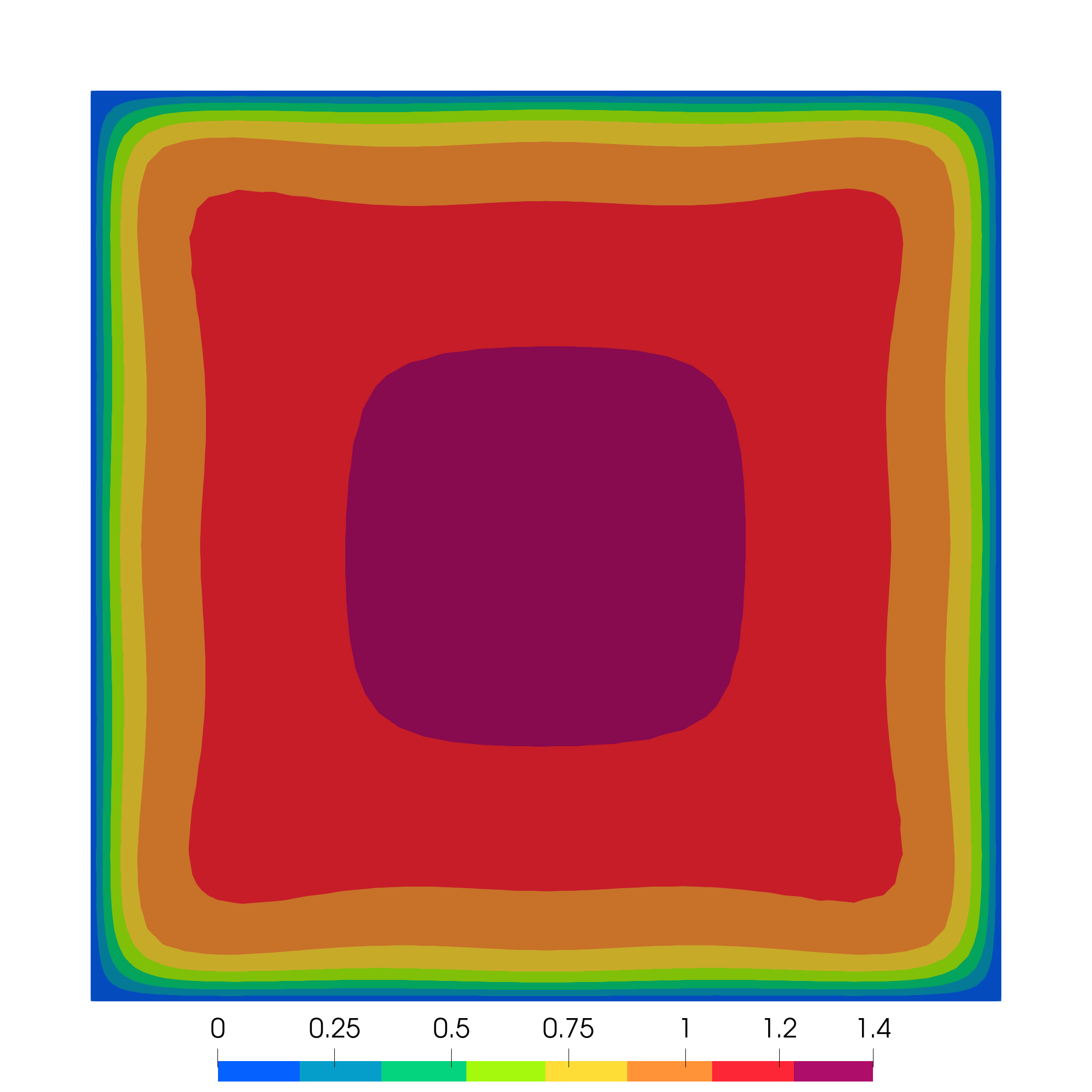}\label{fig:Ux_nutall_rfvNLall_Reb2900}
    \caption{$u_x \ \all$}
    \end{subfigure}
    \hfill
    \begin{subfigure}[t]{0.25\textwidth}
    \includegraphics[width=\textwidth]{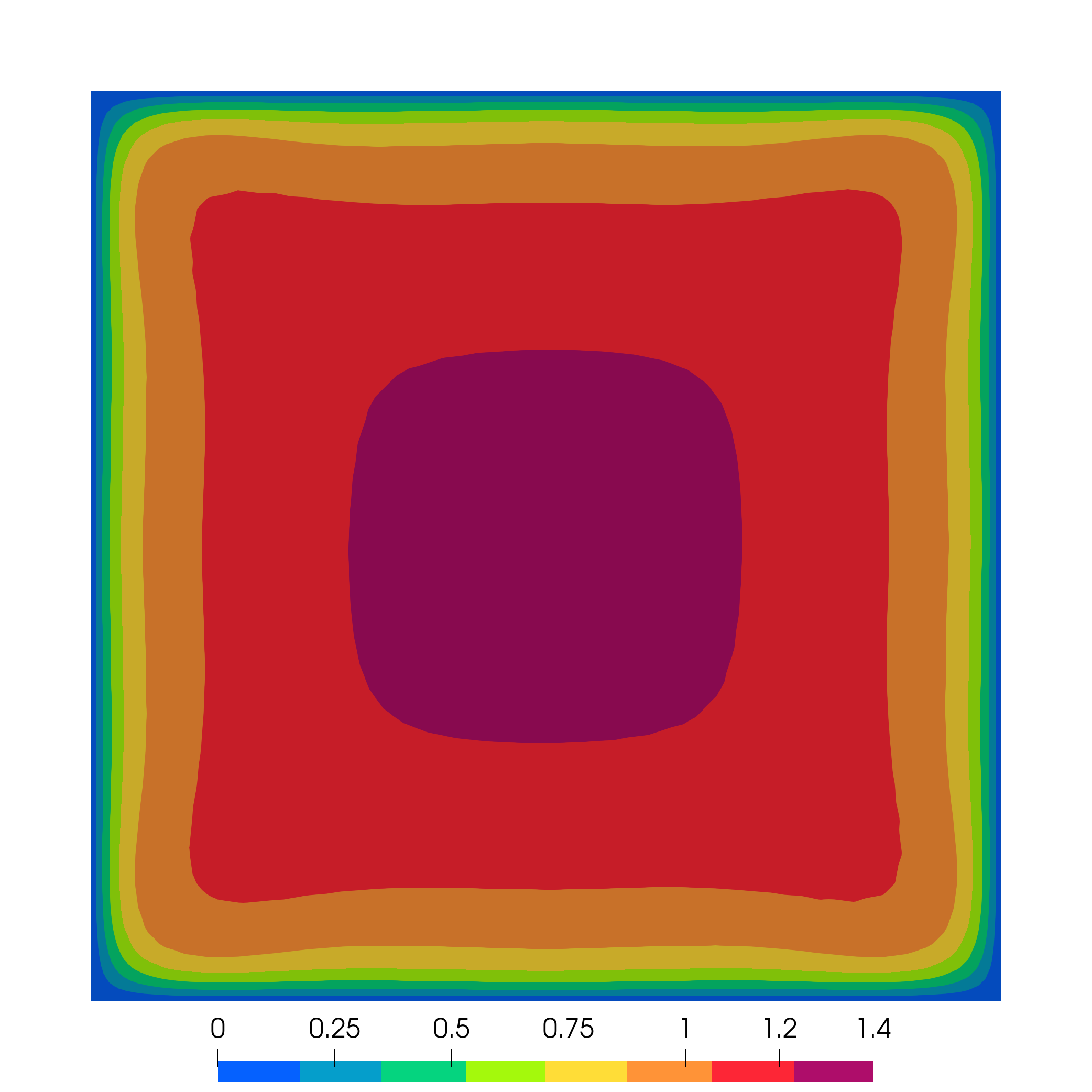}\label{fig:Ux_nutlowhigh_rfvNLlowhigh_Reb2900}
    \caption{$u_x \ \lowhigh$}
    \end{subfigure}

        \begin{subfigure}[t]{0.25\textwidth}
    \includegraphics[width=\textwidth]{images/square_duct/sec_mot_DNS_Re2900}\label{fig:secmot_DNS_2900}
    \caption{$I_s$ DNS}
    \end{subfigure}
    \hfill
    \begin{subfigure}[t]{0.25\textwidth}
    \includegraphics[width=\textwidth]{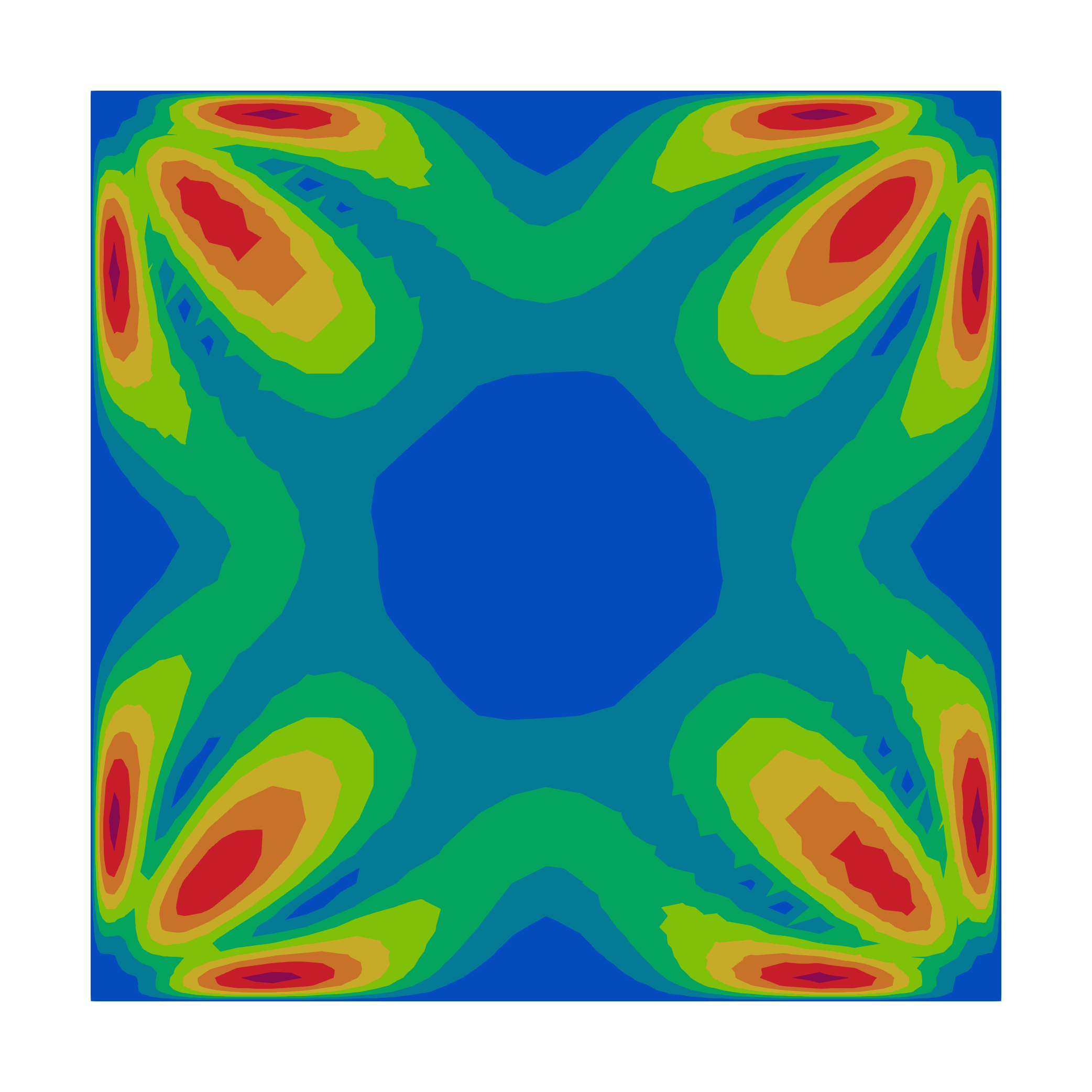}\label{fig:secmot_nutall_rfvNLall_Reb2900}
    \caption{$I_s \ \all$}
    \end{subfigure}
    \hfill
    \begin{subfigure}[t]{0.25\textwidth}
    \includegraphics[width=\textwidth]{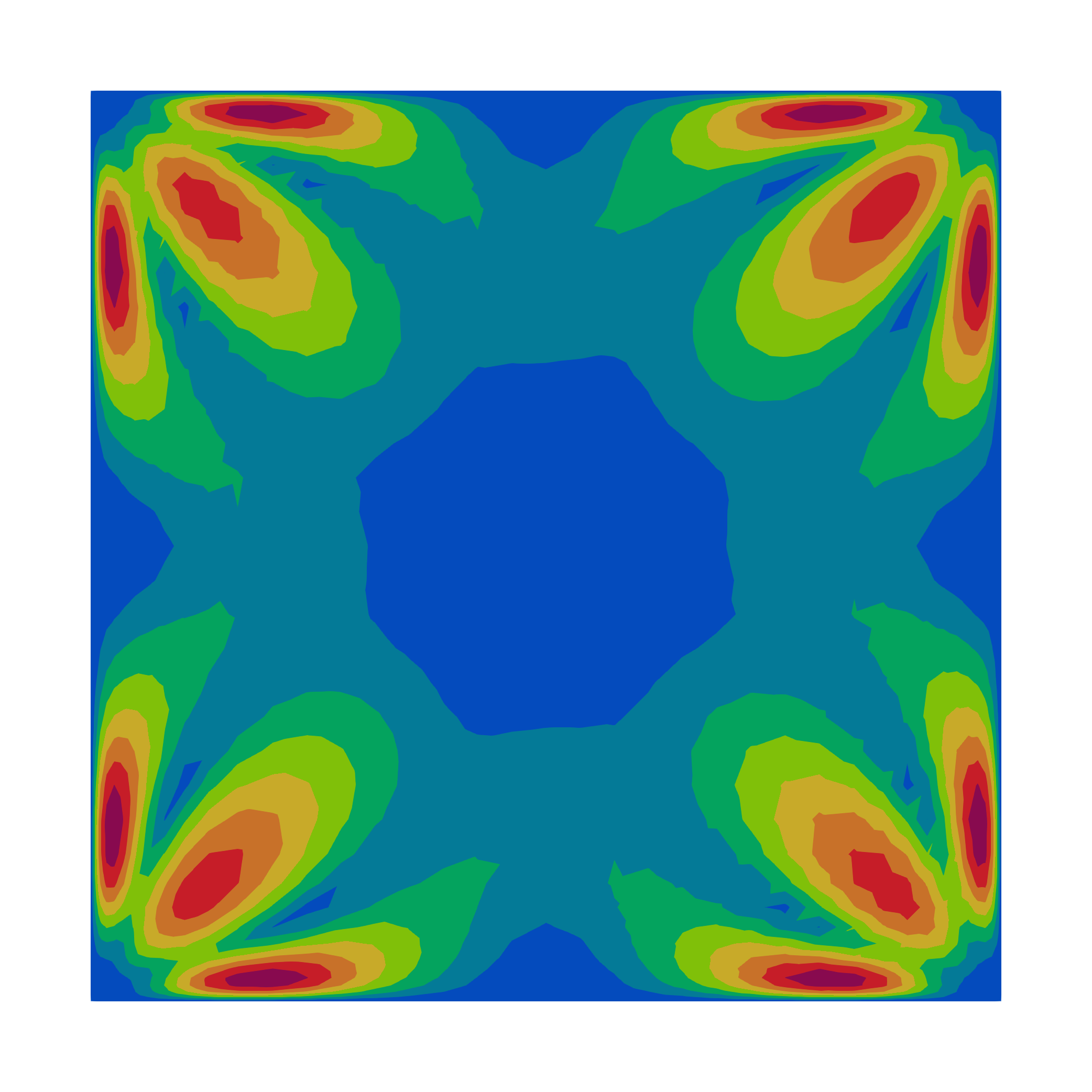}\label{fig:secmot_nutlowhigh_rfvNLlowhigh_Reb2900}
    \caption{$I_s \ \lowhigh$}
    \end{subfigure}
    \caption{Comparison of $u_x/U_b$ and $I_s$ obtained with the $\all$ and $\lowhigh$ models at $Re_b = 2900$ and the DNS references.}\label{fig:Reb2900_DNS_allall_lowhigh}
\end{figure}

To access the FOM accuracy, Figures \ref{fig:Reb1600_DNS_allall_lowhigh} and \ref{fig:Reb2900_DNS_allall_lowhigh} show both the streamwise velocity $u_x / U_b$ and the intensity of the secondary motion $I_s$ for $Re_b = 1600$ and $Re_b = 2900$, i.e., the data kept out during training for both $\all$ and $\lowhigh$ models. The larger discrepancies between the $\all$ and $\lowhigh$ settings are observable for the $Re_b = 1600$ case, where the former poorly represents the maximum isocontours shape of the streamwise velocity and underestimates $I_s$, while the latter systematically better predicts both fields. Conversely, discrepancies between the two settings for $Re_b = 2900$ are less prominent, but noticeable, especially for the $I_s$ field. Also for this case, the $\lowhigh$ model gives closer results to DNS.

\begin{figure}[h]
\centering
    \begin{subfigure}[t]{0.45\textwidth}
    \includegraphics[width=\textwidth]{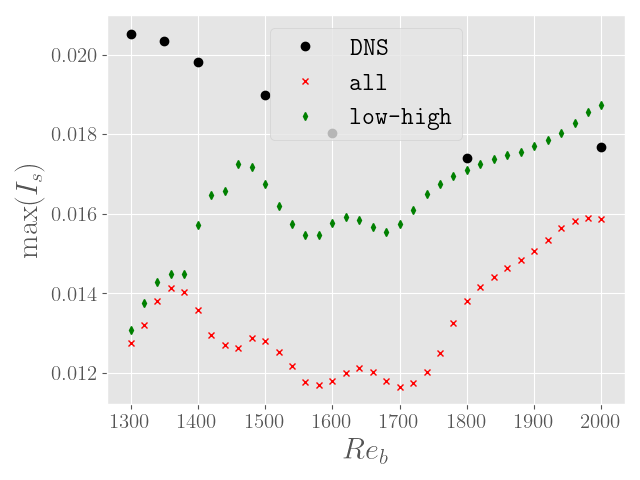}\label{fig:snaps_max_secmot_lowRe}
    \caption{$Re_b \in [1300, 2000]$}
    \end{subfigure}
    \hfill
    \begin{subfigure}[t]{0.45\textwidth}
    \includegraphics[width=\textwidth]{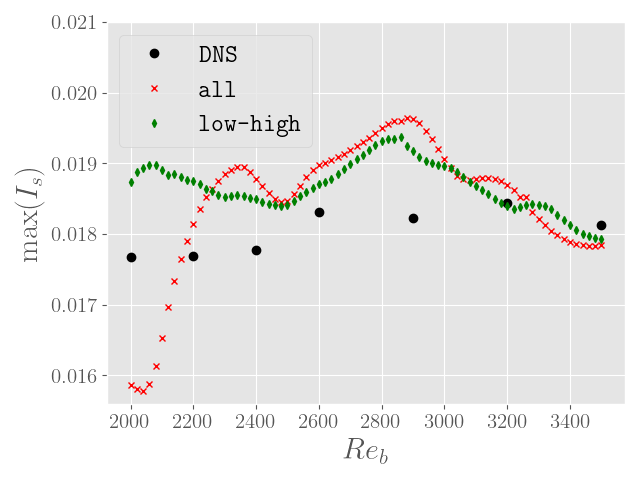}\label{fig:snaps_max_secmot_highRe}
    \caption{$Re_b \in [2000, 3500]$}
    \end{subfigure}
    \caption{Maximum values of $I_s$ along the square section diagonal with respect to $Re_b$. The $\texttt{all}$ and $\texttt{low-high}$ values are compared to DNS.}\label{fig:snaps_max_secmot}
\end{figure}

To give a quantitative measure of the $\all$ and $\lowhigh$ accuracy, Figure \ref{fig:snaps_max_secmot} depicts the maximum values of $I_s$ along the cross-section diagonal while varying $Re_b$ and compares them to the DNS values (i.e., the maxima of curves in Figure \ref{fig:sec_mot_DNS_diags}). For clarity, the values are divided into two plots, one for each flow regime. Regarding the $Re_b \in [1300, 2000]$ regime, neither setting can predict the decreasing trend of the maxima. However, the $\lowhigh$ setting predicts maxima closer to DNS compared to the $\all$ one. Regarding the $Re_b \in [2000, 3500]$ regime, the $\lowhigh$ and $\all$ maxima values are close except for the lowest $Re_b$ values where the $\all$ setting drastically underestimates the maxima, while the $\lowhigh$ one overestimates them.

\begin{figure}[h]
\centering
    \begin{subfigure}[t]{0.45\textwidth}
    \includegraphics[width=\textwidth]{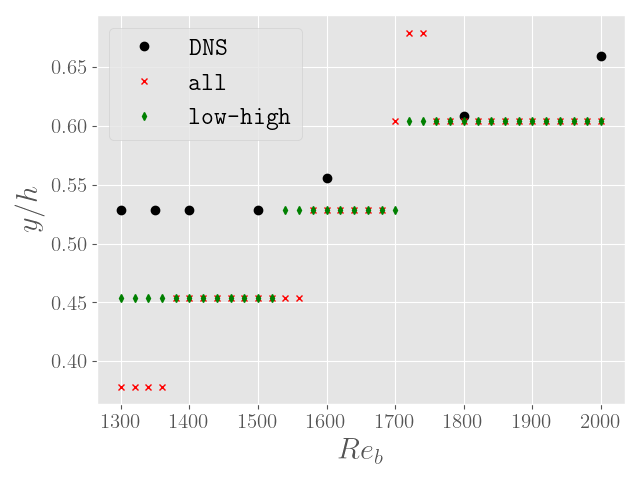}\label{fig:snaps_locmax_secmot_lowRe}
    \caption{$Re_b \in [1300, 2000]$}
    \end{subfigure}
    \hfill
    \begin{subfigure}[t]{0.45\textwidth}
    \includegraphics[width=\textwidth]{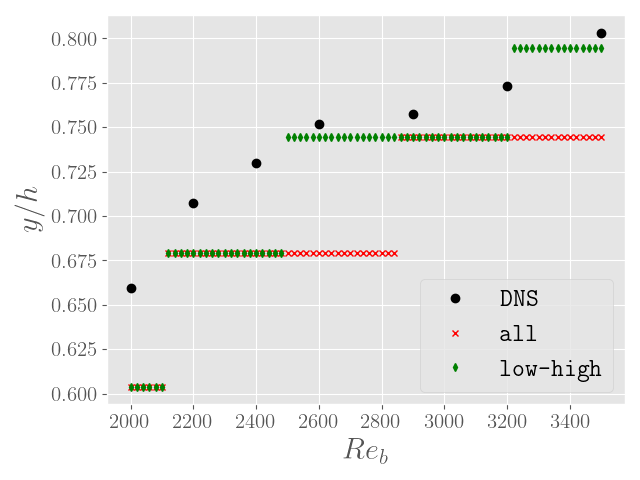}\label{fig:snaps_locmax_secmot_highRe}
    \caption{$Re_b \in [2000, 3500]$}
    \end{subfigure}
    \caption{$y/h$ location of the maximum value of $I_s$ along the upper part ($y/h > 0$) square section diagonal  with respect to $Re_b$. The $\texttt{all}$ and $\texttt{low-high}$ values are compared to the DNS ones.}\label{fig:snaps_locmax_secmot}
\end{figure}

Furthermore, Figure \ref{fig:snaps_locmax_secmot} shows the location of the maxima in terms of the positive coordinate $y/h$. %Being the computational domain discrete, the values exhibit a staircase behaviour \ms{forse questa cosa dello starcase O la spiegherei meglio O la toglierei}. 
For both flow regimes, either the maxima locations of the two models coincide, or the $\lowhigh$ maxima are closer to DNS. Finally, we highlight that for $Re_b = 2900$ the maximum values of $I_s$ are close and their locations are identical for the two settings, in agreement with Figure \ref{fig:Reb2900_DNS_allall_lowhigh}.

\begin{figure}[h]
\centering
\includegraphics[width=0.5\textwidth] {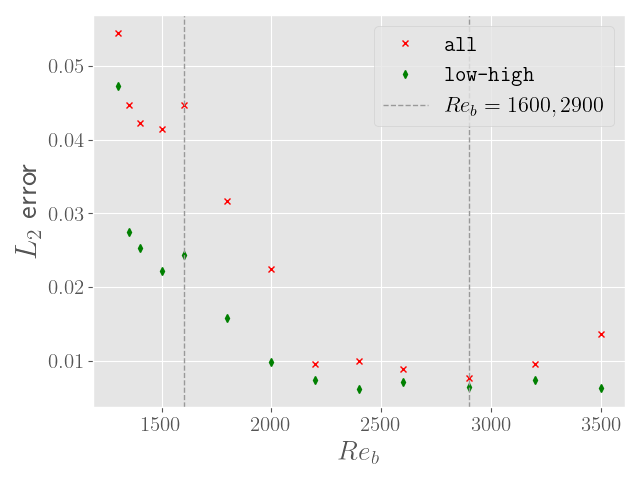}
\caption{Relative $L^2$ error in the velocity prediction for the $\all$ and $\lowhigh$ models using DNS data as reference.}
\label{fig:L2_U_error_vbnn}
\end{figure}

Moreover, Figure \ref{fig:L2_U_error_vbnn} shows the relative $L^2$ error of the two models for the velocity field prediction using DNS data as reference. The $\lowhigh$ model systematically outperforms the $\all$ one both for $Re_b$ training values and for testing values.
Summarizing, %since 
the $\lowhigh$ model 
%has been trained to separately specialize on the two flow regimes, it 
is more accurate than the $\all$ model, in terms of prediction of 
%either equally or better predicts 
the streamwise velocity and secondary motion, both quantitatively and qualitatively. %both in terms of maximum value and maximum location, compared to the $\all$ setting.
For this reason, we decided to use the $\lowhigh$ model to build the ROM space.

\section{ROM results} \label{sec:ROM_results}
This section analyses the results of both the PODG SIMPLE and PODNN approaches in the flow in a square duct setting. To create the reduced models, the $\lowhigh$ high-fidelity $\nut$-VBNN enhanced simulations \red{are used} as snapshots. 

\subsection{Reduced space construction}
We generate simulations for 
$$Re_b \in \{ 1300, 1320, \cdots, 3480, 3500 \},$$ 
%$$Re_b \in \{ 1310, 1330, \cdots, 3470, 3490 \},$$ 
removing the $Re_b$ values contained in the DNS dataset, as \redred{candidate} snapshots for the POD method. \redred{If not explicitly stated in the following, we use all available snapshots for the POD.} To perform a consistent analysis with respect to the previous section, we investigate the possibility of splitting the $Re_b$ interval into two sub-intervals corresponding to the two flow regimes with $Re_b = 2000$ as the threshold value. 
Thus, we define the $\all$-POD approach, with the reduced basis generated using all snapshots, the $\low$-POD approach, using the $Re_b <2000$ snapshots, and the $\high$-POD approach, using the $Re_b > 2000$ snapshots. In particular, we have $N_s = 99$ for the $\all$-POD, $N_s = 30$ for the $\low$-POD, and $N_s = 69$ for the $\high$-POD.

\begin{figure}[h]
\centering
    \begin{subfigure}[t]{0.32\textwidth}
    \includegraphics[width=\textwidth] {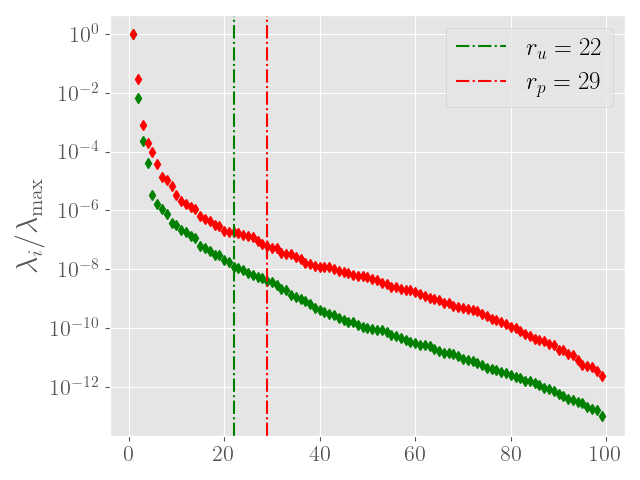}
    \caption{$\all$-POD.}
    \end{subfigure}
    \hfill
    \begin{subfigure}[t]{0.32\textwidth}
    \includegraphics[width=\textwidth] {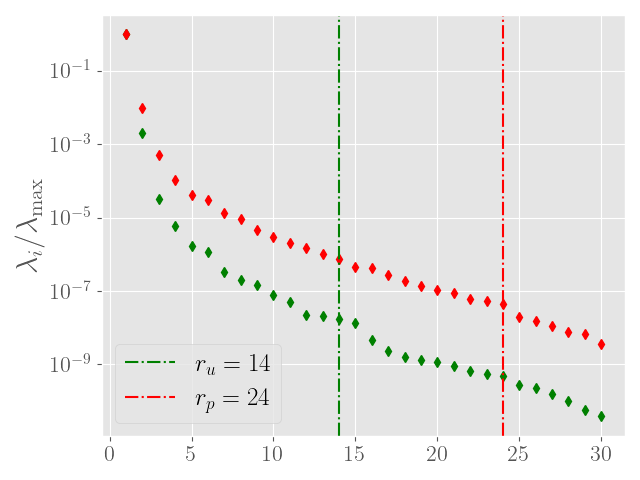}
    \caption{$\low$-POD.}
    \end{subfigure}
    \hfill
    \begin{subfigure}[t]{0.32\textwidth}
    \includegraphics[width=\textwidth] {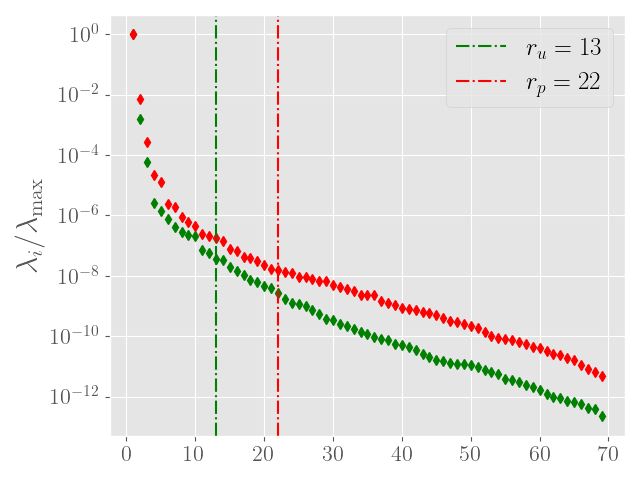}
    \caption{$\high$-POD.}
    \end{subfigure}
\caption{Normalised eigenvalues $\lambda_i/\lambda_{\max}$ of the covariance matrices (green and red for velocity and pressure, respectively).}
\label{fig:eigs}
\end{figure}

Given the positive eigenvalues $\{\lambda_i\}_{i=1}^{N_s}$ of a snapshot covariance matrix, sorted in decreasing order, we select the number of kept modes $r_u$ and $r_p$, for velocity and pressure respectively, for each approach as the smallest $i \in \mathbb{N}$ value such that 
$
\sum_{j=i}^{N_s} \lambda_j / \sum_{j=1}^{N_s} \lambda_j < 10^{-7}.
$
We remark that with $\{\lambda_i\}_{i=1}^{N_s}$, we can refer to velocity or pressure eigenvalues depending on the context.
We obtain $r_u = 22$ and $r_p = 29$, $r_u=14$ and $r_p = 24$ and $r_u=13$ and $r_p = 22$ for the $\all$-, $\low$- and $\high$-POD approaches respectively.   The $\all$-POD requires higher $r_u$ and $r_p$ values to retain the same energy of the other approaches due to the wider phenomenology covered by the snapshots. For completeness, in Figure \ref{fig:eigs}, we show the eigenvalues' decays (green for velocity and red for pressure). 

\subsection{PODNN training}
Depending on the selected reduced basis, the PODNN is trained targeting as outputs the reduced coefficients retrieved from the FOM snapshots using \eqref{eq:norm_system_podnn}. When investigating the flow in a square duct, especially in the data-driven RANS turbulence community, the results analysis is restricted on the velocity field only \cite{Berrone_2022, Wu2019, Wu2018}. Consequently, we train a PODNN for the prediction of the velocity reduced coefficients only, disregarding the pressure contribution. This approach is possible thanks to the non-intrusive nature of the PODNN method that is not affected by the intrinsic coupling between velocity and pressure given by the RANS equations. Specifically, the loss function used to train the PODNN model reads
\begin{equation}
    \mathcal{L}_a(\cdot; \theta_a) = \frac{1}{n \cdot r_u} \sum_{i=1}^n \sum_{j=1}^{r_u}(a_{j,i}-a^{\NN}_{j,i}(\cdot; \theta_a))^2 + \lambda_a \sum_{i=1}^{N_a} (\theta^i_a)^2.
\end{equation}
being $n$ the number of simulations used for training, $a$ the reduced velocity vector obtained by solving \eqref{eq:norml_system_u}, $a^{\NN}$ the predicted reduced velocity coefficients, $\theta_a$ the vector of size $n_a$ containing the weights of the PODNN.

Two models are defined: one to predict the velocity reduced coefficients associated to the $\all$-POD basis and one to predict, through two neural networks, the velocity reduced coefficients associated to the $\low$-POD and $\high$-POD basis, depending if $Re_b < 2000$ or $Re_b > 2000$. We refer to the former as $\all$ and to the second as $\lowhigh$ setting. For all settings, in the hyperparameters tuning stage and for the final hyperparameters choice, fifteen trainings with random weights initialization are performed and the best one in terms of testing error using the DNS $Re_b$ values is kept. For additional insights of the final architectures, we refer to Appendix \ref{sec:PODNN_hyper}.

\subsection{ROM results analysis}
\begin{figure}[h]
\centering
    \begin{subfigure}[t]{0.25\textwidth}
    \includegraphics[width=\textwidth]{images/Ux_OF/nutlowhigh_rfvNLlowhigh/Reb1600.png}\label{fig:Ux_FOM_Reb1600}
    \caption{$u_x$ FOM}
    \end{subfigure}
    \hfill
    \begin{subfigure}[t]{0.25\textwidth}
    \includegraphics[width=\textwidth]{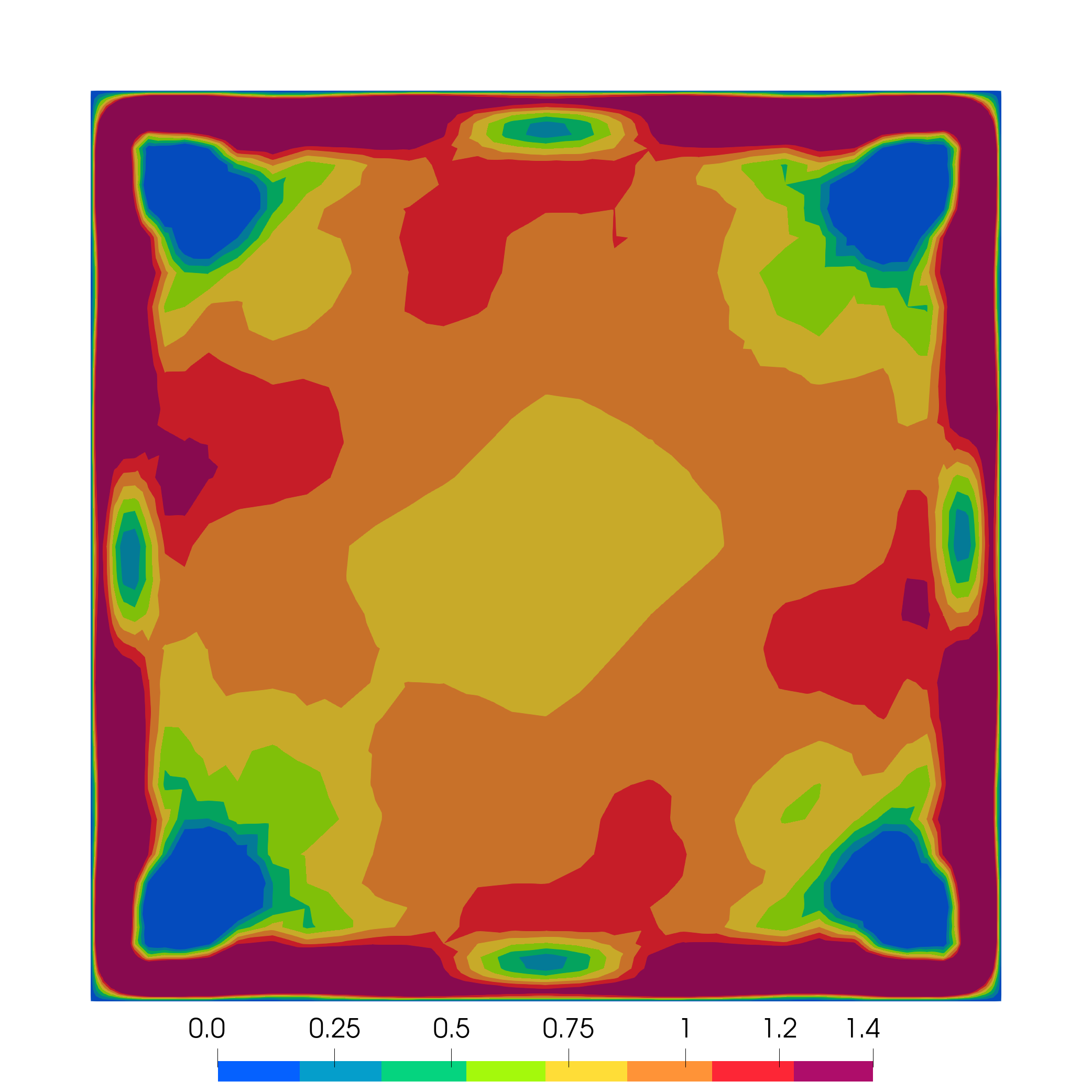}\label{fig:Ux_PODG_Reb1600}
    \caption{$u_x$ PODG SIMPLE}
    \end{subfigure}
    \hfill
    \begin{subfigure}[t]{0.25\textwidth}
    \includegraphics[width=\textwidth]{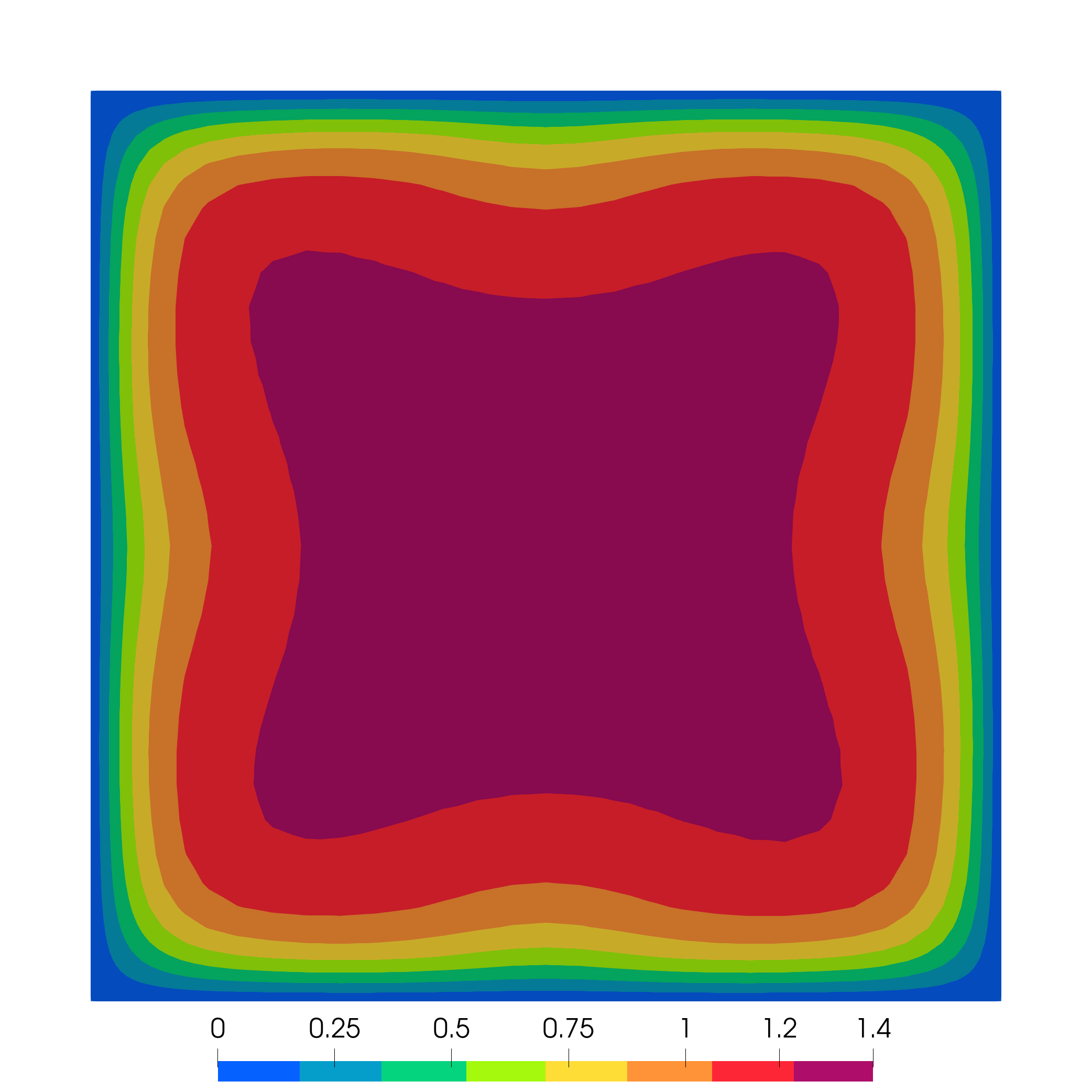}\label{fig:Ux_PODNN_Reb1600}
    \caption{$u_x$ PODNN}
    \hfill
    \end{subfigure}

    \begin{subfigure}[t]{0.25\textwidth}
    \includegraphics[width=\textwidth]{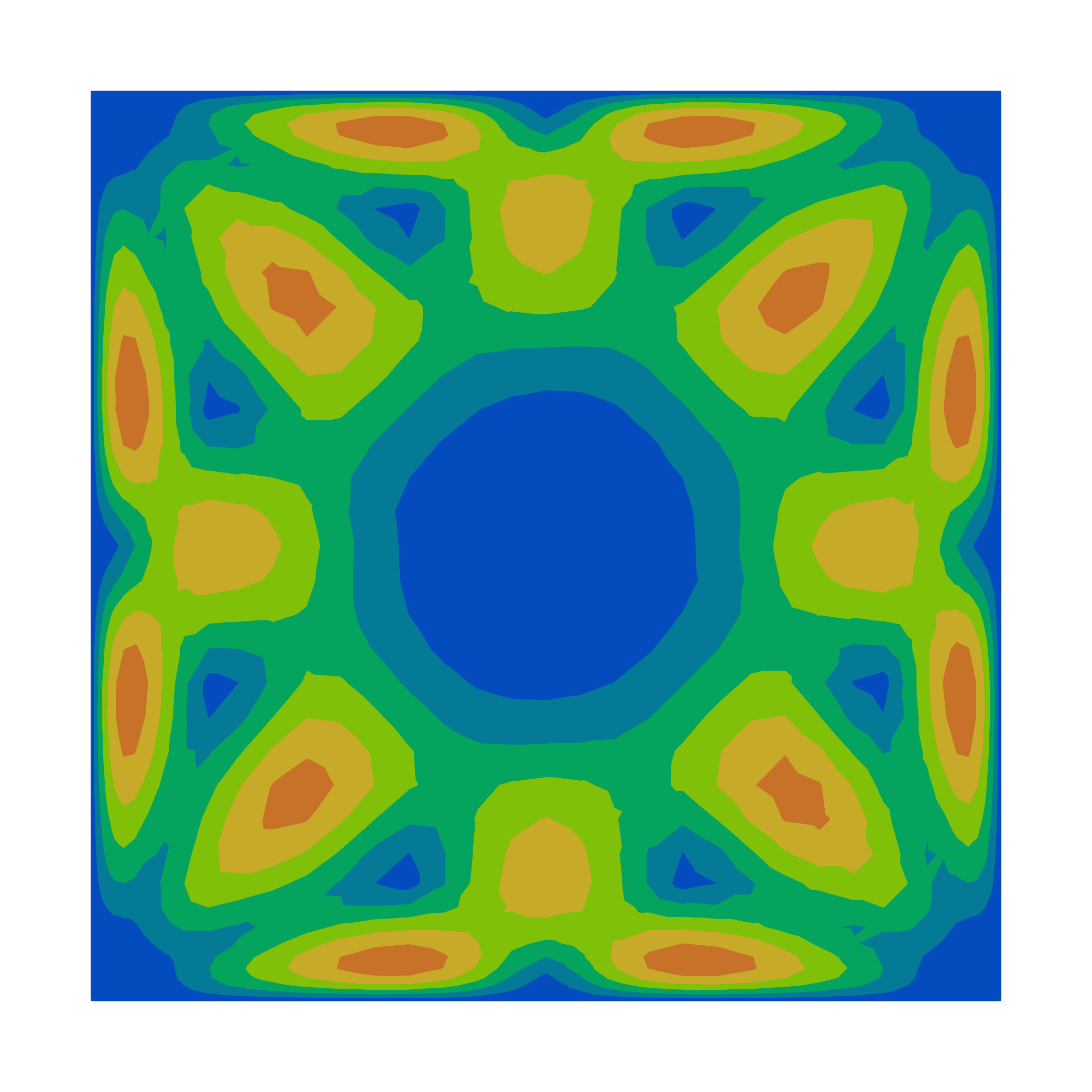}\label{fig:secmot_FOM_Reb1600}
    \caption{$I_s$ FOM}
    \end{subfigure}
    \hfill
    \begin{subfigure}[t]{0.25\textwidth}
    \includegraphics[width=\textwidth]{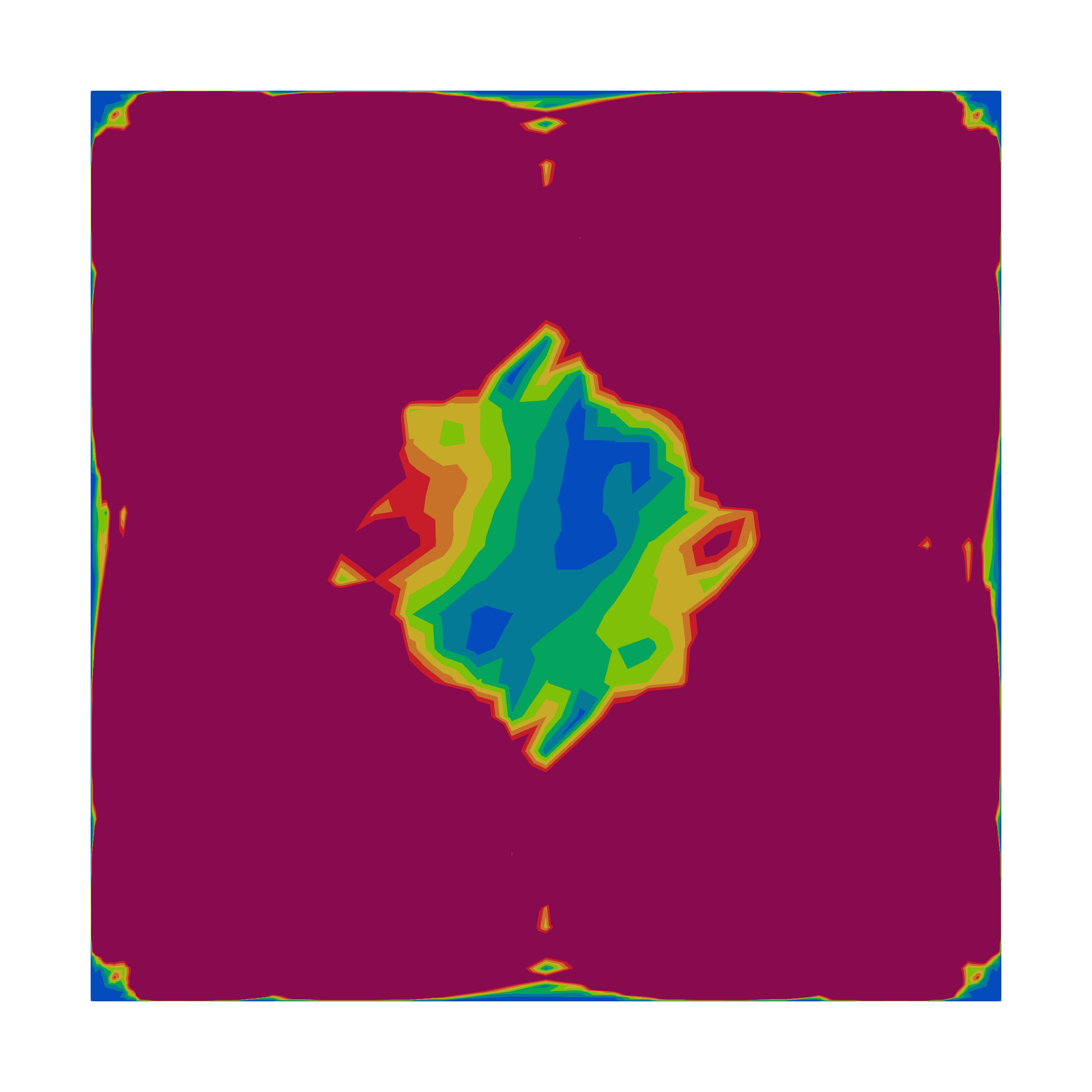}\label{fig:Ux_PODG_Reb1600}
    \caption{$I_s$ PODG SIMPLE}
    \end{subfigure}
    \hfill
    \begin{subfigure}[t]{0.25\textwidth}
    \includegraphics[width=\textwidth]{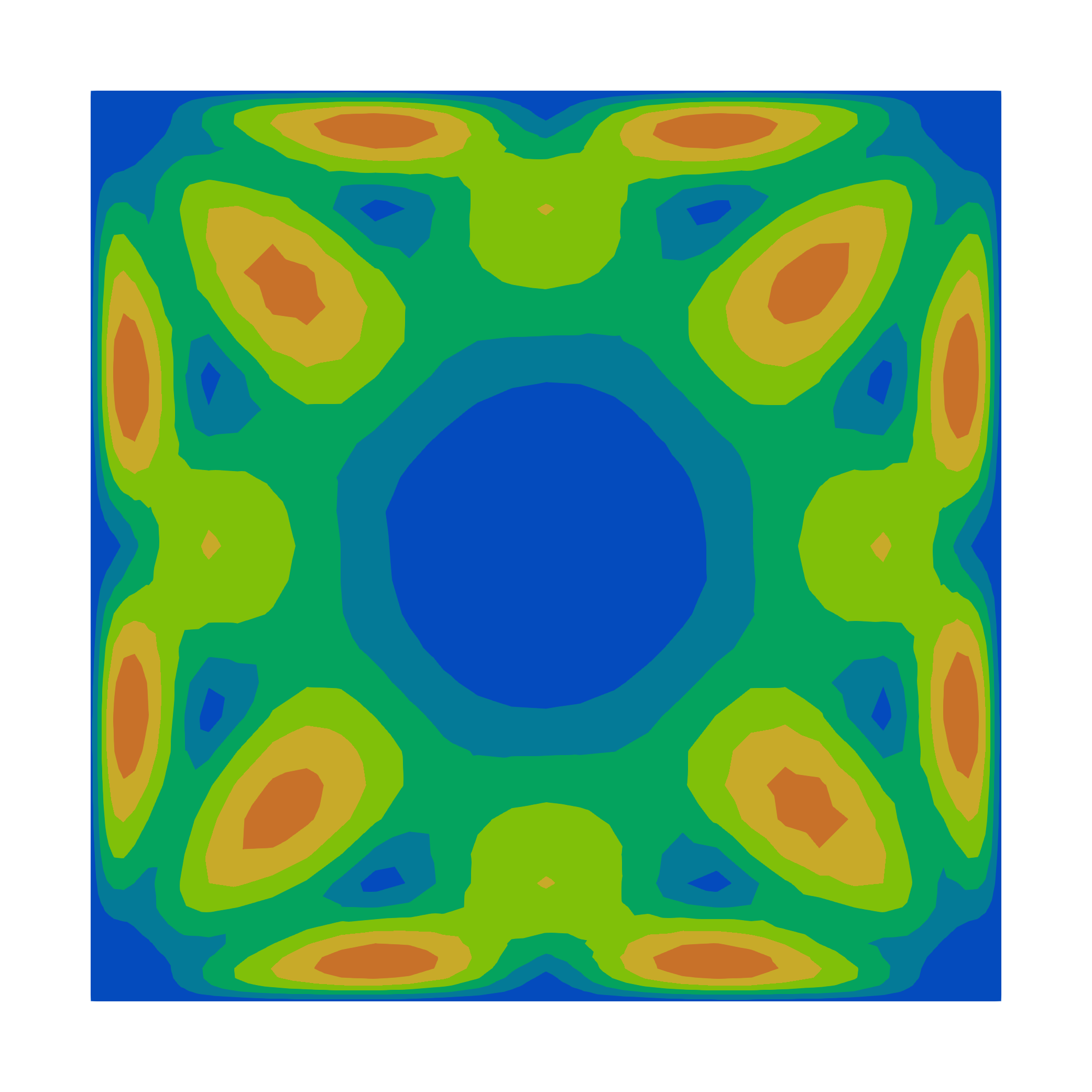}\label{fig:Ux_PODNN_Reb1600}
    \caption{$I_s$ PODNN}
    \hfill
    \end{subfigure}
    \caption{Comparison of $u_x/U_b$ and $I_s$ obtained with the FOM, the PODG SIMPLE and PODNN models at $Re_b = 1600$.}\label{fig:Reb1600_FOM_ROM}
\end{figure}

\begin{figure}[h]
\centering
    \begin{subfigure}[t]{0.25\textwidth}
    \includegraphics[width=\textwidth]{images/Ux_OF/nutlowhigh_rfvNLlowhigh/Reb2900.png}\label{fig:Ux_FOM_Reb2900}
    \caption{$u_x$ FOM}
    \end{subfigure}
    \hfill
    \begin{subfigure}[t]{0.25\textwidth}
    \includegraphics[width=\textwidth]{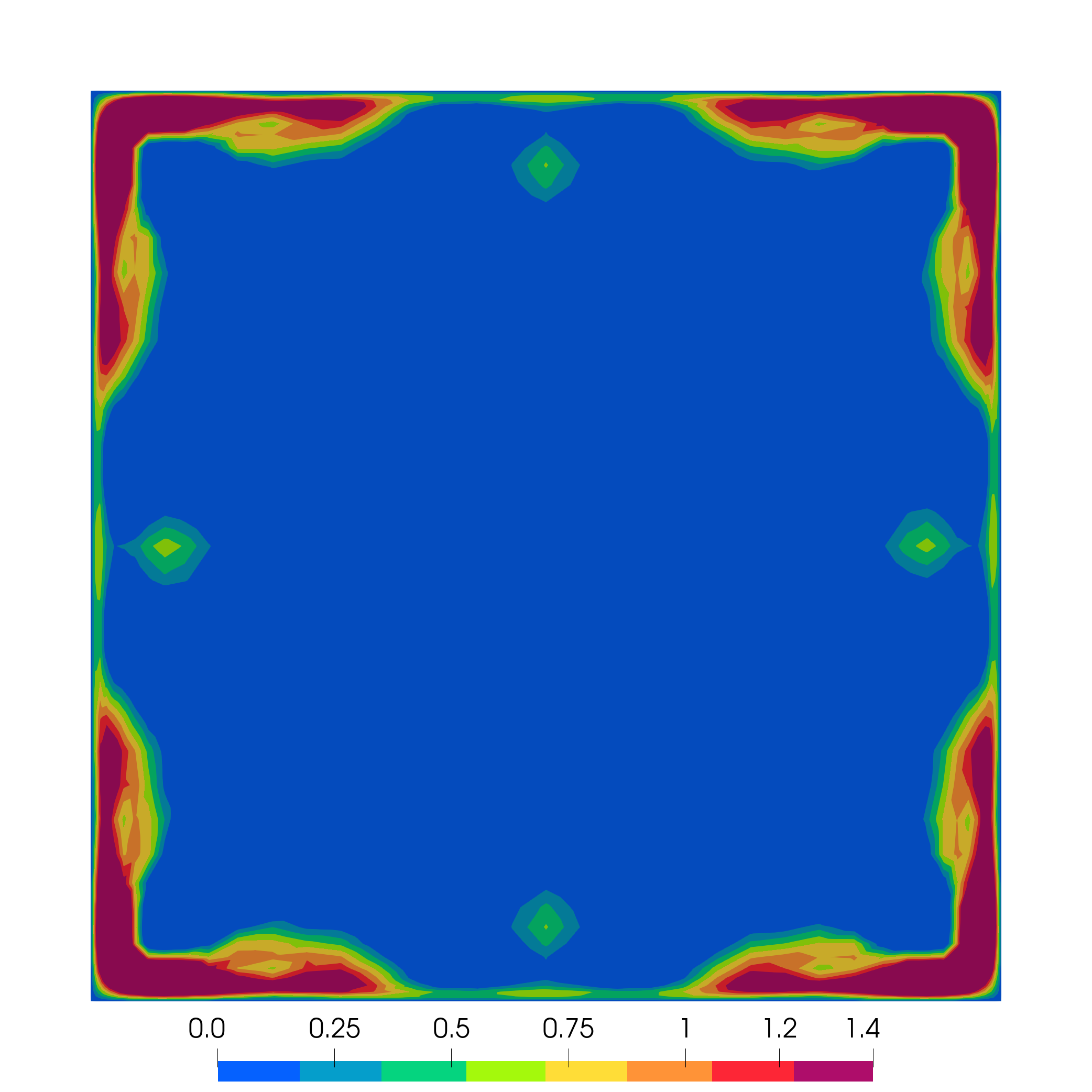}\label{fig:Ux_PODG_Reb2900}
    \caption{$u_x$ PODG SIMPLE}
    \end{subfigure}
    \hfill
    \begin{subfigure}[t]{0.25\textwidth}
    \includegraphics[width=\textwidth]{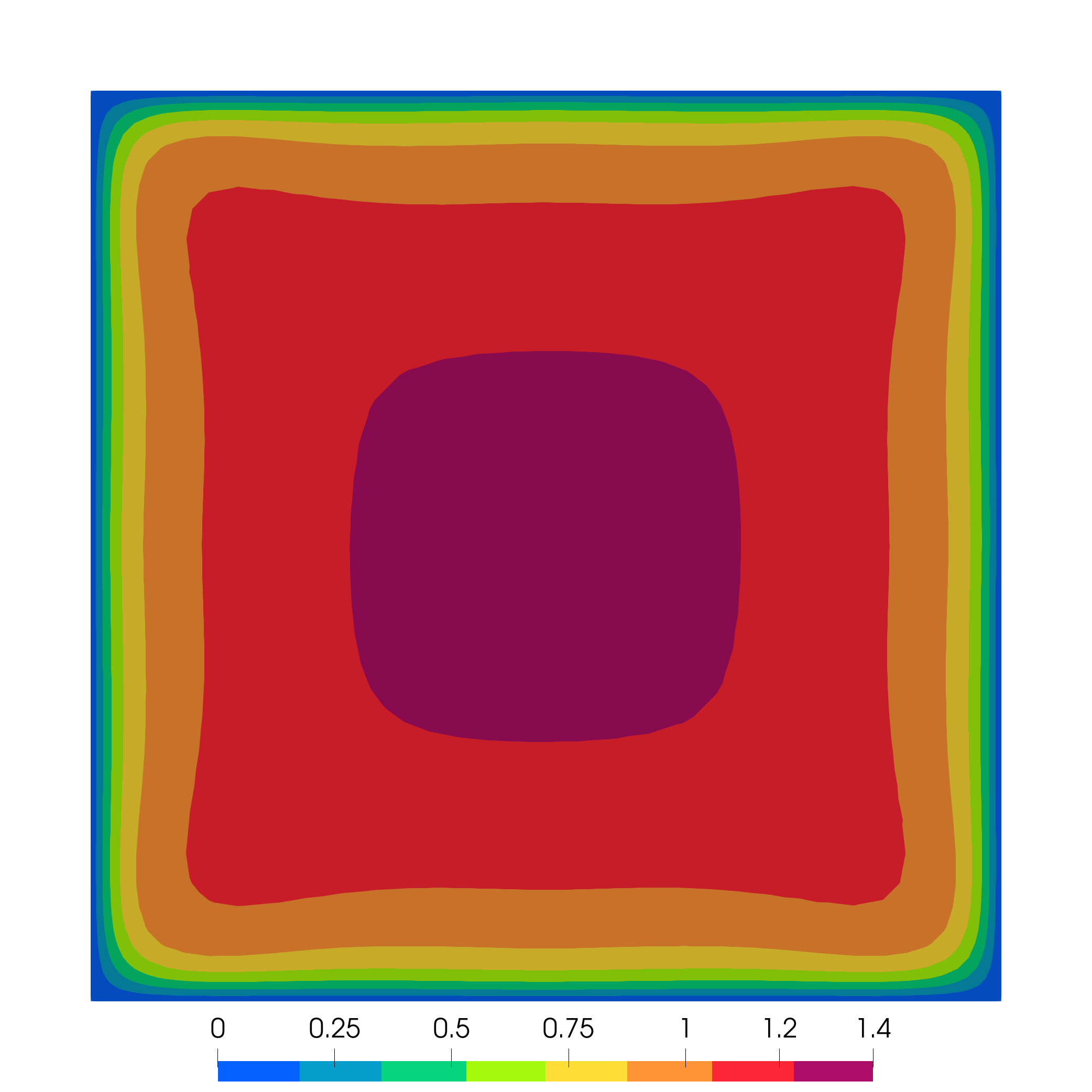}\label{fig:Ux_PODNN_Reb2900}
    \caption{$u_x$ PODNN}
    \hfill
    \end{subfigure}

    \begin{subfigure}[t]{0.25\textwidth}
    \includegraphics[width=\textwidth]{images/secmot_OF/nutlowhigh_rfvNLlowhigh/Reb2900.png}\label{fig:secmot_FOM_Reb2900}
    \caption{$I_s$ FOM}
    \end{subfigure}
    \hfill
    \begin{subfigure}[t]{0.25\textwidth}
    \includegraphics[width=\textwidth]{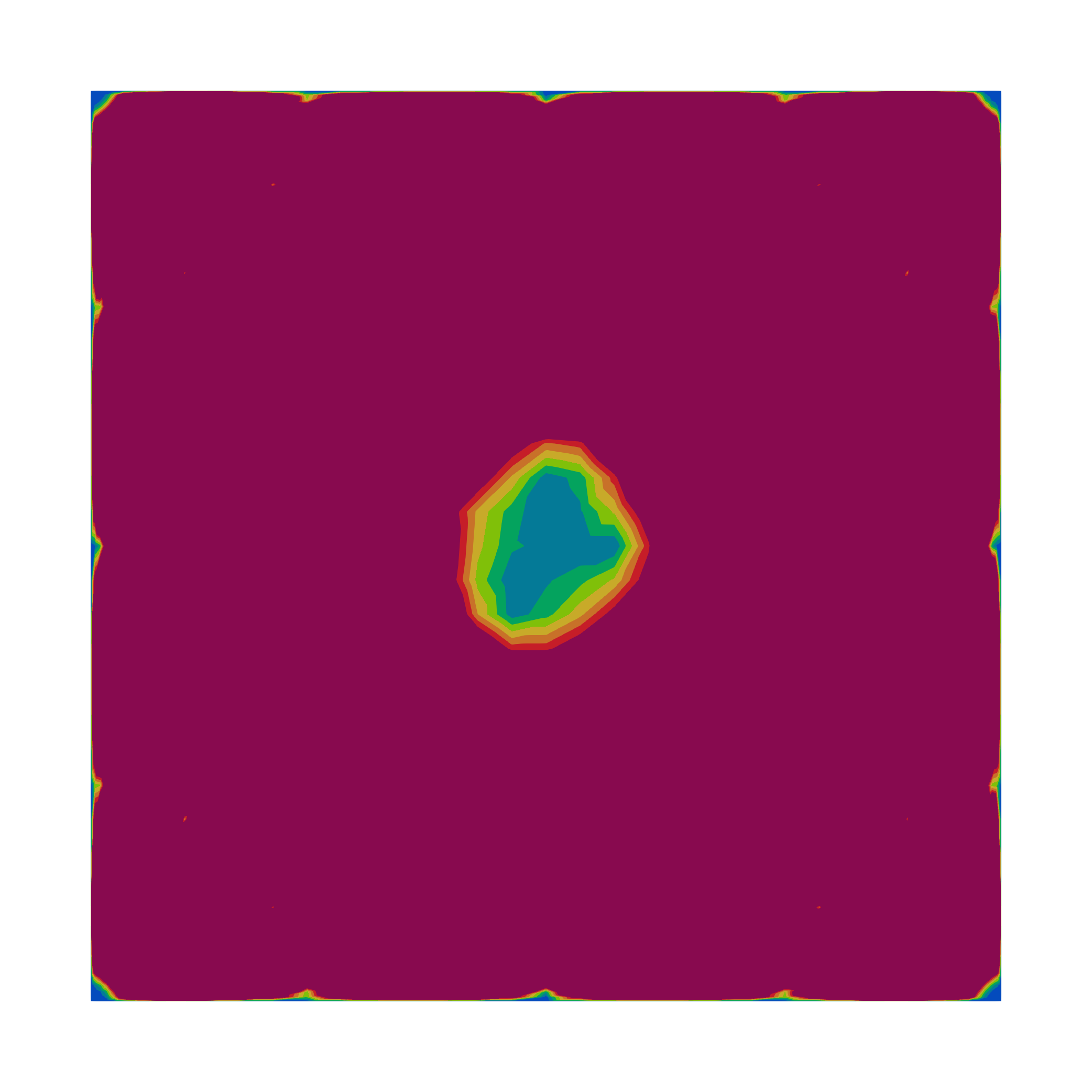}\label{fig:Ux_PODG_Reb2900}
    \caption{$I_s$ PODG SIMPLE}
    \end{subfigure}
    \hfill
    \begin{subfigure}[t]{0.25\textwidth}
    \includegraphics[width=\textwidth]{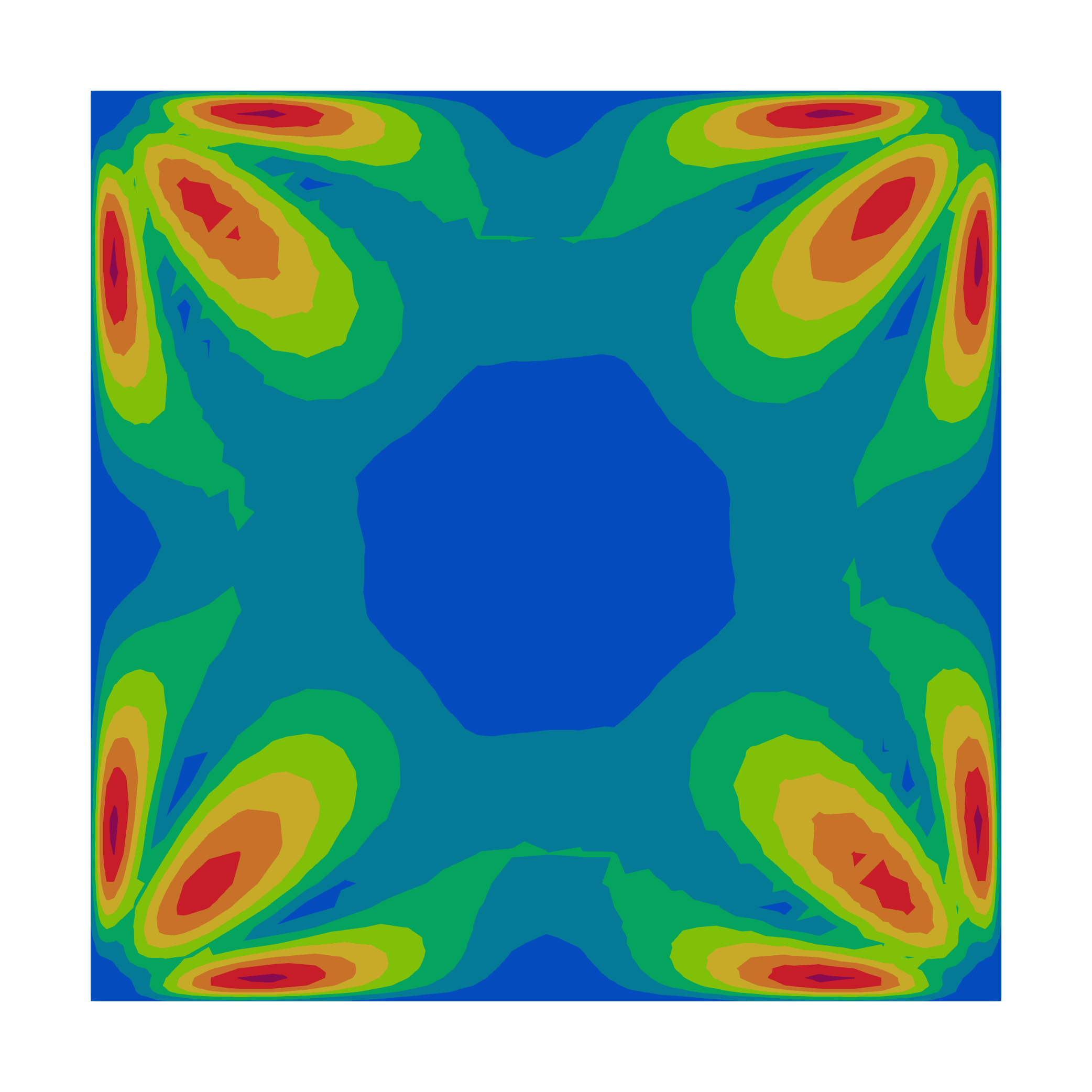}\label{fig:Ux_PODNN_Reb2900}
    \caption{$I_s$ PODNN}
    \hfill
    \end{subfigure}
    \caption{Comparison of $u_x/U_b$ and $I_s$ obtained with the FOM, the PODG SIMPLE and PODNN models at $Re_b = 2900$.}\label{fig:Reb2900_FOM_ROM}
\end{figure}

In this section, we compare the PODG SIMPLE and the PODNN results. For the sake of brevity, we only present the results obtained by choosing the reduced space generated by the $\all$-POD. The same conclusions hold if the $\lowhigh$ approach is investigated. In the PODG SIMPLE method, we use a trivial initial condition for $b$ and $a_i = 0$ for $i=2, \dots, r_u$ and $a_1$ such that the initial reduced velocity $\bm{u}^0_r = a_1 \bm{\varphi}_1$ has streamwise average velocity equal to $U_b$. The PODG SIMPLE method is implemented using the ITHACA-FV library \cite{Stabile2017CAIM,Stabile2017CAF}.\\
Figures \ref{fig:Reb1600_FOM_ROM} and \ref{fig:Reb2900_FOM_ROM} compare $u_x/U_b$ and $I_s$ of the two ROM approaches with respect to the FOM reference both at $Re_b = 1600$ and $Re_b = 2900$. We immediately observe that the PODG SIMPLE approach gives unphysical solutions. In particular, it predicts negative streamwise velocities and the highest streamwise velocities near the wall and drastically overpredicts the secondary motion. In contrast, the PODNN model accurately predicts all flow features both qualitatively and quantitatively for both Reynolds numbers. The difficulty of the PODG SIMPLE algorithm to converge to the correct solution for turbulent flows has been already observed in \cite{Zancanaro_2022_turbulent}, where the authors added snapshots of intermediate iterations of the FOM SIMPLE method to build the snapshots matrix. In this way, the reduced basis is able to drive at the intermediate iterations the ROM algorithm towards the correct solution. Clearly, this approach increases the memory storage requirements and the computational burden of the offline-stage. Because the PODNN method is able to correctly describe all the flow methodology, we decide to focus on this method and to better investigate possible issues mitigation for the PODG SIMPLE algorithm in future works.

\begin{remark}
    The PODG SIMPLE results presented in this section have been obtained projecting the FOM system with the exact $\nut$ and $\rfvNLdim$ fields to be consistent with the FOM procedure. We have also investigated the effect of projecting into the Navier-Stokes equations, thus neglecting the turbulent fields. We observed analogous results in terms of poor flow representation. 
    {We do not present them for the sake of conciseness.}
\end{remark}
The PODNN strategy involves one neural network inference only at inference stage and, consequently, it is computationally cheap. Table \ref{tab:times} shows the average execution times for the LS $k-\varepsilon$ RANS model, for the $\nut$-VBNN model, taking into account only the simulation time with inferred $\nut$ and $\rfvNLdim$ fields using as initial condition the LS $k-\varepsilon$ fields, and for the PODNN model. The average is performed over 110 simulations with $Re_b = 1310, 1330, \dots, 3490$. The speedup is defined as the ratio of the total time to run a FOM simulation, i.e., the sum of the first two columns of Table \ref{tab:times}, divided by the PODNN inference time. From a FOM perspective, using the $\nut$-VBNN increases the execution time by 18$\%$ compared to running the LS $k-\varepsilon$ model only. From a ROM perspective, the PODNN approach gives an average speedup of $O(10^5)$ with respect to FOM, thus justifying the PODNN choice.
\begin{table}[h!]
\centering
\caption{Average time over 110 simulations for the $k-\varepsilon$ RANS model, the LS $\nut$-VBNN enhanced RANS model (with LS $k-\varepsilon$ RANS results as initial condition) and for the PODNN. The speedup of the PODNN is computed using the sum of the first two columns.} 
\renewcommand{\arraystretch}{1.2}
\begin{tblr}{
  columns={halign=c},
  hline{1,1} = {-}{},
  hline{1,2} = {-}{},
  hline{1,3} = {-}{},
  vline{1,1} = {-}{},
  vline{1,2} = {-}{},
  vline{1,3} = {-}{},
  vline{1,4} = {-}{},
  vline{1,5} = {-}{},
}
$k-\varepsilon$ time [$s$] & $\nut$-VBNN time [$s$] & PODNN time [s] & speedup  \\ 
1.38 e2 & 2.51 e1 & 7.72 e-4 &  2.12 e5 \\ 
\end{tblr}
\label{tab:times}
\end{table}

\redred{
\subsection{The effect of the reduced dimension on the PODNN}
In the previous Section, the PODNN has proved to be the suitable ROM framework for the investigated flow. Hence, in this section and the following ones, we deepen our analysis on the PODNN, both to further increase its accuracy and to better understand possible limitations.

Specifically, in this Section, we discuss the role of the reduced dimension $r_u$ in the PODNN approximation.
To this end, we study five PODNN settings with
different outputs dimensions, defined by $r_u \in \{ 6, 10, 14, 18, 22 \}.$

The PODNNs are tested on unseen parameters $Re_b \in \{1310, 1330, \dots, 3470, 3490\}$. Figure \ref{fig:PODNN_red_dim} compares the $L^2$ relative error of the predicted velocity fields with respect to the FOM approximation. Specifically, the left panel compares, for each setting, the average error over the fifteen trainings, while the right panel shows the error of the best models in terms of average error over the parameters. 
The best choices correspond to $r_u = 6$ and $r_u = 10$. Oppositely, the worst settings are the ones with highest $r_u$ values. This phenomenon occurs on all the parametric interval, but it is more evident for the lowest $Re_b$ values and for $2000 < Re_b < 2900$. This behaviour is in contrast to standard intrusive ROM methods \cite{Quarteroni2015-bh}, but it is not unexpected for the PODNN.
Indeed, adding outputs to the PODNN for fixed inputs, yields to a more challenging training which can lead to an increased prediction error.
\begin{figure}
\centering
    \begin{subfigure}[t]{0.45\textwidth}
    \includegraphics[width=\textwidth] {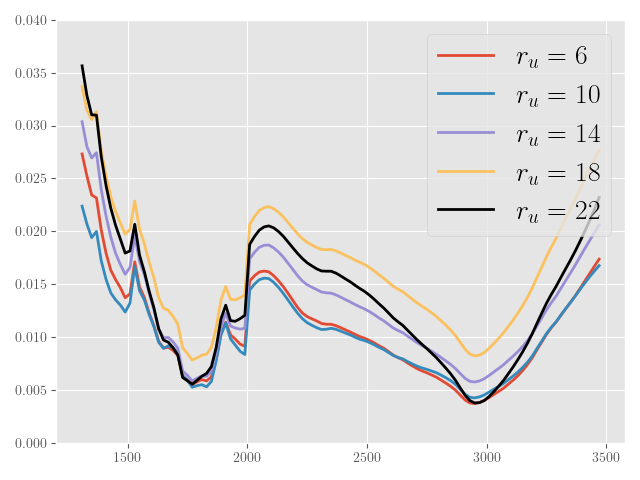}
    \caption{} \label{fig:PODNN_red_dim_avg}
    \end{subfigure}
    \hfill
    \begin{subfigure}[t]{0.45\textwidth}
    \includegraphics[width=\textwidth] {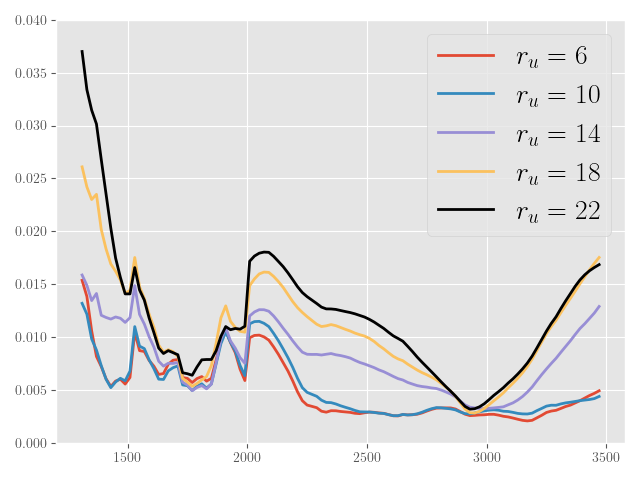}
    \caption{} \label{fig:PODNN_red_dim_best}
    \end{subfigure}
\caption{\redred{Comparison of the average relative $L^2$ errors, on the left, and the relative $L^2$ error of the best model, on the right, for the PODNN for different reduced dimensions $r_u$.}}
\label{fig:PODNN_red_dim}
\end{figure}
\subsection{The effect of the dataset size for the PODNN}
In this Section we investigate the role of the dataset size on the PODNN performance. With this aim, we define three snapshots datasets: 
\begin{itemize}
    \item the original $Re_b \in \{ 1320, 1340, \cdots, 3460, 3480 \}$ one, corresponding to $N_s = 99$;
    \item one with $Re_b \in \{ 1340, 1380, \dots, 3420, 3460\}$, corresponding to $N_s = 53$;
    \item one with $Re_b \in \{1320, 1420, \dots, 3320, 3420 \}$, corresponding to $N_s = 22$. 
\end{itemize}
For a fair comparison, we fix $r_u=6$, guided by the best model in the previous analysis.

It is clear that same $r_u$ yields to different reduced coefficients in the three settings, since they are related to POD compressions applied to different snapshots matrices. 
Thus, using less snapshots not only affects the dataset size for the PODNN training, but also the quality of the reduced basis used in the POD procedure. Figure \ref{fig:PODNN_data_size} compares the relative $L^2$ error for the different dataset sizes. Again, the left panel compares the average error over the fifteen trainings, while the right panel shows the error of the best models. For $Re_b > 1700$, models trained using $N_s = 22$ provide on average systematically less accurate results, while its best model has comparable results to the best two of the other approaches for $Re_b < 2000$, but higher errors elsewhere. The $N_s=99$ and $N_s=53$ models have similar accuracy, with the former approach being slightly more accurate for some $Re_b$ values. We can conclude that using only $N_s = 22$ snapshots negatively affects the accuracy, even if the overall average relative error is always smaller than $2\%$, while $N_s=53$ seems to be a good choice to balance dataset size and accuracy.
\begin{figure}
\centering
    \begin{subfigure}[t]{0.45\textwidth}
    \includegraphics[width=\textwidth] {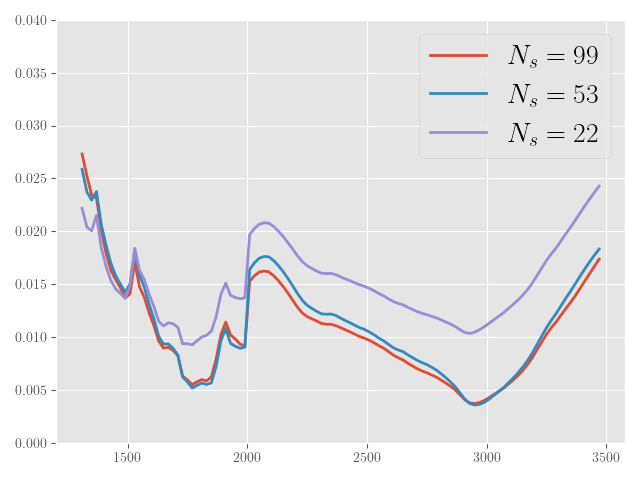}
    \caption{} \label{fig:PODNN_dataset_size_avg}
    \end{subfigure}
    \hfill
    \begin{subfigure}[t]{0.45\textwidth}
    \includegraphics[width=\textwidth] {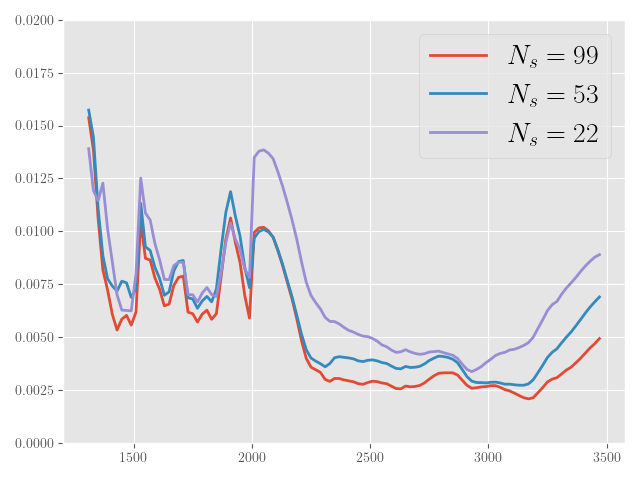}
    \caption{} \label{fig:PODNN_dataset_size_best}
    \end{subfigure}
\caption{\redred{Comparison of the average relative $L^2$ errors, on the left, and the relative $L^2$ error of the best model, on the right, for the PODNN for different snapshots dataset size $N_s$.}}
\label{fig:PODNN_data_size}
\end{figure}
}

\subsection{The effect of training the PODNN on separate flow regimes}
\begin{figure}
\centering
    \begin{subfigure}[t]{0.45\textwidth}
    \includegraphics[width=\textwidth] {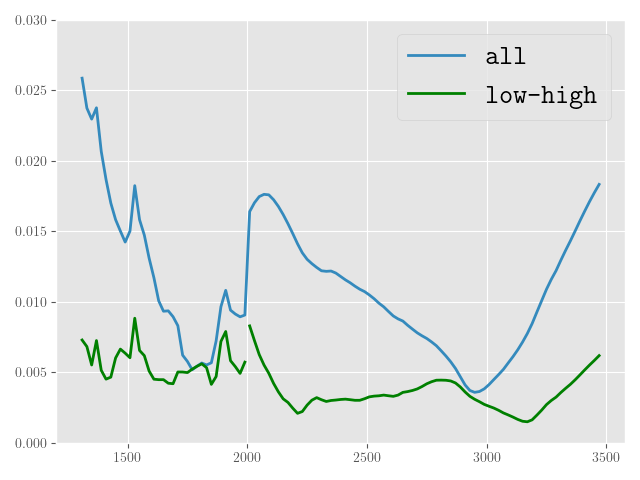}
    \caption{} \label{fig:PODNN_error_no_overlap}
    \end{subfigure}
    \hfill
    \begin{subfigure}[t]{0.45\textwidth}
    \includegraphics[width=\textwidth] {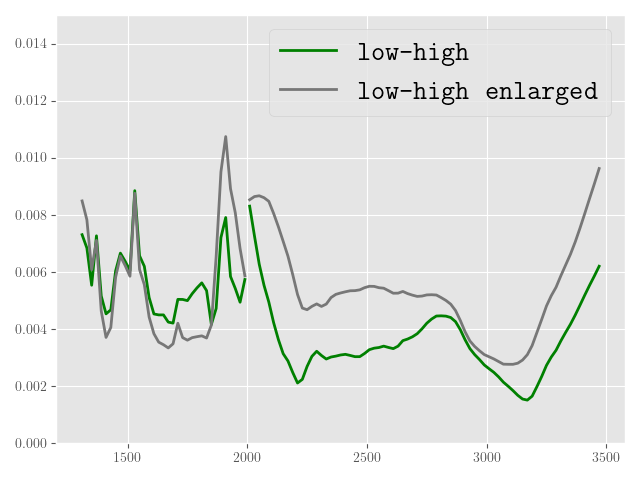}
    \caption{} \label{fig:PODNN_error_with_overlap}
    \end{subfigure}
\caption{\redred{Comparison of the average relative $L^2$ errors for the PODNN inferred $\vel$ fields. On the left, the $\all$ and $\lowhigh$ approaches are compared. On the right, the $\lowhigh$ and $\lowhigh \texttt{ enlarged}$ ones are compared.}}
\label{fig:PODNN_error_two_intervals}
\end{figure}

As for the $\nut$-VBNN model at the FOM level, we investigate the effect of training the PODNN separately depending on the flow regime. To do that, we compute the reduced coefficients for the training using the $\low$-POD or $\high$-POD reduced basis, depending on the $Re_b$ number. Then, two neural networks are trained, one for each flow regime. We denote the resulting model consisting of two trained networks as $\lowhigh$. Analogously, we denote the PODNN model trained using all data as $\all$. For the sake of simplicity, the same hyperparameters are selected for all networks. \redred{Because of the results obtained in the previous sections, to provide a fair comparison between models, we select $r_u = 6$ for all approaches. Furthermore, we select $N_s=53$ for the $\all$ approach. These snapshots are split in the two regimes for the $\lowhigh$ approach.} As before, fifteen trainings are performed for each neural network. Figure \ref{fig:PODNN_error_no_overlap} shows the average $L^2$ error with respect to the FOM $\nut$-VBNN model across the fifteen trainings for both the $\all$ and the $\lowhigh$ predicted velocity fields. We point out that the error has a discontinuity at $Re_b = 2000$ even for the $\all$ model. This is due to using as FOM the $\lowhigh$ $\nut$-VBNN model that uses two distinct models depending on the interval. From the Figure, it appears that using two neural networks, each of them focusing on a specific flow regime, extremely increases the accuracy. As a matter of fact, the maximum $\all$ error is greater than \redred{2.5}$\%$ while the higher error of the $\lowhigh$ model is smaller than \redred{1}$\%$. \redred{For the sake of brevity, we do not show the comparison between best models because similar conclusions can be drawn.}\\
Finally, we also investigate the effect of training the two neural networks using two $Re_b$ intervals that partially overlap\redred{, aiming to reduce the accuracy jump of the $\lowhigh$ method across $Re_b=2000$}. In particular, the $\low$ model is trained using data for $Re_b \leq 2200$ while the $\high$ model with data for $Re_b \geq 1800$. The averaged error of this approach called \texttt{low-high enlarged} is shown in Figure \ref{fig:PODNN_error_with_overlap} and compared to the \texttt{low-high} one. \redred{For both flow regimes, this strategy significantly increases the error in the $1800 < Re_b < 2800$ interval and in the $Re_b > 3000$ one. Consequently, such approach is not suggested.}

\section{Conclusions and perspectives} \label{sec:Conclusions}
In this work, we investigate the possibility of combining ML-based RANS closure models and ROM. In particular, the $\nut$-VBNN model \cite{Oberto2025, Berrone_2022} is used to close through two neural networks the RANS equations. Such a model is trained using DNS data to attain high accuracy while maintaining the computational cost of the RANS framework. Once properly trained, the $\nut$-VBNN model is used to generate FOM simulations to retrieve a reduced space in which \blue{to} apply ROM methodologies. In this work, a non-intrusive ROM model, i.e., purely data-driven, called PODNN, and an intrusive one, i.e., projection-based, called PODG SIMPLE are tested. The former uses a neural network to approximate the map from the parameter-space to the reduced coefficients of the reduced solution. The latter mimics the SIMPLE iterative algorithm used at the FOM level to decouple velocity and pressure fields.\\
These methodologies, both at the FOM and the ROM level, are tested in the flow in a square duct with varying bulk Reynolds number $Re_b$ \cite{Pinelli2010}.

To schematically synthesise the results of the present work:
\begin{itemize}
    \item at the FOM level, the $\nut$-VBNN is trained to build an accurate model. In contrast with the common choice in literature, we train our model covering both the transitional regime and the fully turbulent one. In this way, we challenge our ML-based RANS model and investigate the effect of training specialization into the two distinct regimes;
    \item at the ROM level, we compare a non-intrusive model and an intrusive one. The former is well-suited to our setting while the latter provides inaccurate results;
    \item \redred{finally, a further analysis is carried out for the non-intrusive ROM model. Specifically, the role of the reduced dimension size, of the dataset size and of the training splitting into the two flow regimes effects are investigated.}
\end{itemize}
Many are the valuable extensions of this contribution. First, a deeper analysis on how to stabilise the PODG SIMPLE approach for the flow in a square duct and, in general, for turbulent flows is worth investigating. As a matter of fact, it is well known that intrusive approaches for convection dominated and turbulent flows can require ad-hoc and problem-dependent stabilization techniques \cite{Zoccolan2025237,Siena2025}. \blue{Another valuable extension regards unsteady problems described by unsteady RANS (URANS). In this setting, the evolve-filter-relax (EFR) framework \cite{Ivagnes2025, Ervin2012-bs, Quaini2024-af} is an appropriate algorithm to stabilise the PODG problem}. Another line of research is represented by defining closure models at the ROM level to retrieve the $\nut$ and $\rfvNLdim$ reduced fields. Furthermore, it would be of great interest to investigate more complex flow phenomena\red{, with varying domains and higher dimensional parameter spaces} and move to more sophisticated, possibly not based on a linear expansion at the ROM level, non-intrusive models for flows with high Kolmogorov $n$-width.

\section*{Acknowledgments}
All authors are members of the Gruppo Nazionale Calcolo Scientifico-Istituto Nazionale di Alta Matematica (GNCS-INdAM). SB kindly acknowledges financial support by PNRR M4C2 project of CN00000013 National Centre for HPC, Big Data and Quantum Computing (HPC) CUP: E13C22000990001. DO acknowledges the support provided by the European Union-NextGenerationEU, in the framework of the iNEST-Interconnected Nord-Est Innovation Ecosystem (iNEST ECS00000043– CUP G93C22000610007) consortium.
Moreover, this study was carried out within the ``20227K44ME - Full and Reduced order modelling of coupled systems: focus on non-matching methods and automatic learning (FaReX)" project – funded by European Union – Next Generation EU  within the PRIN 2022 program (D.D. 104 - 02/02/2022 Ministero dell’Università e della Ricerca). % This manuscript reflects only the authors’ views and opinions and the Ministry cannot be considered responsible for them. 
DO and MS thank the INdAM-GNCS Project “Metodi numerici efficienti per problemi accoppiati in sistemi complessi” (CUP E53C24001950001). {MS thanks and the ECCOMAS EYIC Grant ``CRAFT: Control and Reg-reduction in Applications for Flow Turbulence".
\bibliographystyle{acm} %ieeetr
\bibliography{biblio}

\appendix
\section{Invariants and vector basis of the $\nut$-VBNN model}\label{sec:inv_vb}
\vspace*{12pt}
In \cite{Oberto2025}, the following assumption is made:
$$
\rfvNL = \rfvNL(\mathbf{s},\mathbf{w},\divs,\gradk, Re_d),
$$
where $\mathbf{s} = \frac{k}{\varepsilon}  \bS$, $\mathbf{w} = \frac{k}{\varepsilon}  \bW$, $\divs = \frac{k^{5/2}}{\varepsilon^2}  \dive \bS$ and $\gradk = \frac{k^{1/2}}{\varepsilon}  \nabla k$ are the dimensionless counterparts of $\bS$, $\bW$, $\dive \bS$ and $\nabla k$,  respectively. Finally, $Re_d = \min(\frac{\sqrt{k} d}{50 \nu},2)$ is the wall-distance based Reynolds number, where $d$ is the wall distance.

With this assumption, we obtain \eqref{eq:rfvNL_lin_comb} \cite{Zheng1994}, where the vector basis reads
\begin{equation*} \label{eq:basis_div_tau}
\begin{split}
&\mathbf{v}_1 = \divs, \quad \mathbf{v}_2 = \mathbf{s} \ \divs, \quad \mathbf{v}_3 = \mathbf{s}^2 \ \divs,
\\
& \mathbf{v}_4 = \mathbf{w} \ \divs, \quad \mathbf{v}_5 = \mathbf{w}^2 \ \divs, \quad \mathbf{v}_6 = (\mathbf{sw}+\mathbf{ws}) \ \divs, \\ 
\noalign{\vskip5pt}
&\mathbf{v}_7 = \gradk, \quad \mathbf{v}_8 = \mathbf{s} \ \gradk, \quad \mathbf{v}_9 = \mathbf{s}^2 \ \gradk,
\\
& \mathbf{v}_{10} = \mathbf{w} \ \gradk, \quad \mathbf{v}_{11} = \mathbf{w}^2 \ \gradk, \quad \mathbf{v}_{12} = (\mathbf{sw}+\mathbf{ws}) \ \gradk,
\end{split}
\end{equation*}
while the invariants are
\begin{equation*} \label{eq:invariants_div_tau}
\begin{split}
&\lambda_1 = (\divs)^T (\divs), \quad \lambda_2 = \tr(\mathbf{s}^2), \quad \lambda_3 = \tr(\mathbf{s}^3), \quad  \lambda_4 = \tr(\mathbf{w}^2), 
\\
&\lambda_5 = \tr(\mathbf{s}\mathbf{w}^2), \quad \lambda_6 = \tr(\mathbf{s}^2\mathbf{w}^2), \quad \lambda_7 = \tr(\mathbf{s}^2\mathbf{w}^2\mathbf{s}\mathbf{w}), \quad \lambda_8 = (\divs)^T \mathbf{s} (\divs), 
\\
&\lambda_9 = (\divs)^T \mathbf{s}^2 (\divs), \quad \lambda_{10} = (\divs)^T \mathbf{w}^2 (\divs), \quad  \lambda_{11} = (\divs)^T \mathbf{s}\mathbf{w} (\divs),
\\
&\lambda_{12} = (\divs)^T \mathbf{s}^2 \mathbf{w} (\divs), \quad \lambda_{13} = (\divs)^T \mathbf{w}\mathbf{s}\mathbf{w}^2 (\divs),\\ 
\noalign{\vskip5pt}
&\lambda_{14} = (\gradk)^T (\gradk), \quad \lambda_{15} = (\gradk)^T \mathbf{s} (\gradk), \quad \lambda_{16} = (\gradk)^T \mathbf{s}^2 (\gradk), \\
& \lambda_{17} = (\gradk)^T \mathbf{w}^2 (\gradk), \quad \lambda_{18} = (\gradk)^T \ \divs,
\quad  \lambda_{19} = (\gradk)^T \mathbf{s}\mathbf{w} (\gradk),
\\
&\lambda_{20} = (\gradk)^T \mathbf{s}^2 \mathbf{w} (\gradk), \quad \lambda_{21} = (\gradk)^T \mathbf{w}\mathbf{s}\mathbf{w}^2 (\gradk), \quad \lambda_{22} = (\gradk)^T \mathbf{s}\mathbf{w} (\divs), \\
&\lambda_{23} = (\gradk)^T \mathbf{s}^2 \mathbf{w} (\divs), \quad \lambda_{24} = (\gradk)^T  \mathbf{w} (\divs), \\
&\lambda_{25} = (\gradk)^T \mathbf{w}\mathbf{s}\mathbf{w}^2 (\divs), \quad \lambda_{26} = (\gradk)^T (\mathbf{s}\mathbf{w} + \mathbf{w}\mathbf{s}) (\divs) ,\\ 
\noalign{\vskip5pt}
&\lambda_{27} = Re_d.
\end{split}
\end{equation*}
The first twenty-six invariants come from the dependencies on $\mathbf{s},\mathbf{w},\widetilde{\nabla \cdot \mathbf{S}},\widetilde{\nabla k}$. We observe that the invariant $\tr(\mathbf{s})$ is neglected because of the incompressibility constraint.\\
Finally, because $\tnut$ is a scalar field, we directly use \eqref{eq:nut_lin_comb} without involving any linear expansion.

\section{The role of $\nut$ and $\bold{t}^\perp$ in the secondary motion prediction} \label{sec:role_nut_rfvNL}
\vspace*{12pt}

In Figure \ref{fig:nut_rans_dns}, we compare the $\nut$ field of the DNS simulation interpolated on the RANS mesh and the one predicted by the standard LS $k-\varepsilon$ for $Re_b = 2900$. The maxima loci of the DNS are not situated at the centre of the section, whereas the LS $k-\varepsilon$ field monotonically increases moving closer to the section centre. Additionally, the LS $k-\varepsilon$ field assumes higher values with respect to the DNS $\nut$. 
\begin{figure}[h]
\centering
    \begin{subfigure}[t]{0.25\textwidth}
    \includegraphics[width=\textwidth] {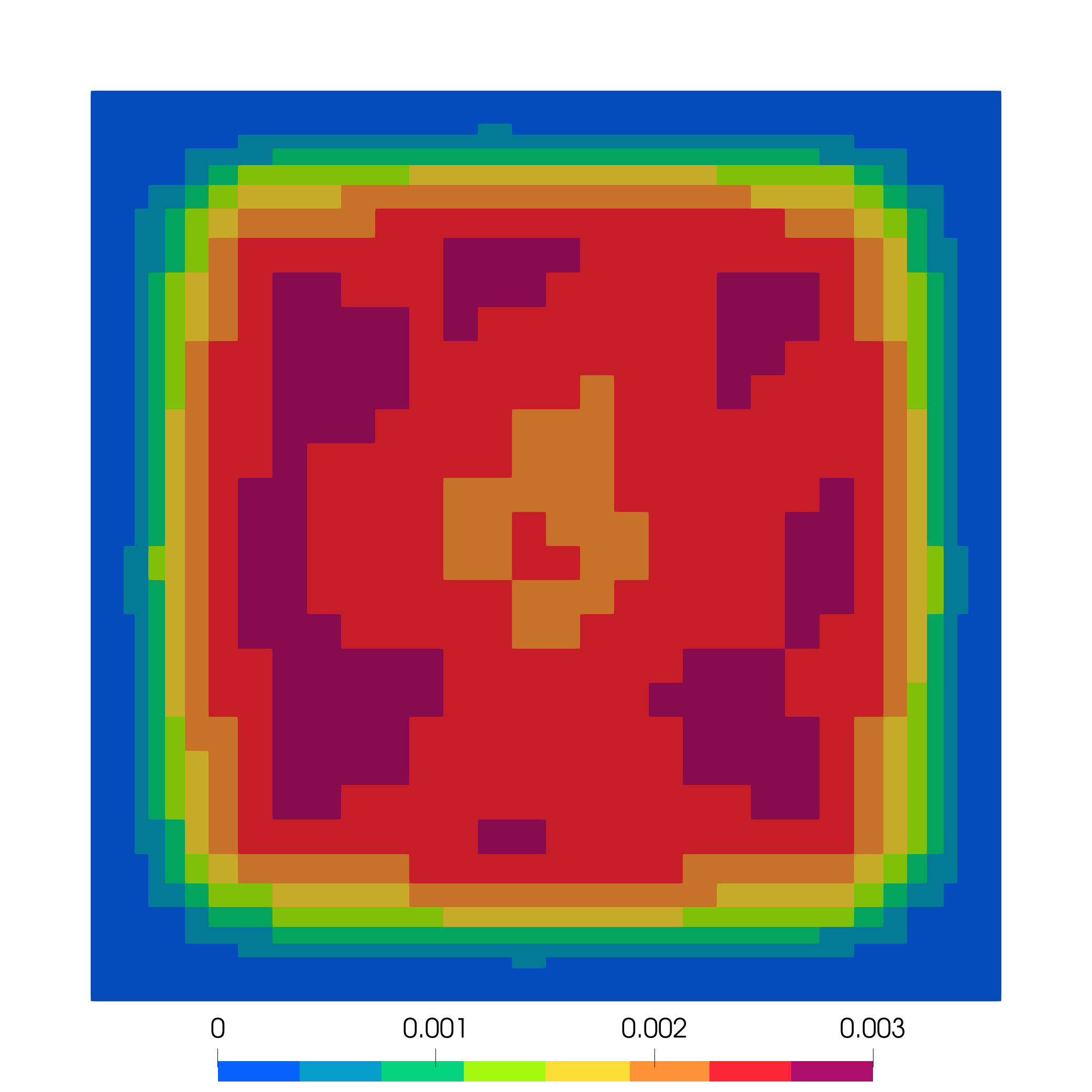}
    \caption{$\nut$ from DNS} \label{fig:snut_DNS_Reb2900_OF}
    \end{subfigure}
    \
    \begin{subfigure}[t]{0.25\textwidth}
    \includegraphics[width=\textwidth] {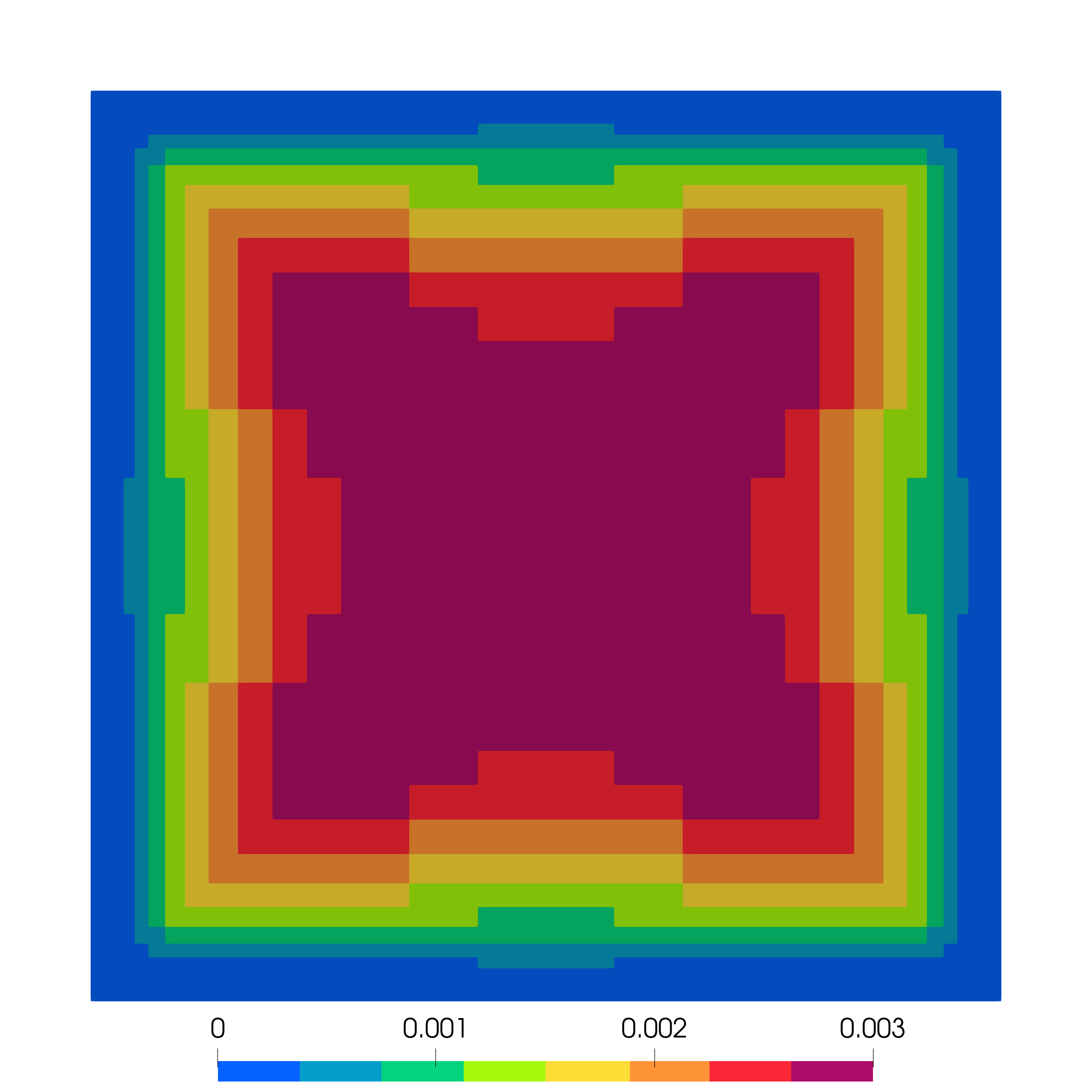}
    \caption{$\nut$ from $k-\varepsilon$ models} \label{fig:nut_kepsilon_Reb2900}
    \end{subfigure}
\caption{Comparison between the DNS $\nut$ (left) and the one predicted by the $k-\varepsilon$ model (right) for $Re_b  = 2900$.}
\label{fig:nut_rans_dns}
\end{figure}
\begin{figure}
\centering
    \begin{subfigure}[t]{0.240\textwidth}
    \includegraphics[width=\textwidth] {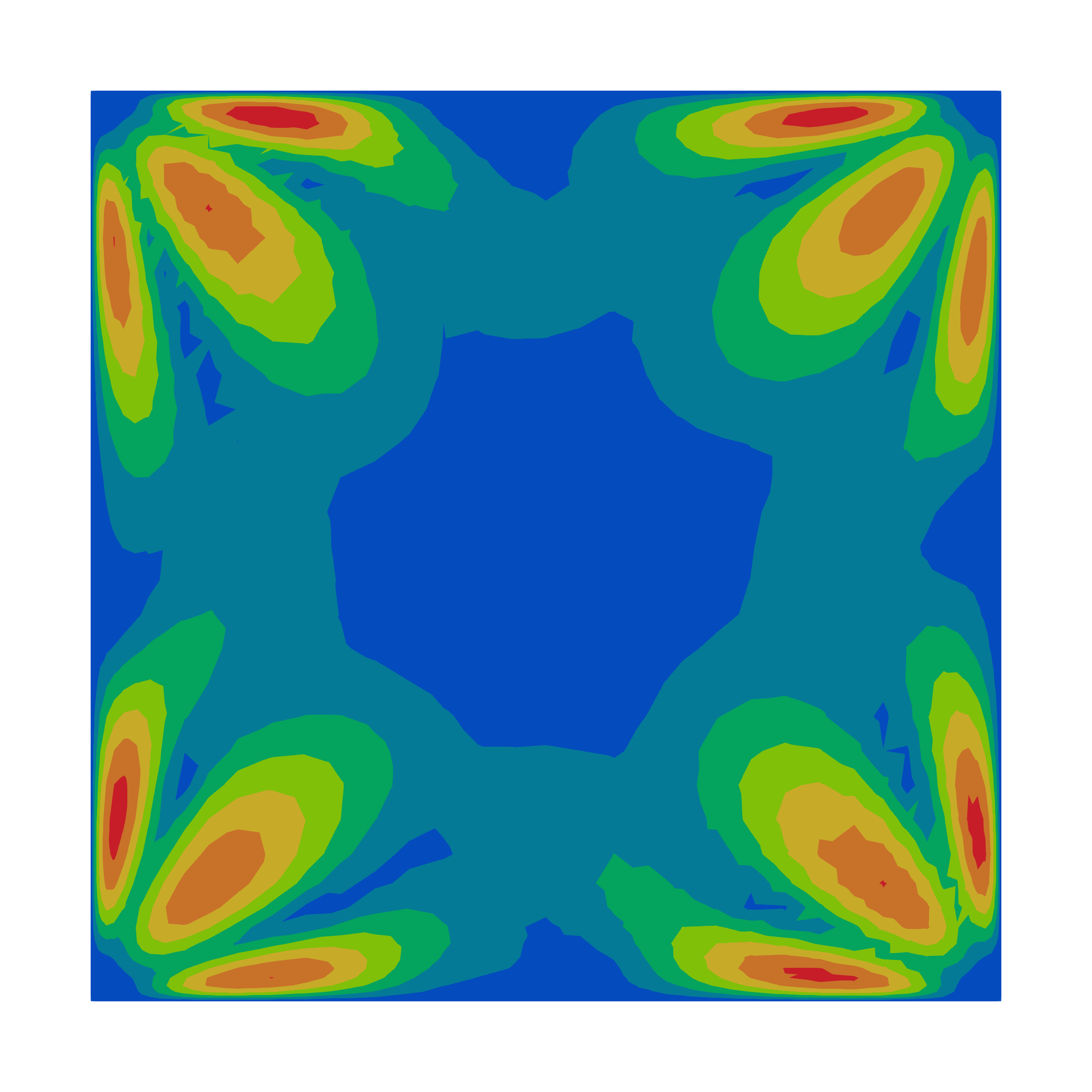}
    \caption{$\nut$ and $\rfvNLdim$ from DNS} \label{fig:secmot_DNS_Reb2900_OF}
    \end{subfigure}
    \hfill
    \begin{subfigure}[t]{0.240\textwidth}
    \includegraphics[width=\textwidth] {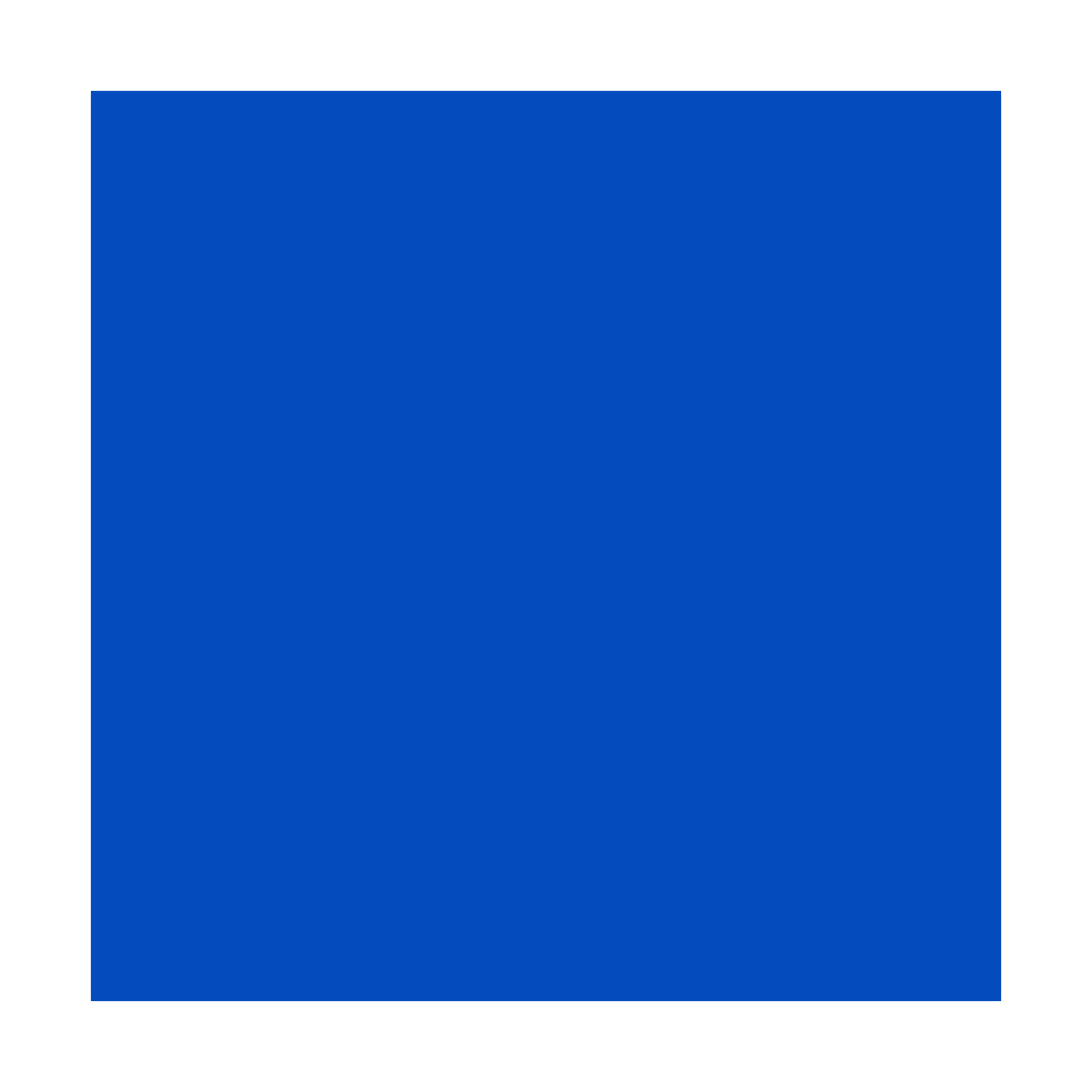}
    \caption{$\nut$ from DNS, $\rfvNLdim = \bm{0}$} \label{fig:sec_mot_DNSnut_norfvNL_Reb2900}
    \end{subfigure}
    \begin{subfigure}[t]{0.240\textwidth}
    \includegraphics[width=\textwidth] {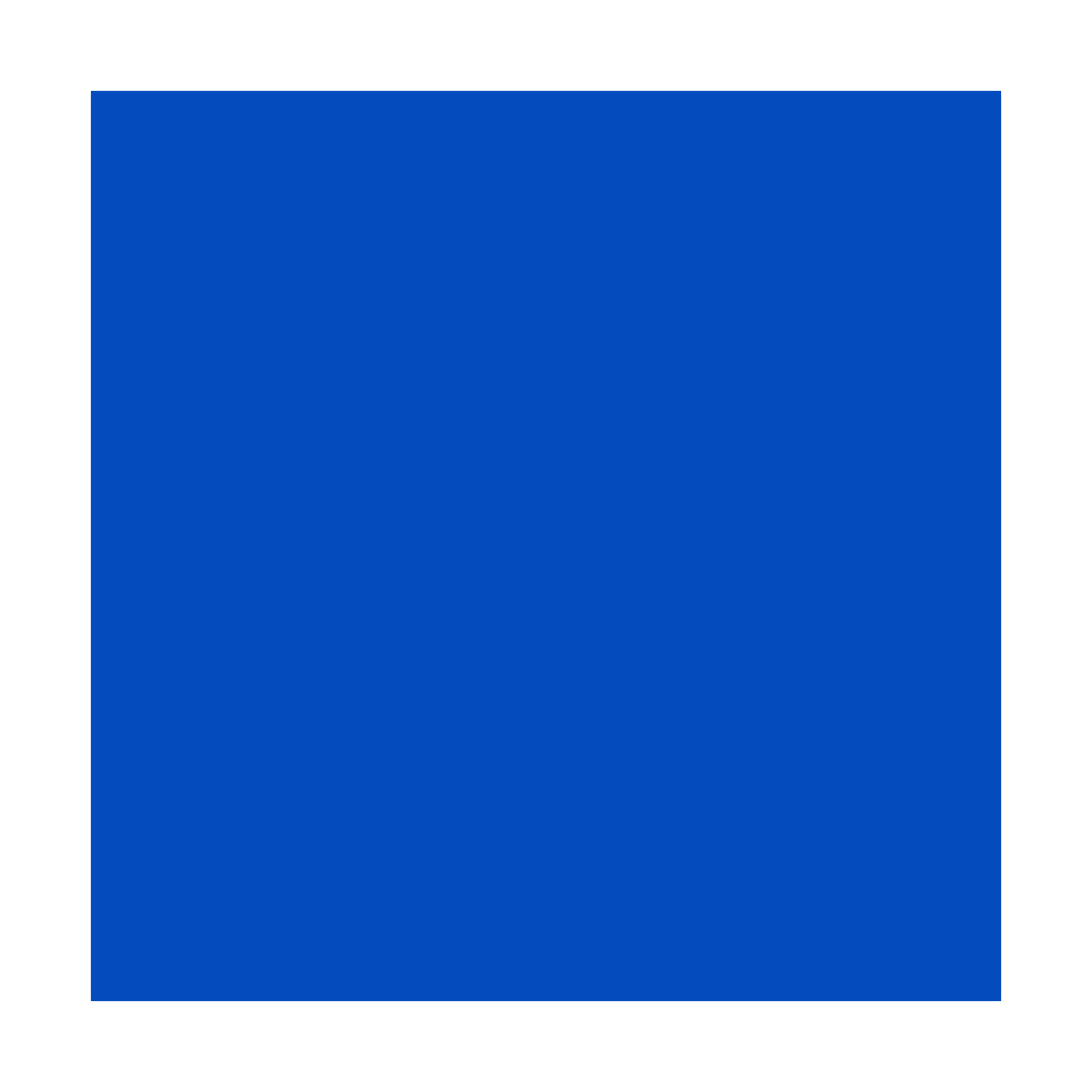}
    \caption{Baseline LS $k-\varepsilon$ model} \label{fig:secmot_kepsilon_Reb2900}
    \end{subfigure}
    \hfill
    \begin{subfigure}[t]{0.240\textwidth}
    \includegraphics[width=\textwidth] {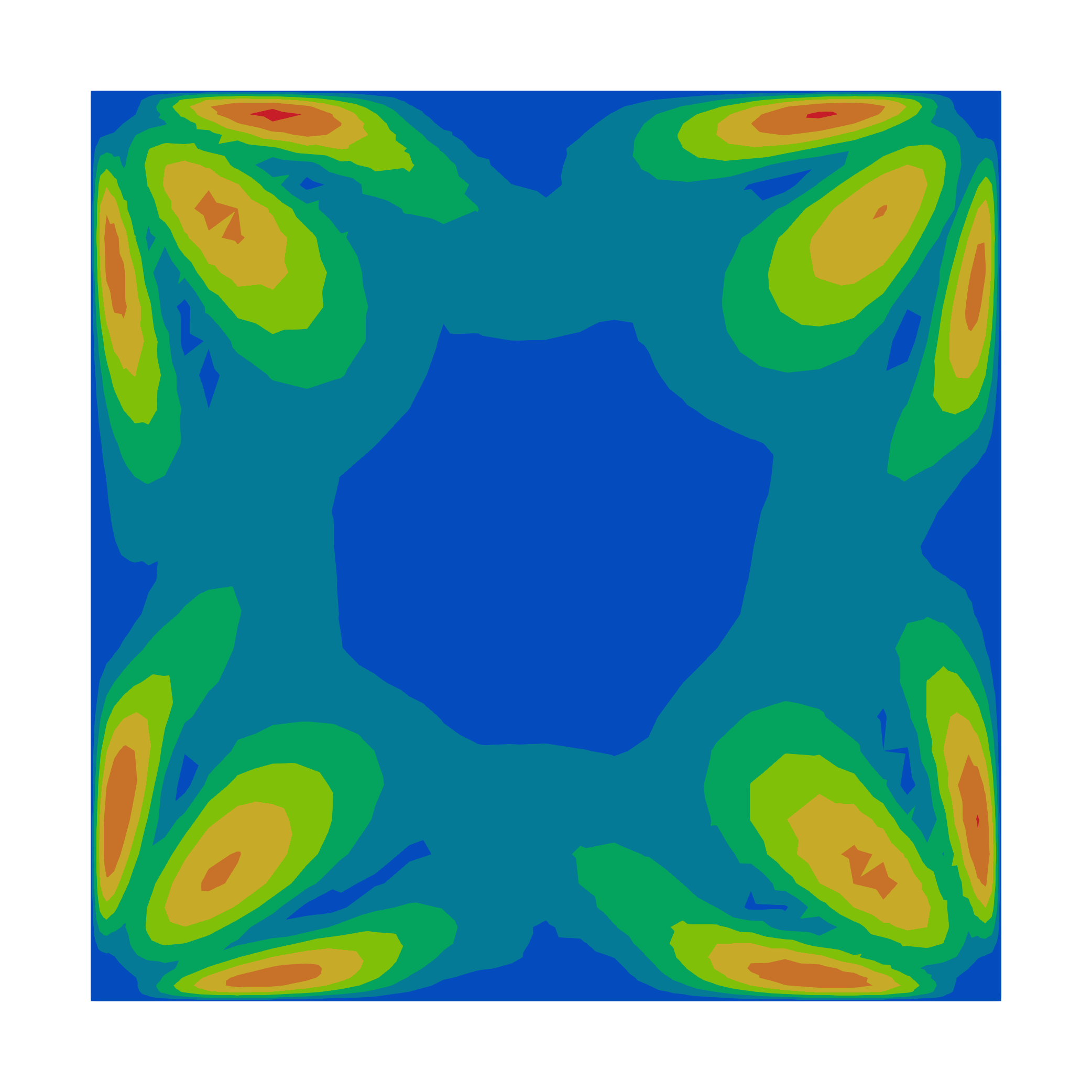}
    \caption{$\nut$ from LS $k-\varepsilon$ model, $\rfvNLdim$ from DNS} \label{fig:sec_mot_RANSnut_DNSrfvNL_Reb2900}
    \end{subfigure}
\caption{$I_s$ when fixing $\nut$ and/or $\rfvNLdim$ at $Re_b = 2900$. The same colour scale of Figure \ref{fig:sec_mot_DNS} has been used for all figures.}
\label{fig:sec_mot_fixing_nut_rfvNLdim}
\end{figure}
To understand how the $\nut$ discrepancy affects the predicted secondary motion, in Figure \ref{fig:sec_mot_fixing_nut_rfvNLdim} we show $I_s$ at $Re_b = 2900$ when using both $\nut$ and $\rfvNLdim$ from DNS, the DNS $\nut$ only, the LS $k-\varepsilon$ $\nut$ and, finally, the LS $k-\varepsilon$ $\nut$ and the DNS $\rfvNL$. All the DNS fields have been interpolated on the RANS mesh. We deduce that the secondary motion is correctly predicted only when including the $\rfvNLdim$ term in the RANS equations, while the accuracy of $\nut$ only slightly affects its magnitude. This behaviour is expected: standard linear models, that do not consider any $\rfvNLdim$ term, systematically do not predict secondary motion.\\

An accurate secondary motion prediction is important to correctly describe the streamwise velocity. In particular, Figure \ref{fig:Ux_fixing_nut_rfvNLdim} displays the corresponding predicted $u_x/U_b$ fields. When secondary motion is predicted, i.e., Figures \ref{fig:Ux_DNS_Reb2900_OF} and \ref{fig:Ux_RANSnut_DNSrfvNL_Reb2900}, the contour lines of $u_x/U_b$ have a square-like shape, while such contour lines have a circle-like shape when secondary motion is not predicted. This behaviour is again expected: the secondary motion transports flow from the central region, with higher $u_x/U_b$, toward the corners, thus increasing the streamwise velocity along the diagonals.
\begin{figure}
\centering
    \begin{subfigure}[t]{0.240\textwidth}
    \includegraphics[width=\textwidth] {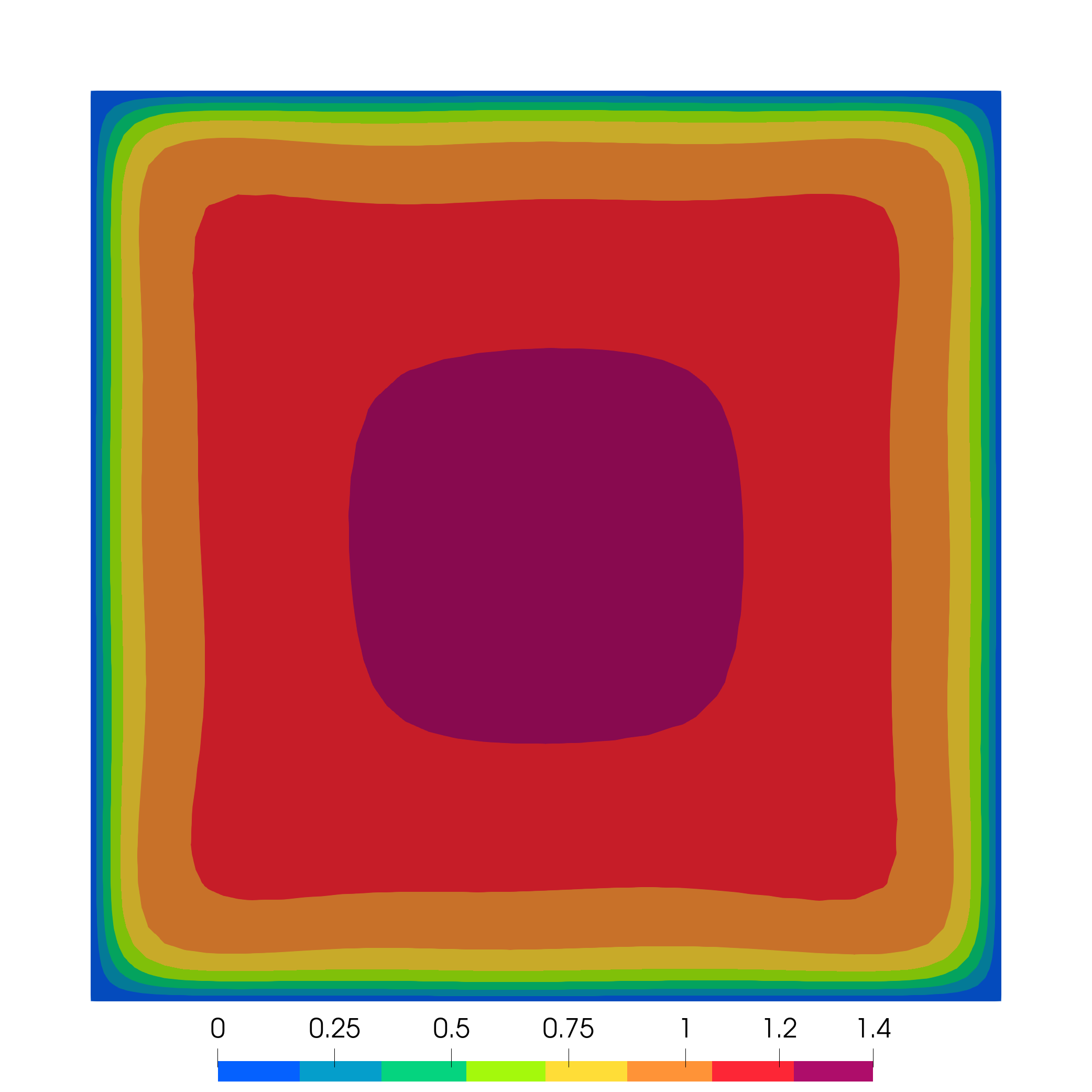}
    \caption{$\nut$ and $\rfvNLdim$ from DNS} \label{fig:Ux_DNS_Reb2900_OF}
    \end{subfigure}
    \hfill
    \begin{subfigure}[t]{0.240\textwidth}
    \includegraphics[width=\textwidth] {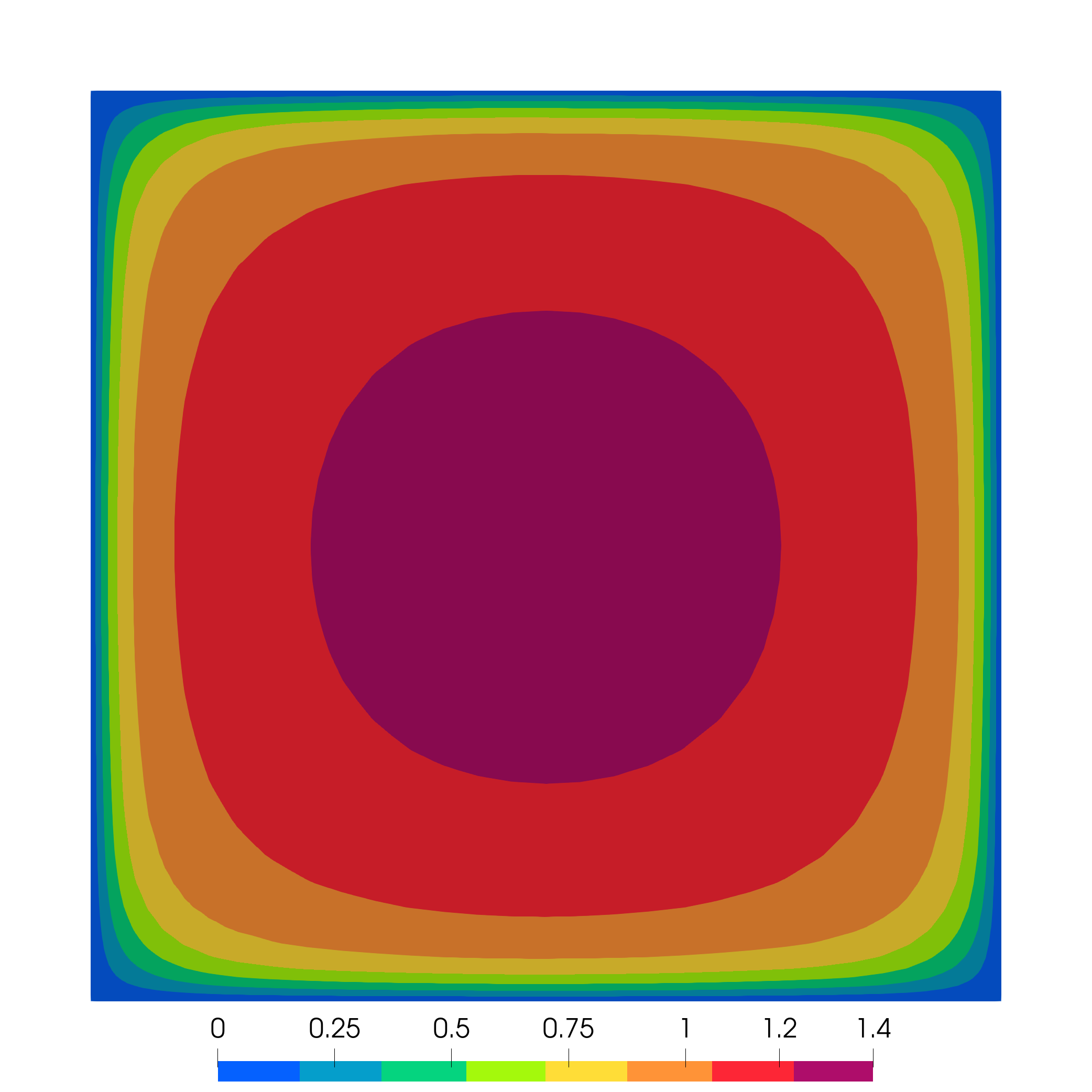}
    \caption{$\nut$ from DNS, $\rfvNLdim = \bm{0}$} \label{fig:Ux_DNSnut_norfvNL_Reb2900}
    \end{subfigure}
    \begin{subfigure}[t]{0.240\textwidth}
    \includegraphics[width=\textwidth] {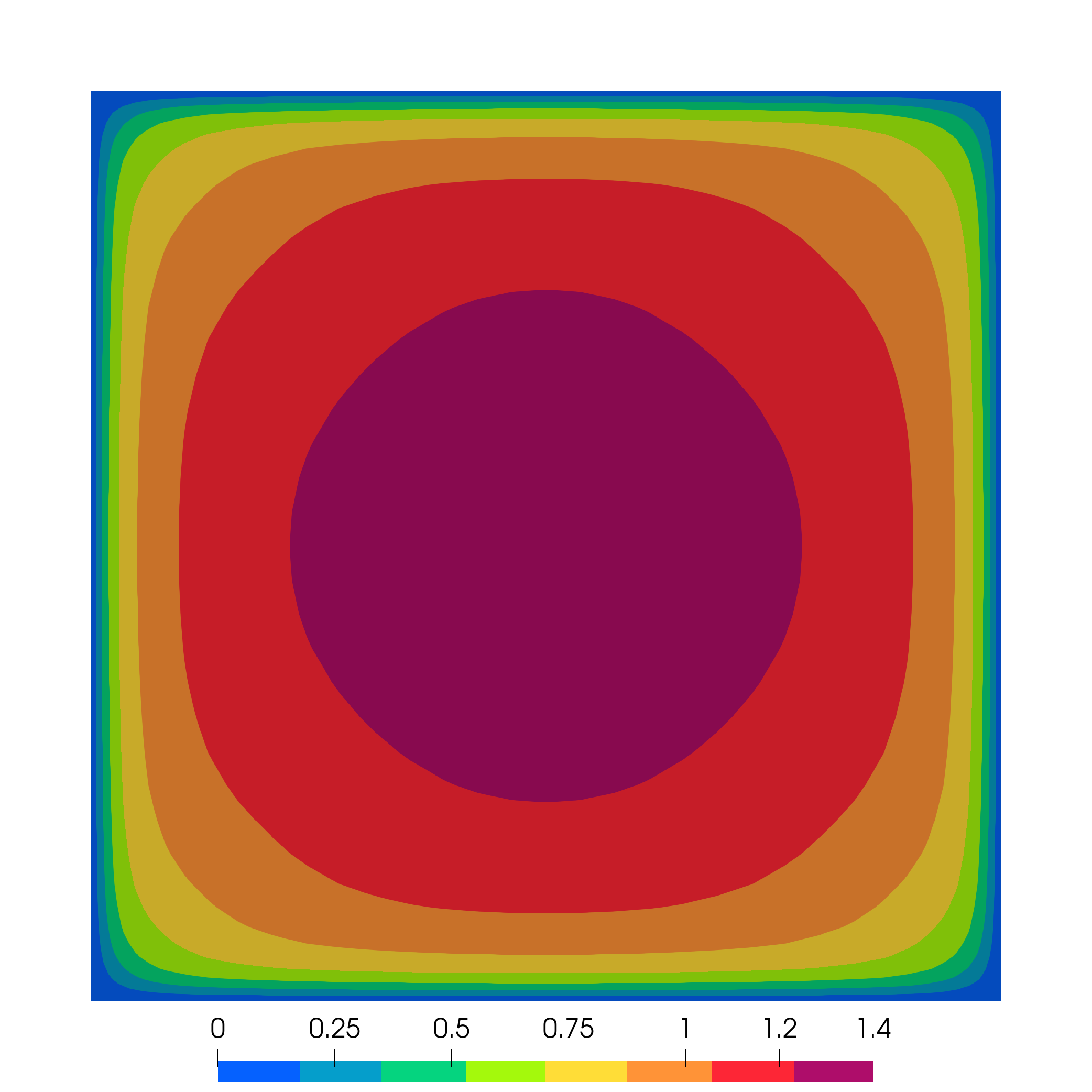}
    \caption{Baseline LS $k-\varepsilon$ model} \label{fig:Ux_kepsilon_Reb2900}
    \end{subfigure}
    \hfill
    \begin{subfigure}[t]{0.240\textwidth}
    \includegraphics[width=\textwidth] {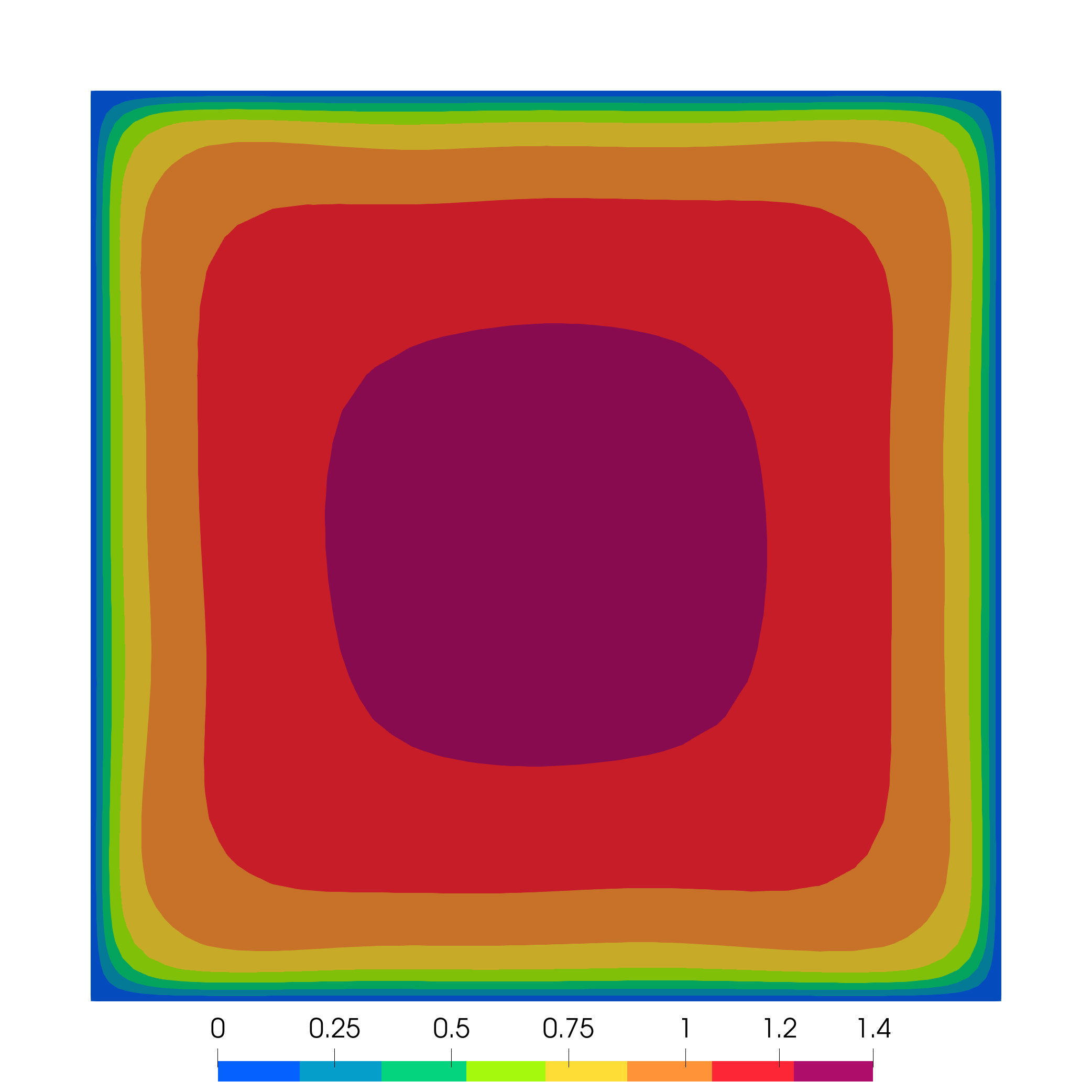}
    \caption{$\nut$ from LS $k-\varepsilon$ model, $\rfvNLdim$ from DNS} \label{fig:Ux_RANSnut_DNSrfvNL_Reb2900}
    \end{subfigure}
\caption{$u_x/U_b$ when fixing $\nut$ and/or $\rfvNLdim$ from DNS at $Re_b = 2900$.}
\label{fig:Ux_fixing_nut_rfvNLdim}
\end{figure}

\section{The regularization role for the $\nut$-VBNN model} \label{sec:reg_factor_FOM}
\vspace*{12pt}
It is well established in the ML community that the regularization factor plays a key role in the neural networks' prediction capability: a too low value can lead to overfitting, while a too high value can lead to underfitting  \cite{Goodfellow_2016}. Despite its importance, to the best of the authors' knowledge, no particular attention to this factor has been extensively commented on in the data-driven turbulence community when dealing with the duct flow case. To have a quantitative analysis on its relevance, we select for both $\lambda_{\tnut}$ and $\lambda_{\rfvNL}$ the values $\{0, 10^{-5}, 10^{-4}, 10^{-3}\}$ and compute the MSE for the predicted $\tnut$ and $\rfvNL$ with respect to DNS fields for each $Re_b$. Moreover, for each setting, we compute the average MSE across the fifteen performed trainings and compute the standard deviation (std). 
Figure \ref{fig:nut_err_vs_Reb_avg_std} depicts these quantities with respect to $Re_b$ in terms of the prediction of $\tnut$.  % and making it necessary to perform more training with fixed parameters.
%
%%%%%%%%%%%%%%%%%%%%%%%%%%%%%%%%
%                              %
%  nut avg and std comparison  %
%                              %
%%%%%%%%%%%%%%%%%%%%%%%%%%%%%%%%
%
\begin{figure}[h!]
\centering
    \begin{subfigure}[t]{0.32\textwidth}
    \includegraphics[width=\textwidth]{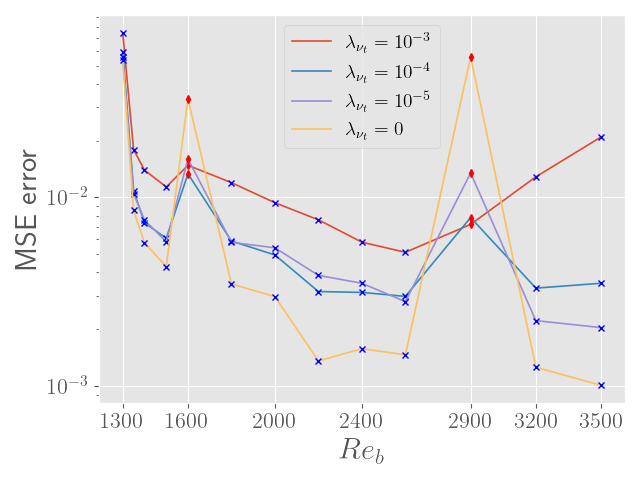} 
    \caption{$\all$, $\nut$ avg error} \label{fig:nut_err_vs_Reb_avg_all}
    \end{subfigure}
    \hfill
    \begin{subfigure}[t]{0.32\textwidth}
    \includegraphics[width=\textwidth]{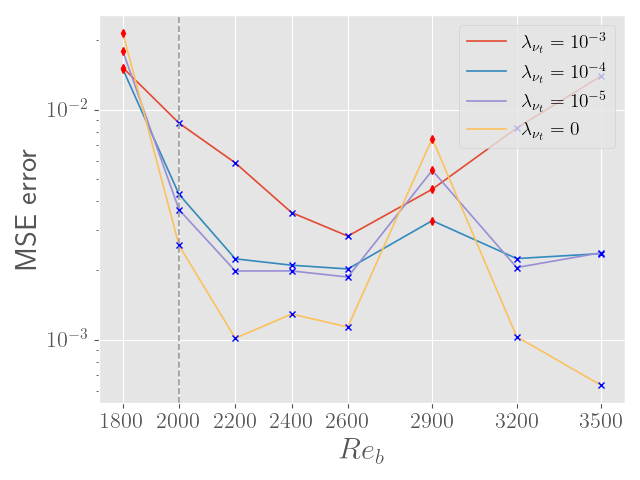}
    \caption{$\high$, $\nut$ avg error}
    \end{subfigure}
    \hfill
    \begin{subfigure}[t]{0.32\textwidth}
    \includegraphics[width=\textwidth]{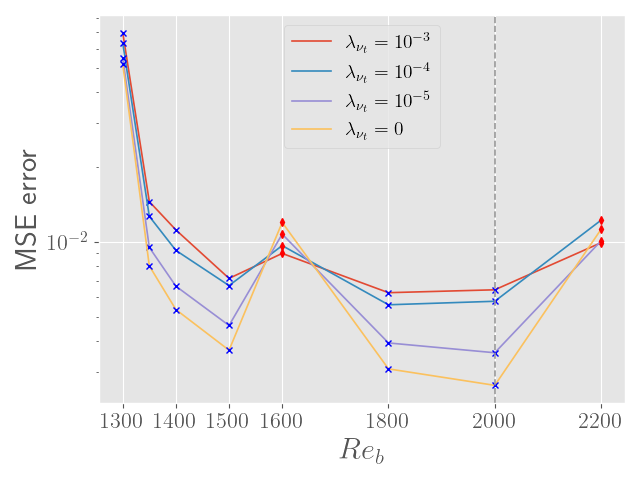}
    \caption{$\low$, $\nut$ avg error}
    \end{subfigure}
    
    \begin{subfigure}[t]{0.32\textwidth}
    \includegraphics[width=\textwidth]{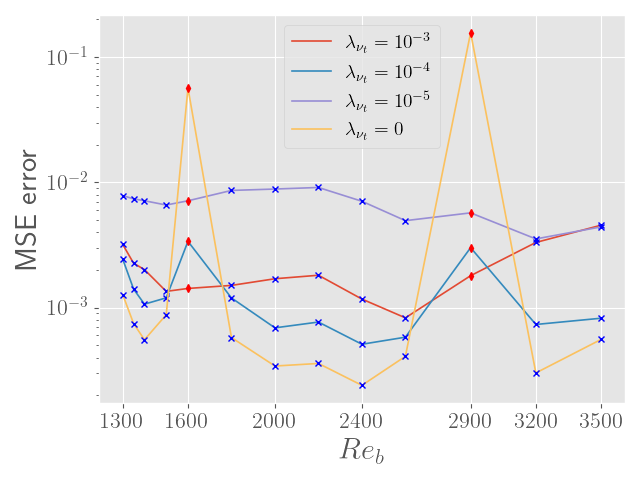}
    \caption{$\all$, $\nut$ std error}
    \end{subfigure}
    \hfill
    \begin{subfigure}[t]{0.32\textwidth}
    \includegraphics[width=\textwidth]{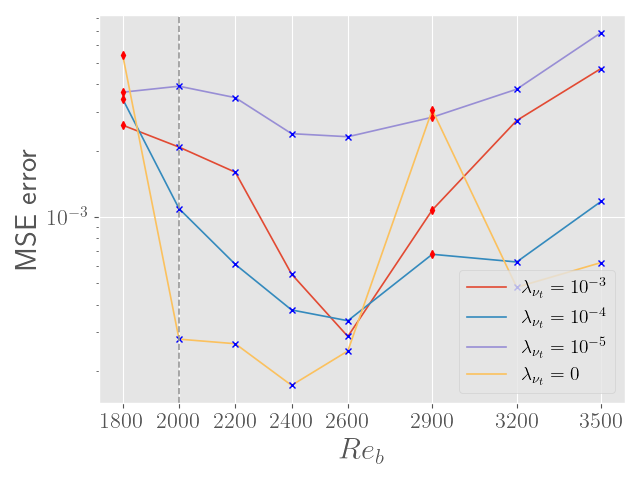}
    \caption{$\high$, $\nut$ std error}
    \end{subfigure}
    \hfill
    \begin{subfigure}[t]{0.32\textwidth}
    \includegraphics[width=\textwidth]{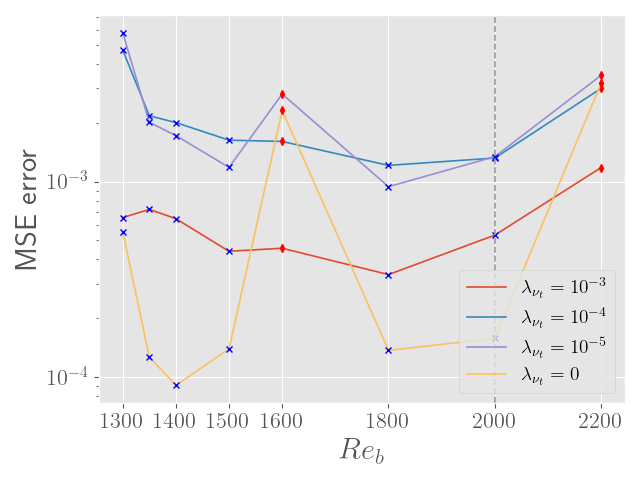}
    \caption{$\low$, $\nut$ std error}
    \end{subfigure}
    \caption{Comparison of average and std MSE with respect to $Re_b$ for the $\tnut$ prediction with different values of $\lambda_{\tnut}$.}\label{fig:nut_err_vs_Reb_avg_std}
\end{figure}
%
%%%%%%%%%%%%%%%%%%%%%%%%%%%%%%%%%%
%                                %
%  rfvNL avg and std comparison  %
%                                %
%%%%%%%%%%%%%%%%%%%%%%%%%%%%%%%%%%
%
\begin{figure}[h!]
\centering
    \begin{subfigure}[t]{0.32\textwidth}
    \includegraphics[width=\textwidth]{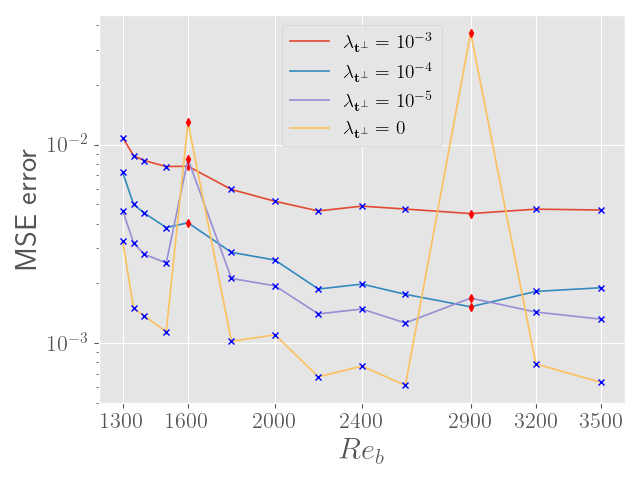}
    \caption{$\all$, $\rfvNL$ avg error}
    \end{subfigure}
    \hfill
    \begin{subfigure}[t]{0.32\textwidth}
    \includegraphics[width=\textwidth]{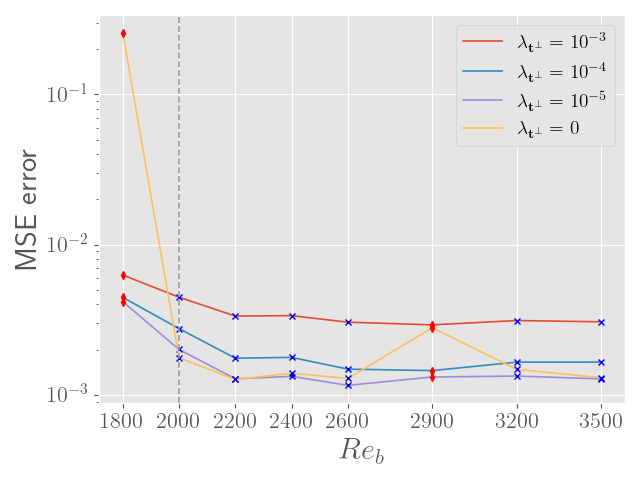}
    \caption{$\high$, $\rfvNL$ avg error}
    \end{subfigure}
    \hfill
    \begin{subfigure}[t]{0.32\textwidth}
    \includegraphics[width=\textwidth]{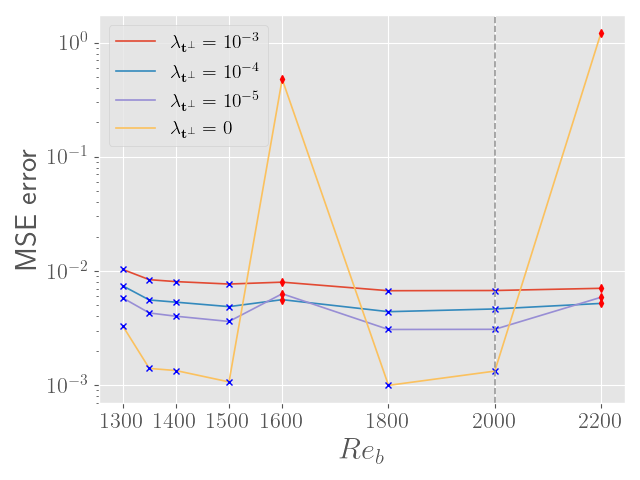}
    \caption{$\low$, $\rfvNL$ avg error}
    \end{subfigure}
    
    \begin{subfigure}[t]{0.32\textwidth}
    \includegraphics[width=\textwidth]{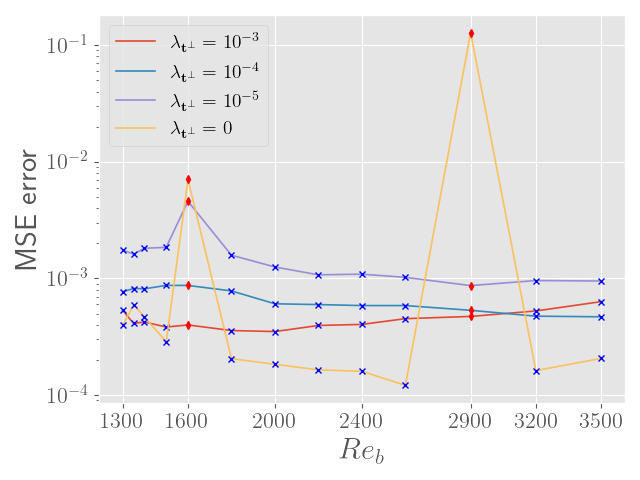}
    \caption{$\all$, $\rfvNL$ std error}
    \end{subfigure}
    \hfill
    \begin{subfigure}[t]{0.32\textwidth}
    \includegraphics[width=\textwidth]{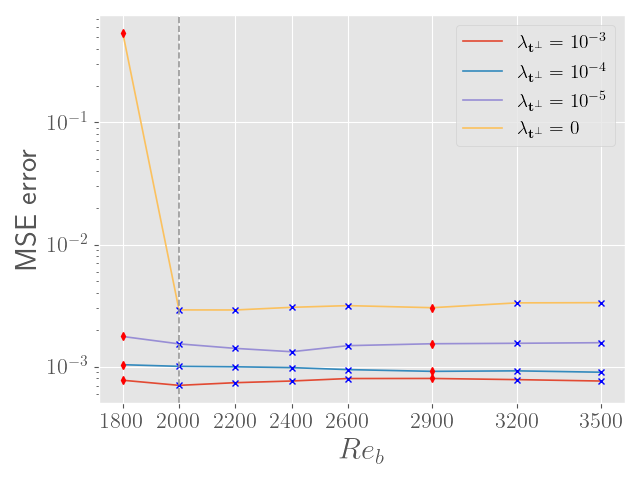}
    \caption{$\high$, $\rfvNL$ std error}
    \end{subfigure}
    \hfill
    \begin{subfigure}[t]{0.32\textwidth}
    \includegraphics[width=\textwidth]{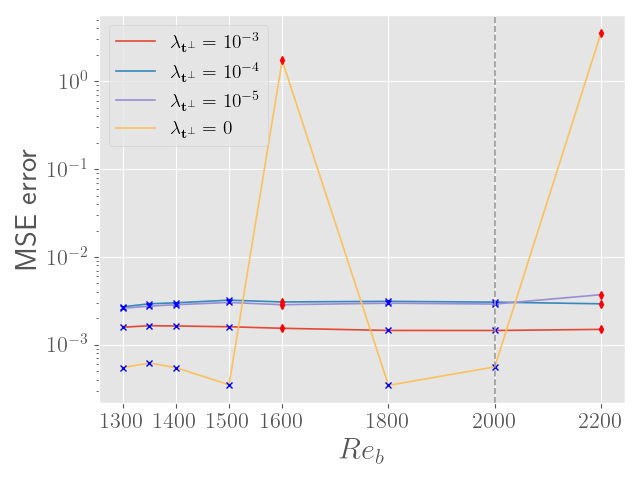}
    \caption{$\low$, $\rfvNL$ std error}
    \end{subfigure}  
    \caption{Comparison of average and std MSE with respect to $Re_b$ for the $\rfvNL$ prediction with different values of $\lambda_{\rfvNL}$.}\label{fig:rfvNL_err_vs_Reb_avg_std}
\end{figure}
As expected, the lower $\lambda_{\tnut}$, the lower the average error and the standard deviation when predicting for training $Re_b$ values. However, a lower training error does not necessarily translate to a lower test error. As a matter of fact, the larger testing errors are always obtained when no regularization is considered during the training. Furthermore, scarce regularization increases the testing error variability, too, and makes the testing ability of the single trained network less reliable.
The same analysis is performed for the $\lambda_{\rfvNL}$ effect on the inferred $\rfvNL$ field. Figure \ref{fig:rfvNL_err_vs_Reb_avg_std} depicts the mean and standard deviation of the MSE. Similar conclusions can be drawn.
To generate the $\nut$-VBNN enhanced simulations, for each $Re_b$ interval choice, we first select either $\lambda_{\tnut}$ or $\lambda_{\rfvNL}$ as the one with lowest average testing error.  Then, among the fifteen networks trained with the selected regularization factors, we pick the best neural network in terms of testing error.

\section{$\nut$-VBNN hyperparameters}\label{sec:nut_VBNN_hyper}
The hyperparameters for both the $\tilde{\nu}_t$ and $\rfvNL$ models are synthesised in Table \ref{tab:nut_vbnn_hyper}, where \texttt{n\_inp} denotes the number of inputs, \texttt{n\_out} the number of outputs, \texttt{n\_hid} the number of hidden layers, \texttt{n\_nodes} the number of nodes per hidden layer and \texttt{lr} the starting learning rate. Their values are consistent with the literature \cite{Berrone_2022,OBERTO2025_PH,Ling2016}. Except for the regularization factor, whose values are deeply investigated in Section \ref{sec:FOM_results} and Appendix \ref{sec:reg_factor_FOM}, all the hyperparameters are the same in $\all$, $\low$, and $\high$ settings. All networks are optimised using the Adam \cite{Adam} algorithm, using a starting learning rate that is decreased during the training in case of stagnation of the validation loss. Mini-batching is performed with batch size equal to sixty-four.
\begin{table}[h]
\centering
\caption{Hyperparameters for the $\nut$-VBNN neural networks.} 
\renewcommand{\arraystretch}{1.2}
\begin{tblr}{
  columns={halign=c},
  hline{1,1} = {-}{},
  hline{1,2} = {-}{},
  hline{1,4} = {-}{},
  vline{1,1} = {-}{},
  vline{1,2} = {-}{},
  vline{1,7} = {-}{},
}
field & \texttt{n\_inp} & \texttt{n\_out} & \texttt{n\_hid} & \texttt{n\_nodes} & \texttt{lr}  \\ 
$\tnut$ & 27 & 1 & 8 & 30 & 5e-3\\ 
$\rfvNL$ & 27 & 12 & 8 & 30 & 5e-3 \\ 
\end{tblr}
\label{tab:nut_vbnn_hyper}
\end{table}

\section{PODNN hyperparameters}  \label{sec:PODNN_hyper}
Table \ref{tab:podnn_hyper} shows the hyperparameters of the PODNN. The number of outputs depend on the number of predicted reduced coefficients and, consequently, it changes in the $\all$, $\low$ and $\high$ settings. The reported hyperparameter values have been selected after a trial-and-error stage. The networks are trained using the Adam algorithm with a scheduled learning rate that decreases during the training. Because of the limited amount of data, the training is carried out for a maximum of $1.5 \cdot 10^4$ epochs, depending on whether the validation loss function stagnates before or not.
\begin{table}[h]
\centering
\caption{Hyperparameters for the PODNN model. The same notation of Table \ref{tab:nut_vbnn_hyper} is used. Additionally, $\lambda$ denotes the regularization factor.} 
\renewcommand{\arraystretch}{1.2}
\begin{tblr}{
  columns={halign=c},
  hline{1,1} = {-}{},
  hline{1,2} = {-}{},
  hline{1,3} = {-}{},
  vline{1,1} = {-}{},
  vline{1,2} = {-}{},
  vline{1,8} = {-}{},
}
field & \texttt{n\_inp} & \texttt{n\_out} & \texttt{n\_hid} & \texttt{n\_nodes} & \texttt{lr} & $\lambda$  \\ 
$a$ & 1 & $\redred{r_u}$ & 4 & 50 & 1e-3 & 1e-3\\ 
\end{tblr}
\label{tab:podnn_hyper}
\end{table}

\end{document}